\documentclass[10pt]{article}
\pdfoutput=1

\usepackage{amsthm,amsfonts,amsmath}
\usepackage{stackrel,amssymb}
\usepackage{pxfonts}
\usepackage{euscript}
\usepackage{hyperref}
\usepackage{xcolor}
\usepackage{tcolorbox}

\usepackage{xypic}
\usepackage[all,2cell,arrow,matrix]{xy} \UseAllTwocells \SilentMatrices
\usepackage{graphicx}
\usepackage{verbatim}
\usepackage{leftidx}

\usepackage{epstopdf} % for pdflatex
\usepackage{enumerate}
\usepackage{bbold}

\usepackage{sectsty}
\sectionfont{\Large}
\newcommand\void[1]       {}

\newcommand{\leftrarrows}{\mathrel{\raise.75ex\hbox{\oalign{%
  $\scriptstyle\leftarrow$\cr
  \vrule width0pt height.5ex$\hfil\scriptstyle\relbar$\cr}}}}
\newcommand{\lrightarrows}{\mathrel{\raise.75ex\hbox{\oalign{%
  $\scriptstyle\relbar$\hfil\cr
  $\scriptstyle\vrule width0pt height.5ex\smash\rightarrow$\cr}}}}
\newcommand{\Rrelbar}{\mathrel{\raise.75ex\hbox{\oalign{%
  $\scriptstyle\relbar$\cr
  \vrule width0pt height.5ex$\scriptstyle\relbar$}}}}

\makeatletter
\def\leftrightarrowsfill@{\arrowfill@\leftrarrows\Rrelbar\lrightarrows}
\newcommand{\xleftrightarrows}[2][]{\ext@arrow 3399\leftrightarrowsfill@{#1}{#2}}
\makeatother

\theoremstyle{definition}

\newtheorem{thm}{Theorem}[section]

\newtheorem{cor}[thm]{Corollary}
\newtheorem{lem}[thm]{Lemma}

\newtheorem{conj}[thm]{Conjecture}
\theoremstyle{definition}
\newtheorem{defn}[thm]{Definition}
\newtheorem{expl}[thm]{Example}
\newtheorem{rem}[thm]{Remark}
\newtheorem{pthm}[thm]{Theorem$^{\mathrm{ph}}$}

\numberwithin{equation}{section}
\numberwithin{thm}{section}

\newcommand\nn             {\nonumber \\}
\newcommand\be            {\begin{equation}}
\newcommand\ee            {\end{equation}}
\newcommand\bea           {\begin{eqnarray}}
\newcommand\eea         {\end{eqnarray}}
\newcommand\bnu          {\begin{enumerate}}
\newcommand\enu          {\end{enumerate}}

\allowdisplaybreaks[1]

\newlength{\fighskip} \fighskip=2pt
\newlength{\figvskip} \figvskip=3pt

\newcommand{\pf}{\begin{proof}}
\newcommand{\epf}{\end{proof}}

\newcommand\Cb            {\mathbb{C}}
\newcommand\Nb            {\mathbb{N}}

\newcommand\Rb            {\mathbb{R}}
\newcommand\Zb            {\mathbb{Z}}
\newcommand\Z             {\mathfrak{Z}}

\newcommand\CA           {\EuScript{A}}
\newcommand\CB           {\EuScript{B}}
\newcommand\CC           {\EuScript{C}}
\newcommand\CD           {\EuScript{D}}
\newcommand\CE          {\EuScript{E}}
\newcommand\CF          {\EuScript{F}}

\newcommand\CL          {\EuScript{L}}
\newcommand\CM          {\EuScript{M}}
\newcommand\CN         {\EuScript{N}}
\newcommand\CO         {\EuScript{O}}

\newcommand\CQ         {\EuScript{Q}}

\newcommand\CX         {\EuScript{X}}
\newcommand\CY         {\EuScript{Y}}
\newcommand\CZ         {\EuScript{Z}}

\newcommand\CBs{{\EuScript{B}^\sharp}}
\newcommand\CCs{{\EuScript{C}^\sharp}}
\newcommand\CDs{{\EuScript{D}^\sharp}}

\newcommand\CMs{{\EuScript{M}^\sharp}}
\newcommand\CNs{{\EuScript{N}^\sharp}}

\newcommand\CXs{{\EuScript{X}^\sharp}}

\newcommand{\FZ}{\mathfrak{Z}}

\newcommand{\ising}{\mathbf{Is}}
\newcommand{\toric}{\mathbf{Toric}}

 \DeclareMathOperator{\Hom}{Hom}

 \DeclareMathOperator{\Id}{Id}
 \DeclareMathOperator{\id}{id}

 \DeclareMathOperator{\ev}{ev}

 \DeclareMathOperator{\fun}{Fun}

 \DeclareMathOperator{\Mod}{Mod}

\newcommand{\rev}{\mathrm{rev}}

\newcommand{\one}{\mathbf1}

\newcommand\bh{\mathbf{H}}

\newcommand\forget  {\mathbf{f}}
\newcommand\cl  {\mathrm{bulk}}
\newcommand\op {\mathrm{bdy}}

\begin{document}

\begin{center} \LARGE
A mathematical theory of gapless edges of 2d topological orders. Part I
\end{center}

\vskip 1em
\begin{center}
{\large
Liang Kong$^{a}$,\,
Hao Zheng$^{a,b}$\,
~\footnote{Emails:
{\tt  kongl@sustech.edu.cn, zhengh@sustech.edu.cn}}}
\\[1em]
$^a$ Shenzhen Institute for Quantum Science and Engineering,\\
and Department of Physics,\\
Southern University of Science and Technology, Shenzhen 518055, China 
\\[0.7em]
$^b$ Department of Mathematics, Peking University\\
Beijing 100871, China
\end{center}

\vskip 3.4em

\begin{abstract}
This is the first part of a two-part work on a unified mathematical theory of gapped and gapless edges of 2d topological orders. We analyze all the possible observables on the 1+1D world sheet of a chiral gapless edge of a 2d topological order, and show that these observables form an enriched unitary fusion category, the Drinfeld center of which is precisely the unitary modular tensor category associated to the bulk. This mathematical description of a chiral gapless edge automatically includes that of a gapped edge (i.e. a unitary fusion category) as a special case. Therefore, we obtain a unified mathematical description and a classification of both gapped and chiral gapless edges of a given 2d topological order. In the process of our analysis, we encounter an interesting and  reoccurring phenomenon: spatial fusion anomaly, which leads us to propose the Principle of Universality at RG fixed points for all quantum field theories. Our theory also implies that all chiral gapless edges can be obtained from a so-called topological Wick rotations. This fact leads us to propose, at the end of this work, a surprising correspondence between gapped and gapless phases in all dimensions. 
%We will study non-chiral gapless edges and 0d defects in Part II. 
\end{abstract}

%It was known that a 2d topological order is described by a pair $(\CC, c)$, where $\CC$ is a unitary modular tensor category (UMTC) and $c$ is the chiral central charge. 

\tableofcontents
\vspace{2cm}

\section{Introduction}
Topological phases of matter have attracted a lot of attentions in recent years among physicists because they go beyond Landau’s paradigm of phases and phase transitions (see a recent review \cite{wen4} and references therein). In this work and part II \cite{kz4}, we develop a unified mathematical theory of the gapped and gapless edges of 2d topological orders. Some results of this two-part work were announced in \cite{kz3} without providing the details. In Part I, we focus on chiral gapless edges. 

Throughout this work, we use ``$n$d'' to mean the spatial dimension and ``$n+$1D'' to mean the spacetime dimension, and by a ``2d topological order'', we mean an anomaly-free 2d topological order without symmetry. We use ``Theorem'' to represent a mathematical (rigorous) result and ``Theorem$^{\mathrm{ph}}$'' to summarize important physical (unrigorous) results.

\medskip
Topological orders are the universal classes of gapped local Hamiltonian lattice models at zero temperature. Since the system is gapped, correlation functions decay exponentially. In the long wave length limit, the only local observables are topological excitations. It was known that a 2d topological order is determined by its (particle-like) topological excitations uniquely up to $E_8$ states. Mathematically, the fusion-braiding properties of the topological excitations in a 2d topological order can be described by a unitary modular tensor category (UMTC) (see for example \cite{frs89,fg} and a review \cite[Appendix\,E]{kitaev}). Therefore, a 2d topological order is described mathematically by a pair $(\CC,c)$, where $\CC$ is the UMTC of topological excitations and $c$ is the chiral central charge. If a topological order $(\CC,0)$ admits a gapped edge, it is called a non-chiral topological order. The mathematical theory of gapped edges is known. More precisely, a gapped edge can be mathematically described by a unitary fusion category (UFC) $\CM$ such that its Drinfeld center $\FZ(\CM)$ coincides with $\CC$ \cite{kk,fsv,anyon}. The fact that the bulk phase is determined by an edge as its Drinfeld center is called the {\it boundary-bulk relation}. 

When a 2d topological order is chiral, it has topologically protected gapless edges \cite{halperin,wen1,moore-read,wen2} (see reviews \cite{wen2,wen3,nssfs} and references therein). Observables on a gapless edge is significantly richer than those on a gapped edge because gapless edge modes are described by a 1+1D conformal field theory (CFT) \cite{bpz,moore-seiberg}, the mathematical structures of which are much richer than that of a UFC \cite{segal,moore-seiberg,longo-rehren,huang-book}. It seems that a categorical description of a gapless edge as simple as that of a gapped edge is impossible. 
%As far as we known, no one has seriously considered if a categorical description of a gapless edge is possible. 

\smallskip
Nevertheless, in the last 30 years, the mathematical theory of boundary-bulk (or open-closed) CFT's has been successfully developed at least in three different approaches:
\bnu
\item the conformal-net approach (see \cite{longo-rehren,rehren1,rehren2,klm} and references therein), 
\item the 2+1D-TQFT approach (see \cite{fffs,fs1,frs1,fjfrs} and references therein), 
\item the vertex-operator-algebra approach (see \cite{huang-cft, osvoa,ffa,kong-cardy} and references therein). 
\enu
These mathematical developments have revealed a universal phenomenon: the mathematical structures of a boundary-bulk CFT can be split into two parts. 
\bnu
\item One part consists of a chiral algebra $V$ (or a conformal net), also called a vertex operator algebra (VOA) in mathematics (see for example \cite{ll}), such that the category $\Mod_V$ of $V$-modules is a modular tensor category \cite{huang-mtc}. 
\item The other parts are purely categorical structures, including certain algebras in $\Mod_V$ and an algebra in the Drinfeld center $\FZ(\Mod_V)$ of $\Mod_V$ (see Theorem\,\ref{thm:bcft-1} and \ref{thm:bcft-3}). 
\enu
This suggests that it might be possible to describe a chiral gapless edge of a 2d topological order $(\CC,c)$ by a pair $(V,\CXs)$, where $V$ is a VOA and $\CXs$ is a purely categorical structure that can be constructed from $\Mod_V,\CC$ and perhaps some additional categorical data. The main goal of this paper is to show that this is indeed possible. More precisely, the main result of this work, summarized in Theorem$^{\mathrm{ph}}$\,\ref{thm:main}, says that $\CXs$ is an $\Mod_V$-enriched unitary fusion category (Definition\,\ref{def:en-UFC}) satisfying some additional properties. 

\medskip

We explain the layout of this paper. In Section\,\ref{sec:gapped-edge}, we briefly review the mathematical theory of 2d topological orders and that of gapped edges. We emphasize previously overlooked details, such as the so-called spatial fusion anomalies, so that it makes our study of chiral gapless edge looks like a natural continuation. In Section\,\ref{sec:gapless-edge-1}, we carefully describe all possible observables (at a RG fixed point) on the 1+1D world sheet of a chiral gapless edge of a 2d topological order $(\CC,c)$. In particular, we show that the observables on the 1+1D world sheet of a chiral gapless edge include a family of topological edge excitations, 0+1D boundary CFT's \cite{cardy1,cardy2,cl}, 0D domain walls between boundary CFT's \cite{ffrs} and two kinds of fusions among domain walls. These boundary CFT's and domain walls are required to preserve a chiral symmetry given by a VOA $V$ such that $\Mod_V$ is assumed to be a UMTC. This symmetry condition allows us to describe all boundary CFT's, domain walls and their fusions as objects or morphisms in $\Mod_V$. As a consequence, all observables organize themselves into a single categorical structure $\CXs$ called an $\Mod_V$-enriched monoidal category. Therefore, a chiral gapless edge can be described by a pair $(V,\CXs)$. 

In order to further explore the additional hidden structures in $\CXs$ (in Section\,\ref{sec:gapless-edge-II}), we need nearly all important mathematical results of rational CFT's in last 30 years. These results are unknown to most working physicists especially to those in the field of condensed matter physics. A briefly review of these results is necessary. In Section\,\ref{sec:RCFT}, we review the mathematical theory of boundary-bulk rational CFT's in the VOA approach \cite{huang-cft,zhu,hl1,ll,huang-mtc,kong-cardy}. This mathematical theory is not only a rigorous version of the physical theory but also a reformulation in terms of new and efficient categorical language, which, together with a classification result, play a crucial role in this work. In particular, in Section\,\ref{sec:def-RCFT}, we recall a Segal-type mathematical definition of a boundary-bulk CFT. In Section\,\ref{sec:cf-RCFT}, we recall the classification theory of boundary-bulk rational CFT's \cite{longo-rehren,rehren1,rehren2,frs1,fjfrs,kong-cardy,kr2}. In Section\,\ref{sec:internal-hom}, we explain the notion of an internal hom. In Section\,\ref{sec:unitary-RCFT}, we discuss issues related to unitary CFT's and show that boundary CFT's and domain walls among them can all be constructed from internal homs, and summarize all useful results in Theorem$^{\mathrm{ph}}$\,\ref{thm:bcft-3}. 

After the preparation in Section\,\ref{sec:RCFT}, we are able to give a natural and explicit construction of chiral gapless edges in Section\,\ref{sec:can-edge}. It turns out that the enriched monoidal categories appearing there are only  special cases of the so-called canonical construction (see Theorem\,\ref{thm:SC}) \cite{MP}. In Section\,\ref{sec:general-edge}, we will construct more general gapless edges by fusing the gapless edges constructed in Section\,\ref{sec:can-edge} with gapped domain walls, or equivalently, by the so-called {\it topological Wick rotations}. We will leave a loophole of our reasoning in Section\,\ref{sec:general-edge} and fix it in Section\,\ref{sec:RG}. Interestingly, these mathematical constructions automatically include all the gapped edges as special cases. Therefore, we obtain a unified mathematical theory of both gapped and chiral gapless edges.

In Section\,\ref{sec:gapless-edge-II}, we continue our exploration of the additional hidden structures in $\CXs$. In particular, 
%in Section\,\ref{sec:underlying-cat}, we reveal the hidden relations between $\Mod_V$ and the underlying category $\CX$ of $\CXs$; and in Section\,\ref{sec:MP}, 
in Section\,\ref{sec:MP}, using the classification theory of boundary-bulk rational CFT's summarized in Theorem$^{\mathrm{ph}}$\,\ref{thm:bcft-3}, we reveal the hidden relations between $\Mod_V$ and the underlying category $\CX$ of $\CXs$, and show that $\CXs$ is an $\Mod_V$-enriched unitary fusion category (see Definition\,\ref{def:en-UFC}).  
%by using a mathematical result of Morrison and Penneys \cite{MP}. 
In Section\,\ref{sec:classification}, using the boundary-bulk relation \cite{kong-wen-zheng-2} and the results in \cite{kz2}, we obtain a precise and complete mathematical description and a classification theory of chiral gapless edges of 2d topological orders. This is the main result of this paper, and is summarized in Theorem$^{\mathrm{ph}}$\,\ref{thm:main}. In the process of deriving the classification result, we will also discuss and propose a definition of a phase transition between two gapless phases via topological Wick rotations. In Section\,\ref{sec:RG}, motivated by a recurring phenomenon in this work, we propose a very general principle: 
\begin{quote}
{\bf Principle of Universality at RG fixed points}: A physical theory at a RG fixed point always satisfies a proper universal property in the mathematical sense. 
\end{quote}
We use it to fix the last loophole of our reasoning introduced in Section\,\ref{sec:general-edge}. This principle also provides a mathematical formulation of the spatial fusion anomaly.

In Section\,\ref{sec:outlooks}, we provide some outlooks for the study of gapless boundaries of higher dimensional topological orders. In particular, we will propose a surprising correspondence between gapped phases and gapless phases in all dimensions. In Appendix, we provide the mathematical definitions of some categorical notions in the enriched settings. 

\medskip
It is worthwhile to point out what is new in this paper. The main result Theorem$^{\mathrm{ph}}$\,\ref{thm:main} was first announced in \cite{kz3} without providing any proofs. This paper contain many missing details and a  complete proof. All physical (mathematically unrigorous) arguments used in the proof are explicitly spelled out. They are the No-Go Theorem (see Section\,\ref{sec:edge-bcft}), Naturality Principle in physics (see Section\,\ref{sec:MP}), the generalization of the mathematically rigorous Theorem\,\ref{thm:bcft-2} to unitary cases (see Theorem${}^{\mathrm{ph}}$\,\ref{thm:bcft-3}), our definition of a purely edge phase transition (see Section\,\ref{sec:classification}) and Principle of Universality at RG fixed points (see Section\,\ref{sec:RG}). All the rest steps in the proof are mathematically rigorous. In this work, we also introduce a few brand new physical concepts for the first time, including spatial fusion anomaly, topological Wick rotation, a model-independent definition of a phase transition between gapless edges (see Section\,\ref{sec:classification}), Principle of Universality and Gapped-Gapless Correspondence (see Section\,\ref{sec:outlooks}).  

In Part II \cite{kz4}, we will develop a mathematical theory of non-chiral gapless edges and 0d defects on a gapless edge. We will also give explicit calculations of various dimensional reduction processes and a complete mathematical description of boundary-bulk relation including both gapped and gapless edges. It is also worthwhile to mention that our theory of gapless edge provides a mathematical description of the critical points of topological phase transitions on the edges of 2d topological orders. An example was explained in \cite{cjkyz}. 
%Throughout this paper, we adopt the convention to call ``a boundary'' of a 2d topological phase of matter as ``an edge''. The term ``a boundary'' is used for all dimensions. In particular, in this work, ``a boundary'' mainly refers to ``a 0d boundary of a 1d phase''  or ``the boundary condition of a 0+1D boundary CFT''.  

%\medskip
%\noindent {\bf Assumptions}: We assume that all UMTC's can be realized as the categories of modules over some VOA's. 

\medskip
\noindent {\bf Acknowledgement}: We would like to thank Ian Affleck, Meng Cheng, Terry Gannon, Xiao-Gang Wen and Wei Yuan for helpful discussions. We thank the referee for many important suggestions for improvement. Both LK and HZ are supported by the Science, Technology and Innovation Commission of Shenzhen Municipality (Grant Nos. ZDSYS20170303165926217 and JCYJ20170412152620376) and Guangdong Innovative and Entrepreneurial Research Team Program (Grant No. 2016ZT06D348), and by NSFC under Grant No. 11071134. LK is also supported by NSFC under Grant No. 11971219. HZ is also supported by NSFC under Grant No. 11871078.

\section{Gapped defects of 2d topological orders} \label{sec:gapped-edge}

In this section, we review some of the basic facts of 2d topological orders, including the categorical description of its particle-like topological excitations, that of gapped edges and the boundary-bulk relation.

\subsection{Particle-like topological excitations}

An $n$d topological order is called {\it anomaly free} if it can be realized by an $n$d local Hamiltonian lattice model, and is called {\it anomalous} if otherwise \cite{kong-wen}. %For example, two well-known gapped edges of the 2d toric code model are anomalous 1d topological orders \cite{bk}. 

\medskip
Before the discovery of fractional quantum Hall effect, a 2d gapped local Hamiltonian lattice model at zero temperature was viewed as ``trivial'' because all the correlation functions decay exponentially. It seems that there is no ``local observables" in the long wave length limit. It was realized latter, however, the ground state degeneracy (GSD) of such 2d systems on surfaces with non-trivial topology, such as spheres with holes, torus, higher genus surfaces (with edges), etc., are non-trivial and different for different systems \cite{taw,ntw,nw}. It means that there are different kinds of ``trivialness'' of such ``trivial'' systems, which are, therefore, non-trivial. 

Note that the GSD is a global observable \cite{ai}. But the notion of a phase or a topological order is a local concept that is defined on an open disk (in the infinite system size limit). It was realized later that the local observables that are responsible for the non-trivial global observable GSD are the particle-like topological excitations (or anyons).

\medskip
It is well known that all particle-like topological excitations form a mathematical structure called a unitary modular tensor category (UMTC) (see a review \cite[Appendix\,E]{kitaev}). The physical meanings of some (not all) ingredients of a UMTC $\CC$ are precisely the observables in the long wave length limit, and are explained below. 
\bnu

\item An object in $\CC$ represents a particle-like topological excitation (or an anyon). The tensor unit $\one_\CC$ represents the trivial topological excitation. A simple object represents a simple anyon, and a non-simple object, i.e. a direct sum of simple objects (e.g. $\oplus_i x_i$ for simple $x_i$), represents a composite anyon. 

\item For any two anyons $x$ and $y$, a morphism from $x$ to $y$ is an instanton, which is a localized defect on time axis (e.g. at $t=t_0$),  with two boundary conditions $x$ (for $t<t_0$) and $y$ (for $t>t_0$). If $x$ and $y$ are simple, then the space of instantons from the boundary condition $x$ to $y$ is given by $\hom_\CC(x,y)=\delta_{x,y}\Cb$. For two composite anyons $\oplus_n x_n$ and $\oplus_m y_m$, where $x_n, y_m$ are simple, we have
$$
\hom_\CC(\oplus_n x_n, \oplus_m y_m) = \oplus_{n,m} \delta_{x_n,y_m} \Cb. 
$$ 

\item The fusion of instantons along the time axis, called a temporal fusion, defines a composition of morphisms 
\begin{align} \label{eq:order-instantons}
\hom_\CC(y,z) \times \hom_\CC(x,y) &\rightarrow \hom_\CC(x,z) \nn
 (f,g) &\mapsto f\circ g
\end{align}
Our convention is that the time coordinate of $f$ is later than that of $g$. 
There is a distinguished morphism $1_x\in \hom_\CC(x,x)$, which is called the identity morphism, represents the trivial instanton. It is trivial in the sense that $f\circ 1_x = f$ and $1_y \circ f = f$ for $f\in \hom_\CC(x,y)$.

\item The tensor product functor $\otimes: \CC \times \CC \to \CC$ defined by $(x,y) \mapsto x\otimes y$ describes the fusion between two anyons in spatial dimensions, called a spatial fusion. The trivial anyon is denoted by $\one_\CC$. We have $\one_\CC \otimes x \simeq x \simeq x \otimes \one_\CC$. 

\item The spatial fusion $\otimes$ of anyons also induces a spatial fusion of instantons. The naive guess of the space of instantons after the fusion is $\hom_\CC(x,y) \otimes_\Cb \hom_\CC(x',y')$. But this naive guess is wrong because an anyon, viewed as a 0d topological order, is anomalous. The correct one is $\hom_\CC(x\otimes y, x'\otimes y')$. In general, there is a map (induced from the universal property of $\hom_\CC(x\otimes y, x'\otimes y')$)
\be \label{eq:otimes-on-morphisms}
\hom_\CC(x,y) \otimes_\Cb \hom_\CC(x',y') \xrightarrow{\otimes} \hom_\CC(x\otimes y, x'\otimes y') 
\ee
for $x,x',y,y'\in\CC$. Its failure of being an isomorphism is called {\it spatial fusion anomalies}. This is a recurring phenomenon in this work, and will be discussed in Section\,\ref{sec:RG}. It is also worthwhile to point out that the spatial fusion anomaly vanishes when $x=y=x'=y'=\one_\CC$. This result has a non-trivial analogue for chiral gapless edges (see Remark\,\ref{rem:fusion-anomaly-free}).

\item The creation and annihilation of a particle-antiparticle pair are described by the duality morphisms 
\be \label{eq:duality-maps}
u_x: \one_\CC \to x\otimes x^\ast, \quad \quad \quad v_x: x^\ast \otimes x \to \one_\CC, 
\ee 
where $x^\ast$ is the anti-particle of $x$, satisfying some natural properties. Mathematically, this amounts to the rigidity of a UMTC. 

\item Since the associated Hamiltonian model is gapped, one can adiabatically move one anyon around another one. This adiabatic movement defines a braiding isomorphism $c_{x,y}: x\otimes y\to y\otimes x$ for $x,y\in \CC$. The composed morphism $x\otimes y\xrightarrow{c_{x,y}} y\otimes x \xrightarrow{c_{y,x}} x\otimes y$ is called a double braiding, which amounts to adiabatically moving anyon $x$ around $y$ (say clockwise) in a full circle. 
\enu
The coherence conditions satisfied by the above data are all physically obvious. 
The non-degeneracy condition of the double braiding of a UMTC is stated as follows: 
\void{
\bnu
\item[$(\bullet)$] The following matrix (called $S$-matrix), where $i,j$ are simple objects in $\CC$,
\be  
 s_{i,j}   ~=~  \quad
\raisebox{-30pt}{
  \begin{picture}(110,65)
   \put(0,8){\scalebox{.65}{\includegraphics{pic-S-ij-eps-converted-to.pdf}}}
   \put(0,8){
     \setlength{\unitlength}{.75pt}\put(-18,-19){
     \put( 90, 48)       {\scriptsize $ i $}
     \put( 50, 48)      {\scriptsize $ j $}
     }\setlength{\unitlength}{1pt}}
  \end{picture}}
  \quad\quad
  \mbox{is non-degenerate.}
\ee
\enu
This condition is equivalent to the following condition: 
}
\bnu
\item[$(\star)$] For a given simple anyon $x$, if its double braiding with all anyons $y$ (including $x$ itself) is trivial, i.e. $(x\otimes y \xrightarrow{c_{x,y}} y\otimes x \xrightarrow{c_{y,x}} x\otimes y)=\id_{x\otimes y}$, then $x$ is the trivial anyon, i.e. $x \simeq \one_\CC$.  
\enu
The condition $(\star)$ is an anomaly-free condition, which says that all anyons in a topological order should be able to detect themselves via double braidings. If a system of anyons does not satisfy the condition $(\star)$, it must represent an anomalous 2d topological order, which must be a gapped boundary of a non-trivial 3+1D topological order \cite{kong-wen}. 

\begin{rem}
The space of instantons $\hom_\CC(x,y)$ has another physical meaning. It is also the space of ground states of the topological order defined on a 2-sphere together with two anyons $x^\ast, y$. Two physical meanings of the same hom-space is a manifestation of the state-field correspondence of a TQFT or a topological order. 
\end{rem}

For each UMTC $\CC$, one can define a topological central charge $c_\CC^{\mathrm{top}}$, which is defined modulo $8$. We use $\overline{\CC}$ to denote the same monoidal category as $\CC$ but equipped with the braidings given by the anti-braidings in $\CC$. $\overline{\CC}$ is also a UMTC. It was known that the Drinfeld center $\FZ(\CC)$ of $\CC$ is given by $\CC \boxtimes \overline{\CC}$ \cite{mueger2}, where $\boxtimes$ is the Deligne tensor product. The simplest UMTC is the category of finite dimensional Hilbert spaces, denoted by $\bh$.  

\medskip
In the long wave length limit, are there any other physical observables in a 2d topological order? It turns out that there is a special state called the $E_8$-state (see for example \cite{kitaev2}), which is a non-trivial topological order but has no non-trivial anyon. It was known that its 1d edge is gapless and is given by the $E_8$ chiral conformal field theory of central charge $c=8$. It was generally accepted that a 2d topological order is completely determined by its anyons up to the $E_8$-states (i.e. $E_8^{\boxtimes N}$) (see a review \cite{wen5}). We summarize this well-known result below. 
\begin{quote}
A 2d (anomaly-free) topological order can be described mathematically by a pair $(\CC, c)$, where $\CC$ is a UMTC and $c$ is the chiral central charge such that $c_\CC^{\mathrm{top}}=c\, (\mathrm{mod}\,\, 8)$. 
\end{quote}
The pair $(\bh,0)$ describes the trivial 2d topological order. 

\void{
\begin{rem} \label{rem:fh-1}
%A braided monoidal category can be viewed as an $E_2$-algebra or a 2-disk algebra in the 2-category of categories. Therefore, a UMTC $\CC$ describes observables on an open 2-disk (in the large size limit) of a 2d topological order in the long wave length limit. 
Global observables on a closed 2d manifold $\Sigma$ can be obtained by integrating observables on open 2-disks, i.e. $\int_\Sigma \CC$, which is rigorously defined by the theory of factorization homology \cite{}, and is given by
$\int_\Sigma \CC = (\bh, u_\Sigma)$ (see \cite{ai}), where $u_\Sigma$ is a distinguished object in $\bh$, i.e. a finite dimensional Hilbert space that gives the ground state degeneracy (GSD). 
\end{rem}
}

\subsection{1d gapped edges} \label{sec:gapped-edge-2}

%Gapped 1d topological defects in a 2d topological order include gapped edges and gapped domain wall between two potentially different 2d topological orders. By the folding trick, a gapped domain wall between two 2d topological orders $(\CC,c)$ and $(\CD,c)$, where the two chiral central charges are necessarily equal, is the same thing as a gapped edge of the 2d topological order $(\CC\boxtimes \overline{\CD}, 0)$ consisting of decoupled two layers of topological orders. Mathematically, the stacking of two layers of 2d topological orders is described by the Deligne tensor product denoted by $\boxtimes$. Therefore, we will only discuss gapped edges in this subsection. 

%In this case, $\CC$ is given by the Drinfeld center of a unitary fusion category (UFC) $\CM$, i.e. $\CC=\FZ(\CM)$. 

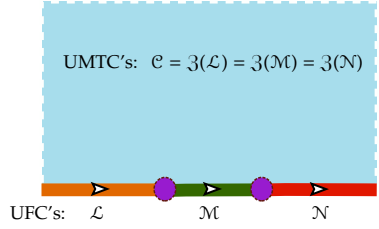
\begin{figure}
$$
\raisebox{-0pt}{
  \begin{picture}(95,80)
   \put(-20,10){\scalebox{1.2}{\includegraphics{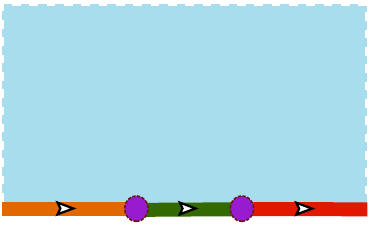}}}
   \put(-20,10){
     \setlength{\unitlength}{.75pt}\put(-18,-19){
     \put(3,12) {\scriptsize UFC's:}
     \put(43, 12)       {\scriptsize $\CL$}
     \put(155, 12)     {\scriptsize $ \CN $}
     \put(98, 12)     {\scriptsize $ \CM $}
     %\put(75,37) {\scriptsize $(\CX,x)$}
     %\put(124,37) {\scriptsize $(\CY,y)$}
     \put(30, 88)     {\scriptsize UMTC's:\,\, $ \CC=\FZ(\CL)=\FZ(\CM)=\FZ(\CN)$}
          }\setlength{\unitlength}{1pt}}
  \end{picture}}
$$
\caption{This picture depicts a 2d topological order $(\CC,0)$, where $\CC$ is a UMTC, with three gapped edges given by UFC's $\CL,\CM,\CN$. The 2d bulk is oriented as the usual $\Rb^2$ with the normal direction pointing out of the paper in readers' direction. The arrows indicate the induced orientation on the edge. %The 0d domain wall between $\CL$ and $\CM$ is described by a pair $(\CX,x)$, where $\CX$ is a $\CL$-$\CM$-bimodule category and $x$ is a distinguished object in $\CX$. The pair $(\CY,y)$ is similar.
}
\label{fig:bulk=center-1}
\end{figure}

A 2d topological order $(\CC,0)$ admitting a gapped edge is called {\it a non-chiral 2d topological order}. As depicted in Figure\,\ref{fig:bulk=center-1}, the bulk phase $(\CC,0)$ might have several different gapped edges, each of which represents a 1d anomalous topological order.  Such an anomalous 1d topological order can be described mathematically by a unitary fusion category (UFC) $\CL$, in which 
\bnu
\item an object represents a particle-like topological edge excitation; 
\item for topological edge excitations $x,y\in \CL$, the space of morphisms $\hom_\CL(x,y)$ is the space of instantons (or localized defects on time axis) with boundary conditions $x$ and $y$;
\item the composition of two morphisms describes the fusion of two instantons along the time axis;  
\item the fusion product $\otimes$ in $\CL$ represents the fusion of two such edge excitations in the spatial dimension according to the orientation of the edge. $\one_\CL$ denote the trivial topological edge excitation, i.e. $\one_\CL \otimes x \simeq x \simeq x\otimes \one_\CL$. 

\item This fusion also induces a fusion of two instantons in the spatial dimension. Again, the naive guess $\hom_\CL(x,y) \otimes_\Cb \hom_\CL(x',y')$ for the space of instantons after the fusion is wrong. The correct one is $\hom_\CL(x\otimes x', y\otimes y')$. In general, there is a map 
\be \label{map-otimes}
\hom_\CL(x,y) \otimes_\Cb \hom_\CL(x',y') \xrightarrow{\otimes} \hom_\CL(x\otimes x', y\otimes y'),  
\ee
for $x,x',y,y'\in \CL$. This map is not an isomorphism in general. Its failure of being an isomorphism is an indication of spatial fusion anomaly. This phenomenon is a recurring theme of this work. We will further explore this phenomenon as a special case of a more general principle in Section\,\ref{sec:RG}. 
\enu
Note that there is no braiding because these topological excitations are restricted on the edge, thus cannot be braided. 

%\begin{rem} A monoidal category can be viewed as an $E_1$-algebra or a 1-disk algebra in the 2-category of categories. Therefore, a UFC $\CL$ describes observables on an open 1-disk (in the large size limit) of a 1d topological order in the long wave length limit. The Drinfeld center of a UFC is also called the center of an $E_1$-algebra. \end{rem}

\medskip
The relation between the boundary phase and the bulk phase, or simply boundary-bulk relation, was known \cite{kk, anyon}. 
\begin{quote}

{\bf Boundary-bulk relation}: The 2d bulk phase is uniquely determined by the anomalous 1d topological order on its boundary (or edge), and the UMTC $\CC$ describing the 2d bulk phase is given by the Drinfeld center of the UFC $\CL$ (or $\CM,\CN$) describing a 1d edge (as depicted in Figure\,\ref{fig:bulk=center-1}), i.e. $\CC\simeq\FZ(\CL)\simeq\FZ(\CM)\simeq\FZ(\CN)$. 

\end{quote}
Mathematically, different gapped edges $\CL, \CM, \CN$ that share the same bulk or Drinfeld center if and only if they are Morita equivalent as UFC's \cite{mueger2,eno2008}. 

\medskip
Importantly, although a UFC $\CL$ determines an anomalous 1d topological order, it does not fix a gapped edge of a 2d topological order completely. The same UFC might be realized as different edges of the same bulk. For example, the $\Zb_2$ 2d topological order has two different gapped edges \cite{bk}, both of which realize the same 1d anomalous topological order defined by the UFC $\mathrm{Rep}(\Zb_2)$. 

The additional physical data that is needed to uniquely determine the edge is the information of how anyons in the bulk can be fused onto the edge. This information is given by a monoidal functor $L: \CC \to \CL$ which factors through the forgetful functor $\forget: \FZ(\CL) \to \CL$, i.e. there exists a braided monoidal functor $L': \CC \to \FZ(\CL)$ such that the following diagram
\be \label{diag:central-functor}
\xymatrix{ \CC \ar[r]^{L'}  \ar[dr]_L& \FZ(\CL) \ar[d]^\forget \\
& \CL
}
\ee
is commutative. Such a functor $L$ is called a central functor \cite{dmno}. 

\begin{rem}
The mathematical theory of 0d domain walls between different gapped edges is also known \cite{kk,anyon,ai}. 
\end{rem}

\subsection{What happens to gapless edges?} \label{sec:gapped-to-gapless}

\begin{figure}
$$
  \raisebox{-0pt}{
  \begin{picture}(95,80)
   \put(-20,10){\scalebox{1.2}{\includegraphics{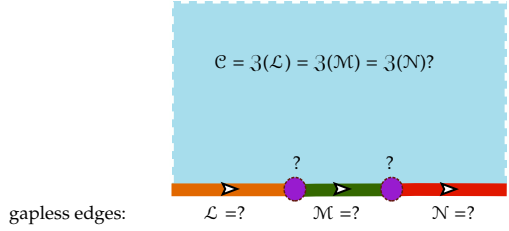}}}
   \put(-20,10){
     \setlength{\unitlength}{.75pt}\put(-18,-19){
     \put(-63,12) {\scriptsize gapless edges:}
     \put(35, 12)       {\scriptsize $ \CL=? $}
     \put(150, 12)     {\scriptsize $ \CN=? $}
     \put(90, 12)     {\scriptsize $ \CM=? $}
     \put(80,37)  {\scriptsize $ ? $}
     \put(127,37)  {\scriptsize $ ? $}
     \put(40, 88)     {\scriptsize $ \CC=\FZ(\CL)=\FZ(\CM)=\FZ(\CN)?$}
          }\setlength{\unitlength}{1pt}}
  \end{picture}}
$$
\caption{Obvious questions about gapless edges of a 2d chiral topological order.}
\label{fig:gapless=?}
\end{figure}

We have seen that we have a beautiful mathematical theory of gapped edges of 2d topological orders. Ironically,  experimentally discovered 2d topological orders, such as those discovered in quantum Hall effects, all have topologically protected gapless edges. The natural question is whether there is a similar mathematical theory for gapless edges.  As depicted in Figure\,\ref{fig:gapless=?}, this question contains at least the following three parts: 
\bnu
\item What is the mathematical description of a gapless edge of $(\CC,c)$? 
\item Does the boundary-bulk relation, i.e. bulk = the center of the edge, still hold? 
\item What is the mathematical description of a 0d gapless domain wall between two gapless 1d edges? 
\enu
In \cite{kong-wen-zheng-2}, using a very formal argument, we have shown that the bulk topological order (in any dimensions) should be given by the center of its boundary, i.e. 
$$\CC\simeq\FZ(\CL)\simeq\FZ(\CM)\simeq\FZ(\CN).$$ 
regardless whether the boundary is gapped or gapless \cite[Remark\, 5.7]{kong-wen-zheng-2}, and regardless what the mathematical description of a gapless boundary is. 

\begin{rem}
A reader might be puzzled by the above statement. How could one know the notion of the center of certain algebraic object before we have a complete mathematical definition of the algebraic object. This is because the notion of center is defined by its universal property, which only depends on the notion of a morphism between two such algebraic objects. In \cite{kong-wen-zheng-2}, the notion of a morphism between two anomalous $n$d topological orders was introduced without knowing how to define an $n$d topological order. 
\end{rem}

This result \cite[Remark\, 5.7]{kong-wen-zheng-2} convinced us that there should be a unified mathematical theory of gapped and gapless edges of 2d topological orders. We will answer the first question for chiral gapless edges in this work, and answers the second and third questions in \cite{kz4}.

\begin{rem}
The result of \cite[Remark\, 5.7]{kong-wen-zheng-2} and the success of this work have non-trivial and exciting predictions for the study of higher dimensional gapped and gapless phases. We will briefly discuss them in Section\,\ref{sec:outlooks}. 
\end{rem}

\section{Chiral gapless edge I} \label{sec:gapless-edge-1}

In this section, we will try to find all observables living on the 1+1D world sheet of a chiral gapless edge of a 2+1D topological order, which is fixed to be $(\CC,c)$. 

\subsection{Vertex operator algebras}  \label{sec:observables}

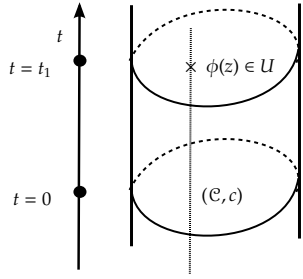
\begin{figure} 
$$
\quad\quad\quad \quad\quad\quad  
 \raisebox{-50pt}{
  \begin{picture}(100,110)
   \put(-40,8){\scalebox{0.5}{\includegraphics{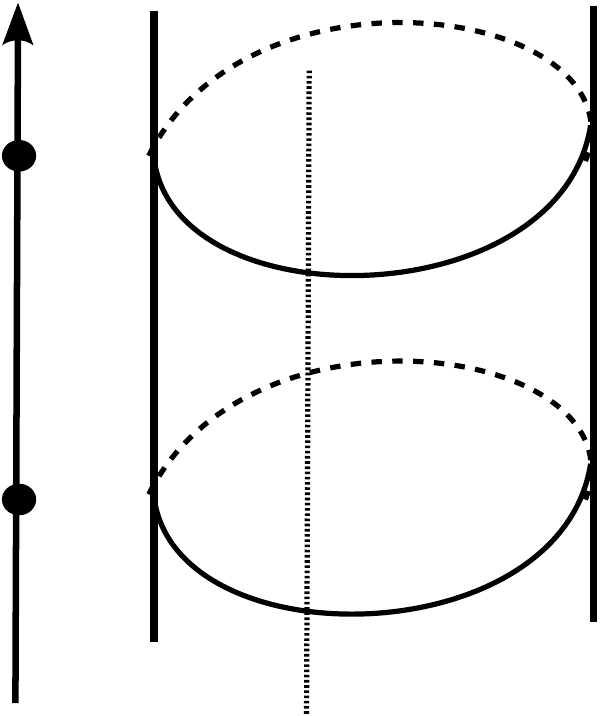}}}
   \put(-40,8){
     \setlength{\unitlength}{.75pt}\put(0,-83){
     \put(-30,118)  {\scriptsize $ t=0 $}
     \put(-32,184)  {\scriptsize $ t=t_1$}
     \put(-8, 200)  {\scriptsize $t$}
     \put(65,120)  {\scriptsize $ (\CC,c) $}
     %\put(78,152)  {\scriptsize $ a \in \CC$}
     %\put(126,180)  {\scriptsize $ A_x $}
     %\put(118,262)  {\scriptsize $A_y$}
     %\put(90,223)   {\scriptsize $M_{x,y}$}
     \put(56.5, 185) {\scriptsize $\times \,\,\, \phi(z) \in U$}
     %\put(75,85) {\scriptsize $U$}
     %\put(125, 105) {$x$}
     }\setlength{\unitlength}{1pt}}
  \end{picture}}
$$
\caption{This picture depicts a 2d topological order $(\CC,c)$ on a 2-disk, together with a 1d gapless edge, propagating in time. A chiral field $\phi(z)\in U$ is depicted on the 1+1D world sheet. 
}
\label{fig:cylinder-0}
\end{figure}

Suppose that the 2d bulk phase $(\CC,c)$ is realized on an open 2-disk as depicted in Figure\,\ref{fig:cylinder-0}. We choose a complex coordinate $z$ on the 1+1D world sheet of the chiral gapless edge (i.e. the 2D boundary of the solid cylinder) such that the time axis is the real axis.

We assume that gapless edge modes are completely chiral. It is  known that the chiral edge modes are states in a chiral conformal field theory with the central charge $c$ (see \cite{halperin,wen1,moore-read,wen2}, reviews in \cite{wen2,wen3,nssfs} and references therein). This chiral CFT is defined on the 1+1D world sheet of the edge. We denote this chiral CFT by $U$. By the state-field correspondence in a 2D CFT, there is a bijective map $Y$ from $U$ to the space of chiral fields, still denoted by $U$: 
$$
Y: \phi \mapsto Y(\phi,z)= \phi(z).
$$ 
The notation $\phi(z)$ is commonly used in physics literature and $Y(\phi,z)$ is commonly used in mathematics literature. We prefer to use the mathematical notation $Y(\phi,z)$ sometimes in this work because the first notion can be quite ambiguous whenever we discuss intertwining operators. We should view these chiral fields as ``local observables'' on the world sheet. Since these chiral fields can live on the entire 1+1D world sheet, they cannot be multi-valued. Therefore, these chiral fields, say $\phi(z)$, contains only integer powers of the complex variable $z$, i.e. 
$$
\phi(z) = \sum_{n\in \Zb} \phi_n z^{-n-1}, 
$$
where $\phi_n\in \mathrm{End}(U)$. Any two chiral fields in $U$ can have the so-called operator product expansion (OPE) as shown below: 
$$
\phi(z_1) \psi(z_2) \sim \frac{(\psi_k\phi) (z_2)}{(z_1-z_2)^{k+1}} + \frac{(\psi_{k-1}\phi) (z_2)}{(z_1-z_2)^{k}} + \cdots , \quad\quad \mbox{for some $k\in \Nb$}. 
$$
This OPE is commutative, i.e. $\phi(z_1) \psi(z_2)   \sim   \psi(z_2) \phi(z_1)$. It provides $U$ with an algebraic structure, which is called a chiral algebra in physics, and is rigorously defined in mathematics under a different name: a vertex operator algebra (VOA) (see for example \cite{ll} for a review). In this work, we use the following working definition of a VOA, which is sometimes called a CFT-type VOA. We avoid to use formal variables that are commonly used in mathematics literature. 

\begin{defn}
A {\em vertex operator algebra} (VOA) $U$ consists of the following data: 
\begin{itemize}

\item a $\Cb$-linear vector space $U$ graded by non-negative integers (called conformal weights), i.e. $U=\oplus_{n\in \Nb} U_{(n)}$, 

\item a distinguished element $\mathbb{1}\in U_{(0)}$ called vacuum state, 

\item a map $Y: (U \otimes_\Cb U) \times \Cb^\times \to \overline{U}:=\prod_{n\in \Nb} U_{(n)}$ called vertex operator
$$ 
(u\otimes_\Cb v,z) \mapsto Y(u,z)v=\sum_{n\in \Zb} u_nv \, z^{-n-1},
$$ 

\item a distinguished element $\omega_U\in U_{(2)}$ called Virasoro element,

%\item $\langle \cdot | \cdot \rangle$ is an inner product (anti-linear on the first variable) normalized to $\langle \mathbb{1} | \mathbb{1} \rangle =1$, 

%\item an antilinear involutive automorphism $\theta: U \to U$ (i.e. $\theta^2=\id_U$), 

\end{itemize}
satisfying the following axioms: 

\bnu

\item $\dim U_{(n)} < \infty$ and $U_{(0)}=\Cb\mathbb{1}$; 

\item if $u\in U_{(k)}$, then $u_n$ maps $U_{(m)}$ into $U_{(m+k -n-1)}$; 

\item $Y(\mathbb{1},z)=\id_U$ and $u_{-1}\one=u$ for $u\in U$;  

\item for $u,v,w\in U$ and $w'\in U':=\oplus_n U_{(n)}^\ast$, the following three expressions
$$
\langle w', \,\, Y(u,z_1)Y(v,z_2) w\rangle, \quad\quad \langle w', \,\, Y(v,z_2)Y(u,z_1)w\rangle, \quad\quad
\langle w', \,\, Y(Y(u,z_1-z_2)v,z_2)w\rangle
$$ 
converge absolutely in three different domains $|z_1|>|z_2|>0$, $|z_2|>|z_1|>0$, $|z_2|>|z_1-z_2|>0$, respectively, to the same rational function $f(z_1,z_2)$ (with poles only at $z_1,z_2=0,\infty$ and $z_1=z_2$); 

\item we have $Y(\omega_U,z) = \sum_{n\in \Zb} L(n) z^{-n-2}$ such that $L(n)$, $n\in \Zb$ generate the famous Virasoro algebra, i.e.
$$[L(m), L(n)]  = (m-n)L(m+n) + \frac{c}{12}(m^3-m)\delta_{m+n,0}$$
where the number $c$ is called the central charge;
%in the sense that they have the same OPE. % and both sides are analytic continuation of each other.  

\item for $u \in U_{(n)}$, we have $L(0) u = n u$; 

\item  for $u\in U$, we have $Y(L(-1)u, z) = \frac{d}{dz} Y(u,z)$.

%\item $\theta$ is anti-unitary, i.e. $\langle \theta w| \theta v\rangle = \langle v | w\rangle$. 

%\item for $u,v,w\in U$, we have $$\langle w | Y(u,z) v\rangle = \langle Y(e^{-zL(1)} z^{-2L(0)} \theta u, -z^{-1}) w| v \rangle, $$ (as a consequence, the grading is orthogonal with respect to the inner product). 

\enu
\end{defn}

\begin{rem}
$Y(\omega_U,z)$ is called the energy momentum tensor in physics. It generates a sub-VOA $\langle \omega_U \rangle$, which is the smallest sub-VOA of $U$. The number $c$ is the same central charge as the one in the 2d bulk phase $(\CC,c)$. 
\end{rem}

\subsection{Open-string vertex operator algebras} \label{sec:osvoa}

\begin{figure}
$$  
\quad\quad\quad \quad\quad
\raisebox{-50pt}{
  \begin{picture}(80,110)
   \put(-40,8){\scalebox{0.45}{\includegraphics{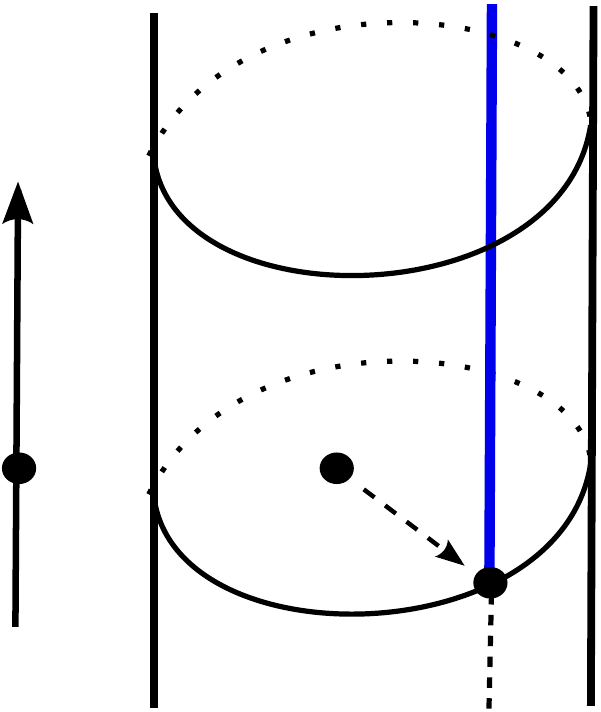}}}
   \put(-40,8){
     \setlength{\unitlength}{.75pt}\put(0,-83){
     \put(-30,133)  {\scriptsize $ t=0 $}
     %\put(-32,219)  {\scriptsize $ t=t_1$}
     %\put(-8, 250)  {\scriptsize $t$}
     %\put(65,120)  {$ \CB $}
     \put(58,135)  {\scriptsize $ a \in \CC$}
     \put(63, 185) {\scriptsize $\phi(r)\,\, \times $ }
     %\put(126,180)  {\scriptsize $ A_x $}
     \put(82,212)  {\scriptsize $A_x = \mbox{a boundary CFT}$}
     %\put(90,223)   {\scriptsize $M_{x,y}$}
     \put(45,85) {\scriptsize $A_{\one}=U$}
     \put(90, 100) {\scriptsize $x$}
     }\setlength{\unitlength}{1pt}}
  \end{picture}}
$$
\caption{This picture depicts a 2d topological order $(\CC,c)$ on a 2-disk, together with a 1d gapless edge, propagating in time. At $t=0$, we move a bulk particle $a$ to the edge. The blue world line is supported on $x$.}
\label{fig:cylinder-Ax}
\end{figure}
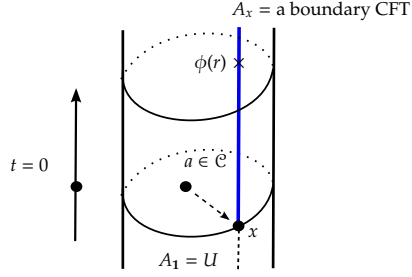

Is this VOA $U$ the only observables on the 1+1D world sheet of the edge? Consider the situation depicted in Figure\,\ref{fig:cylinder-Ax}. At $t=0$, a topological excitation $a$ in the bulk is moved to the edge. It becomes a topological edge excitation $x$. As a typical quantum quenching scenario, this movement suddenly changes the microscopic physics at the location of $x$. As a consequence, the RG flow will drive the world line supported on $x$ to a fixed point theory, which is denoted by $A_x$. In particular, $A_x$ should contain all the physical observables on the world line at the RG fixed point. It is clear that when $a=\one_\CC$, $x$ is the trivial topological edge excitations, denoted by $\one$. Therefore, we must have $A_\one=U$. More general topological edge excitations can be created by inserting impurities from outside. In other words, there are potentially more topological edge excitations than those obtained from moving bulk excitations to the edge. 

\medskip
%It was well known that moving bulk topological excitations to the gapless edge do not break the conformal symmetry on the boundary (see for example \cite{wen1,wenwu1,wwh}). 
At the RG fixed point, the conformal symmetry is restored. The observables in $A_x$ are again chiral fields that can have OPE along the world line. But the chiral fields in $A_x$ are potentially different from those in $U$. By our choice of complex coordinate, the world line (or the time axis) is the real axis. There is no multi-value issue here because the chiral fields are restricted on the real axis. As a consequence, the chiral fields on the world line can have non-integer powers, i.e. 
$$
\phi(r) = \sum_{n\in \mathbb{R}} \phi_n r^{-n-1}, % \in A_x,
$$
Moreover, two chiral fields on the world line can also have OPE as follows: 
$$
\psi(r_1) \phi(r_2) \sim \frac{(\psi_k\phi) (r_2)}{(r_1-r_2)^{k+1}} + \frac{(\psi_{k-1}\phi) (r_2)}{(r_1-r_2)^{k}} + \cdots , \quad\quad \mbox{for some $k\in \mathbb{\Rb}$}. 
$$
This OPE is potentially non-commutative, i.e. 
$$
\psi(r_1) \phi(r_2) \nsim \phi(r_2)\psi(r_1).
$$
More precisely, OPE among all these chiral fields defines an algebraic structure on $A_x$ called an open-string vertex operator algebra (OSVOA) \cite{osvoa}. We recall this notion below. 
\begin{defn}
An {\em open-string vertex operator algebra} (OSVOA) $A$ of central charge $c$ consisting of the following data:
\begin{itemize}
\item an $\Rb$-graded vector space $A=\oplus_{n\in \Rb} A_{(n)}$, 
\item a map $Y_A: (A \otimes_\Cb A) \times \Rb_+ \to \overline{A}:=\prod_{n\in\Rb} A_{(n)}$ defined by $(u\otimes v, r) \mapsto Y_A(u,r)v$,
\item two distinguished elements $\mathbb{1}\in A_{(0)}$ and $\omega_A \in A_{(2)}$,
%\item $\langle \cdot | \cdot \rangle$ is an inner product (anti-linear on the first variable) normalized to $\langle \mathbb{1} | \mathbb{1} \rangle =1$, 
%\item an antilinear involutive automorphism $\theta: A \to A$ (i.e. $\theta^2=\id_U$), 
\end{itemize}
satisfying the following conditions:  
\bnu
\item $\dim A_{(n)} < \infty$ and $A_{(n)}=0$ for $n\ll0$; 

%\item if $u\in U_{(k)}$, then $u_n: U_{(m)} \to U_{(m+k -n-1)}$; 

\item $Y(\mathbb{1},r)=\id_A$ and $u_{-1}\one=u$ for $u\in A$;  

\item for $u_1,\cdots, u_n,w\in A$ and $w'\in A'=\oplus_n A_{(n)}^\ast$, the series: 
$$
\langle w', Y_A(u_1, r_1) \cdots Y_A(u_n,r_n)w\rangle := \sum_{m_1, \cdots, m_{n-1}} \langle w', Y_A(u_1, r_1) P_{m_1} \cdots  P_{m_{n-1}}Y_A(u_n,r_n)w\rangle\, , 
$$
where $P_k: \overline{A} \twoheadrightarrow A_{(k)}$ is the projection operator, converges absolutely when $r_1>\cdots > r_n>0$ and can be extended to a potentially multi-valued complex analytic function in $(\Cb^\times)^n$ with only possible singularities at $z_i=z_j$ for $i,j=1,\cdots, n$ and $i\neq j$; 

\item the following two expressions
$$
\langle w',  \,\, Y_A(u_1,r_1)Y_A(u_2,r_2) w\rangle, \quad\quad \langle w',  \,\, Y_A(Y_A(u,r_1-r_2)v,r_2)w\rangle
$$ 
converge absolutely in two different domains $r_1>r_2>0$ and $r_2>r_1-r_2>0$, respectively, and equal on the intersection of the above two domains; 

\item we have $Y_A(\omega_A,r) = \sum_{n\in \Zb} L(n) r^{-n-2}$ such that $\{L(n)\}_{n\in \Zb}$ generate the Virasoro algebra of central charge $c$;
%in the sense that they have the same OPE. % and both sides are analytic continuation of each other.  

\item for $u \in A_{(n)}$, we have $L(0) u = n u$; 

\item  for $u\in A$, we have $Y_A(L(-1)u, r) = \frac{d}{dr} Y(u,r)$.

\enu
\end{defn}

\medskip
An OSVOA is a non-commutative generalization of the notion of a VOA. In particular, a VOA is automatically an OSVOA. An OSVOA contains a smallest subalgebra $\langle \omega_A \rangle \subset A$ generated by $\omega_A$. Actually, $\langle \omega_A \rangle$ is a VOA. More generally, the following subspace of an OSVOA $A$: 
\be \label{eq:C0A}
C_0(A) = \{ u \in \oplus_{n\in \Zb} A_{(n)} | Y_A(u,r) = \sum_{n\in \Zb} u_n r^{-n-1}, \quad Y_A(v,r)u = e^{rL(-1)} Y(u, -r)v, \forall v\in A \}, 
\ee
%where $Y_A(u,z)v=\sum_{n\in \Zb} u_nv z^{-n-1}$ for $z\in \Cb^\times$, 
defines a subalgebra of $A$ called the {\it meromorphic center} of $A$. Moreover, it is a VOA. The defining property of $C_0(A)$ is equivalent to the following condition: 
\begin{itemize}
\item[$(\bullet)$]  For $u\in C_0(A), v,w\in A$ and $w'\in A'$, there exists a (possibly multi-valued) analytic function
on 
$$
\{ (z_1, z_2) \in \Cb^2 | z_1\neq 0, z_2\neq 0, z_1\neq z_2 \}.
$$
such that it is single valued in $z_1$ and equals to 
$$
\langle w',  \,\, Y_A(u,z_1)Y_A(v,r_2) w\rangle, \quad\quad \langle w',  \,\, Y_A(v,r_2)Y_A(v,z_1)w\rangle, \quad\quad
\langle w',  \,\, Y_A(Y_A(u,z_1-r_2)v,r_2)w\rangle
$$ 
in the domains $|z_1|>r_2>0$, $r_2>|z_1|>0$, $r_2>|z_1-r_2|>0$, respectively; 
\end{itemize}
%(or equivalently, in a more physical language, OPE between the fields in $C_0(A)$ and $A$ are commutative, i.e. $\phi(z_1) \psi(r_2) \sim \psi(r_2) \phi(z_1)$ for $\phi\in C_0(A)$ and $\psi\in A$.) 

\medskip
It is clear that $A$ is a $C_0(A)$-module. For $v\in A$, it can be shown that there exist $v_n$ such that $Y_A(v,r)=\sum_{n\in \Rb} v_n r^{-n-1}$ \cite[Prop.\,1.4]{osvoa}. We introduce a formal vertex operator:
$$Y_A^f (v, x) := \sum_{n\in \Rb} v_n x^{-n-1},$$
where $x$ is a formal variable. %The defining property of $C_0(A)$ or the condition ($\bullet$) is equivalent to the statement. %that the vertex operator $Y_A(-,x)-$ is an intertwining operator of $C_0(A)$ \cite{osvoa}.
\begin{thm}[\cite{osvoa}] \label{thm:intertwining}
The condition ($\bullet$) is equivalent to the statement that the formal vertex operator $Y_A^f(-,x)-$ is an intertwining operator of $C_0(A)$. 
\end{thm}

\subsection{Boundary CFT's and domain walls} \label{sec:edge-bcft}

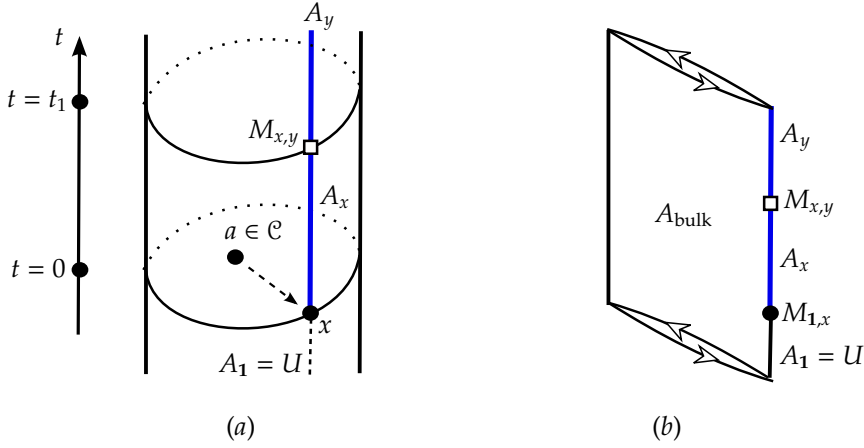
\begin{figure} 
$$
 \raisebox{-30pt}{
  \begin{picture}(130,130)
   \put(-20,8){\scalebox{0.64}{\includegraphics{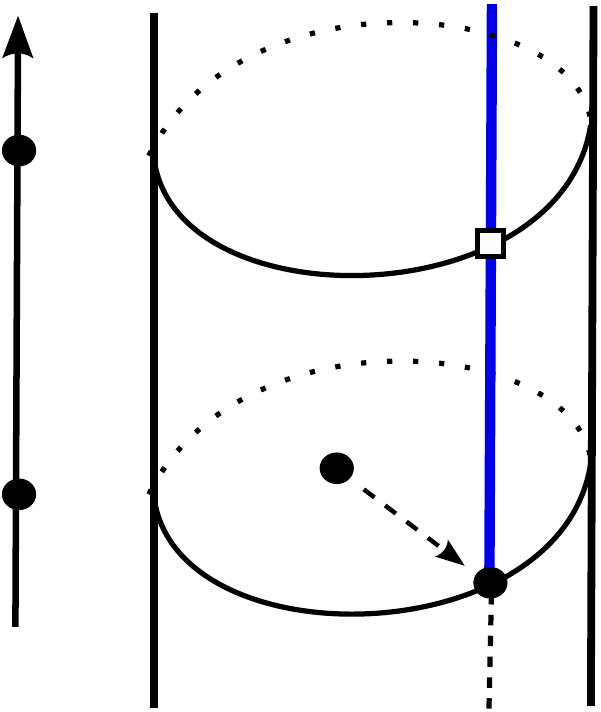}}}
   \put(-20,8){
     \setlength{\unitlength}{.75pt}\put(0,-83){
     \put(-30,133)  {$ t=0 $}
     \put(-32,219)  {$ t=t_1$}
     \put(-8, 250)  {$t$}
     %\put(65,120)  {$ \CB $}
     \put(78,152)  {$ a \in \CC$}
     \put(126,170)  {$ A_x $}
     \put(118,262)  {$A_y$}
     \put(90,203)   {$M_{x,y}$}
     \put(75,85) {$A_{\one}=U$}
     \put(125, 105) {$x$}
     }\setlength{\unitlength}{1pt}}
  \end{picture}}
  \quad\quad\quad\quad\quad
  %\rightsquigarrow\quad\quad\quad \quad
 \raisebox{-30pt}{
  \begin{picture}(70,130)
   \put(0,5){\scalebox{0.6}{\includegraphics{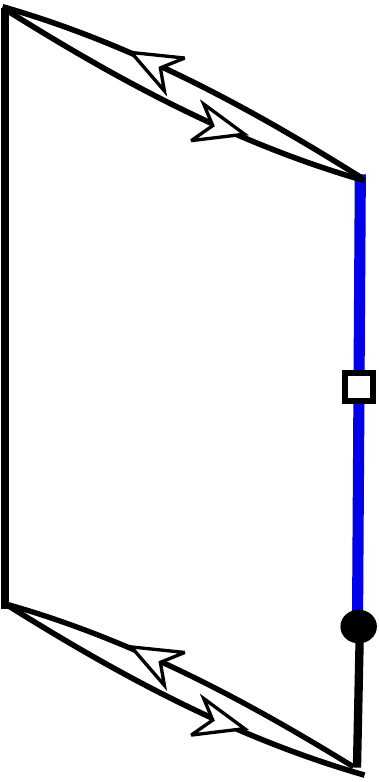}}}
   \put(0,5){
     \setlength{\unitlength}{.75pt}\put(0,-83){
     \put(88,142)  {$ A_x $}
     \put(88,205)  {$A_y$}
     \put(90,172)   {$M_{x,y}$}
     \put(88,93) {$A_{\one}=U$}
     \put(89, 115) {$M_{\one,x}$}
     \put(25,165) {$A_\cl$}
     }\setlength{\unitlength}{1pt}}
  \end{picture}}
$$
$$
(a) \quad\quad\quad\quad\quad\quad\quad\quad\quad\quad\quad\quad\quad\quad\quad (b)
$$
\caption{The picture (a) depicts a 2d topological order $(\CC,c)$ on a 2-disk, together with a 1d gapless edge, propagating in time. At $t=0$, a topological edge excitation $x$ is created. At $t=t_1>0$, the edge excitation is changed to $y$. This change creates a domain wall $M_{x,y}$ between OSVOA's $A_x$ and $A_y$. The picture (b) depicts the 1+1D world sheet obtained by squeezing the picture (a). 
}
\label{fig:cylinder-1}
\end{figure}

More general situations can occur on the same world line as depicted in Figure\,\ref{fig:cylinder-1} (a). More precisely, 
one can insert an impurity at $t=t_1>0$ on the world line so that the topological edge excitation is changed from $x$ to $y$. As a consequence, the OSVOA living on the $\{t>t_1\}$-part of the world line is given by $A_y$. The domain wall between $A_x$ and $A_y$ also contains chiral fields (sometimes called defect fields in CFT's). This domain wall is similar to the instantons on a gapped edge (recall Section\,\ref{sec:gapped-edge-2}). We denote the space of such defect fields by $M_{x,y}$. Then it is clear that the space of defect fields localized around $x$ is $M_{\one,x}$. Moreover, we should have $M_{x,x}=A_x$ and $M_{\one,\one}=A_\one=U$. This space $M_{x,y}$ should also be an $\Rb$-graded vector space, i.e. $M_{x,y}=\oplus_{n\in \Rb} (M_{x,y})_{(n)}$, such that $\dim (M_{x,y})_{(n)} < \infty$. 
%It is clear that chiral fields in $U$ can be fused into $M_{x,y}$ along different paths. Therefore, $M_{x,y}$ should carry a structures of certain ``module structure'' over $U$. We will make this statement precise later.

Similar to the fusion of instantons along the time axis in the gapped edge case, defect fields in different domain walls can also have OPE along the world line. This type of OPE defines a vertex operator:  
\begin{equation} \label{eq:Y-xyz}
Y_{(z,y,x)}: (M_{y,z} \otimes_\Cb M_{x,y}) \times \Rb_+ \to \overline{M}_{x,z}, \quad\quad 
(u \otimes_\Cb v, r) \mapsto Y_{(z,y,x)}(u, r) v. 
\end{equation}
This OPE should also be associative. More precisely, it satisfies the following condition.
\begin{itemize}
\item Associativity of OPE: for $u \in M_{y,z}, v\in M_{x,y}, w\in M_{w,x}$ and $w'\in M_{w,z}'$, the following two expressions 
$$
\langle w', Y_{(z,y,w)}(u,r_1) Y_{(y,x,w)}(v,r_2)w\rangle, \quad\quad 
\langle w', Y_{(z,x,w)}(Y_{(z,y,x)}(u,r_1-r_2)v,r_2)w\rangle
$$
converge absolutely in the domains: $r_1>r_2>0$ and $r_2>r_1-r_2>0$, respectively, and coincide in the domain $r_1>r_2>r_1-r_2>0$. 
\end{itemize}

It is important to note that we have not yet discussed the issue if these $M_{x,y}$ or $A_x$ are really different from $U$. At least, it is clear that there should be some relations among $U$, $A_x$ and $M_{x,y}$. For example, $A_x$ acts on $M_{x,y}$ from one side and $A_y$ acts from the other side, and $U$ acts on $M_{x,y}$ from above (or below) the time axis, and $U$ can be mapped into $A_x$, etc. In order to formulate them precisely, we need first to show that these $A_x$ are more than just OSVOA's. They are actually boundary CFT's \cite{cardy1,cardy2,cl} that share the same bulk 1+1D CFT. Indeed, consider a dimensional reduction process, which starts from Figure\,\ref{fig:cylinder-1} (a), and gradually squeeze the solid cylinder, and  ends up in the 1+1D world line depicted in Figure\,\ref{fig:cylinder-1} (b). The 0+1D boundary of this 1+1D world sheet is precisely given by the world line supported on $x$. Observables on this world line remain the same during this process. It is clear that both chiral and anti-chiral fields live on the resulting 1+1D world sheet. It was shown in \cite[Sec.\,2.1]{cz} that they form a modular invariant bulk CFT $A_{\mathrm{bulk}}$. This fact was emphasized in \cite{rz,levin} as a consequence of the following ``no-go theorem'': {\it Any 1+1D CFT's realized by 1d lattice Hamiltonian models should be modular invariant}. Therefore, the OSVOA's $A_x$ and $A_y$ must be the boundary CFT's sharing the same modular invariant bulk CFT $A_{\mathrm{bulk}}$, and $M_{x,y}$ is a domain wall between them and is also compatible with the same bulk. Mathematically, however, it is not a clear statement because there are different mathematical definitions of a boundary-bulk CFT. Here we prefer to make this statement mathematically precise. We propose a stronger ``no-go theorem''.
\begin{quote}
{\bf No-go Theorem}: A 1+1D boundary-bulk CFT realized by a 1d lattice Hamiltonian model with boundaries should satisfy all the axioms in the mathematical definition of a boundary-bulk CFT given in Definition\,\ref{def:bcft-1} or \ref{def:bcft-2}.  
\end{quote}
In particular, the boundary-bulk CFT should satisfy all the consistence conditions, including the famous modular invariant condition \cite{bpz,moore-seiberg} and Cardy condition \cite{cardy2,cl}.

\begin{rem}
Using boundary CFT's to study 0d defects or impurities in other condensed matter systems has a long history \cite{al1,al2}. But we are not aware of any earlier works mentioning the appearance of the boundary CFT's on the gapless edges of 2d topological orders. That a chiral vertex operator living in $M_{\one,x}$ was known in 90's (see \cite{wenwu1,wwh}). %But it is still far from enough to imply the appearance of the boundary CFT $A_x$. 
\end{rem}

\begin{rem}
Once we established the precise mathematical description of a chiral gapless edge, we can compute this dimensional reduction process and identify precisely which modular invariant bulk CFT is obtained at the end of this process \cite{kz3} (see more details in \cite{kz4}).
\end{rem}

\void{
\begin{rem}
It seems that a topological edge excitation can also be viewed as a boundary condition for a boundary CFT. It turns out that there is a subtle difference between these two notions. More precisely, the category of topological edge excitations and that of boundary conditions are slightly different. Their relation will be clarified in Section\,\ref{sec:underlying-cat}. 
\end{rem}
}

In summary, observables on the 1+1D world sheet of the edge of a 2d topological order form a ``not-yet-categorical'' structure: 
\bnu
\item objects are topological edge excitations: $x, y, \cdots$;
\item the space of morphisms are $M_{x,y}$ (boundary CFT's and domain walls); 
\item there is a map given by the associative OPE $Y_{(z,y,x)}: (M_{y,z} \otimes_\Cb M_{x,y}) \times \Rb_+ \to \overline{M}_{x,z}$. 
\enu
We can not yet claim it is a categorical structure not only because the OPE is not yet a composition map but also because the ``identity morphisms'' are missing. We will discuss this important missing data in the next subsection. 

\subsection{Chiral symmetries} 

The no-go theorem demands $A_x$ to be a boundary CFT, which requires many additional structures beyond that of an OSVOA. One of them is called the boundary condition of a boundary CFT. It says that, as a boundary CFT, $A_x$ should satisfy a conformal invariant boundary condition (as the minimal requirement) or a $V$-invariant boundary condition in general, where $V$ is called the chiral symmetry. We explain this in detail in this subsection. %Other additional structures on $A_x$ will be discussed in Section\,\ref{sec:RCFT}. 

\begin{figure} 
$$
 \raisebox{-50pt}{
  \begin{picture}(100,140)
   \put(-60,8){\scalebox{0.6}{\includegraphics{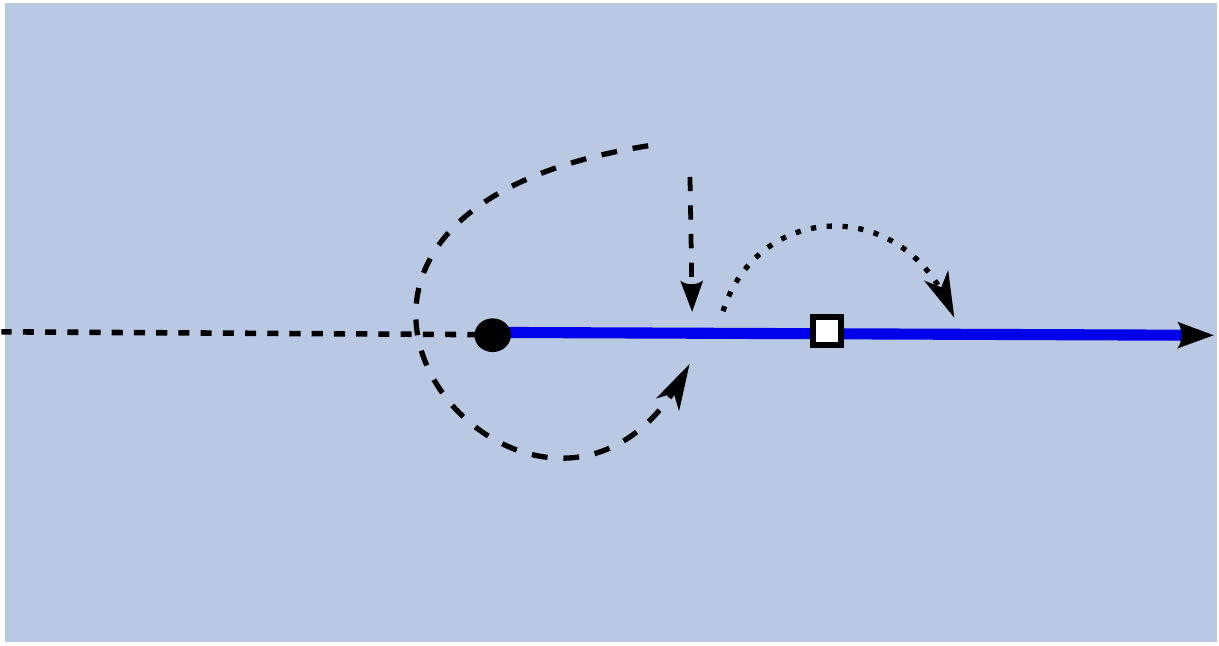}}}
   \put(-60,8){
     \setlength{\unitlength}{.75pt}\put(0,0){
     \put(108,57)  {$ t=0 $}
     \put(275,57)  {$t$}
     \put(102,110)  {\scriptsize $\gamma_2$}
     \put(145,95)  {\scriptsize $\gamma_1$}
     \put(185, 105) {\scriptsize $\gamma_3$}
     %\put(78,152)  {$ x \in \CB$}
     %\put(88,145)  {$ [x,x] $}
     \put(156,113) {\scriptsize $\times$}
     \put(150,125) {$\phi(z)\in V \subset U$}
     \put(185,53)   {$Y_{A_x}(v,t)\in A_x$}
     \put(110, 80) {$M_{\one,x}$}
     }\setlength{\unitlength}{1pt}}
  \end{picture}}
$$
\caption{This picture depicts how the chiral fields in $U$ can be mapped to $A_x$. 
}
\label{fig:chiral-symmetry}
\end{figure}
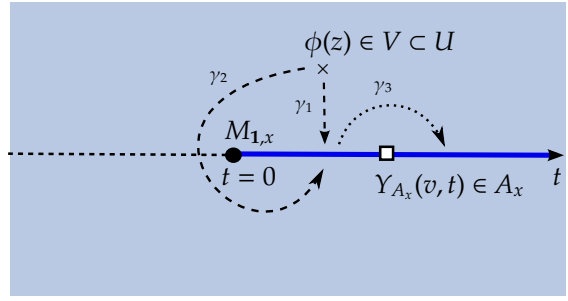

\medskip
Consider moving chiral fields in $U$ into those in $A_x$ along a path $\gamma$ (e.g. $\gamma_1, \gamma_2$ in Figure\,\ref{fig:chiral-symmetry}). This process defines a linear map $\iota_\gamma: U \to A_x$, which is clearly independent of the homotopy type of the path. Therefore, as shown in Figure\,\ref{fig:chiral-symmetry}, there are essentially two independent ways of mapping $U$ into $A_x$ along the paths $\gamma_1, \gamma_2$, respectively, i.e. $\iota_{\gamma_1}, \iota_{\gamma_2}: U\to A_x$. It is clear that the vacuum state and the OPE of the chiral fields must be preserved in these processes. Therefore, $\iota_{\gamma_1}, \iota_{\gamma_2}$ are two OSVOA homomorphisms. The minimal requirement for $A_x$ to give a consistent boundary CFT is that $\iota_\gamma$ should satisfy the following condition:  
\begin{itemize}
\item {\it Conformal-invariant boundary condition} \cite{cardy1,ocfa}: 
\bnu
\item $\iota_{\gamma_1} |_{\langle \omega_U \rangle} = \iota_{\gamma_2} |_{\langle \omega_U \rangle}$; 
\item $\iota_{\gamma_1} |_{\langle \omega_U \rangle}: \langle \omega_U \rangle \to \langle \omega_{A_x} \rangle$ defines an isomorphism of VOA. 
\enu
\end{itemize}
Let $V$ be a sub-VOA of $U$, i.e. $\langle \omega_U \rangle \subset V \subset U$. In most general situation, we expect the following condition to be true. 
\begin{itemize}
\item {\it $V$-invariant boundary condition} \cite[Definition\,1.25]{ocfa}:
\bnu
\item $\iota_{\gamma_1} |_V = \iota_{\gamma_2} |_V$; 
\item $\iota_{\gamma_1} |_V: V \to A_x$ is injective. 
\enu
We denote that path independent map by $\iota_x: V \hookrightarrow A_x$ 
\end{itemize}
In other words, chiral fields in $V$ can move transparently on the entirely 1+1D world (except those 0D walls $M_{x,y}$ for $x\neq y$). If the $V$-invariant boundary condition holds for all $A_x$, then, for a sub-VOA $V'$ of $V$, the $V'$-invariant boundary condition holds automatically. If we require the set of topological edge excitations to be fixed, then there will be the maximal sub-VOA $V$ in $U$ that $A_x$ is $V$-invariant for all $x$. This VOA $V$ will be called the {\it chiral symmetry} of the edge. From now on, we take $V$ to be the chiral symmetry of the edge. 

\begin{rem} \label{rem:breaking-cs-1}
If we demands $V' \subsetneq V$ to be the chiral symmetry, then the set of topological edge excitations can be enlarged accordingly. We will discuss this issue in Section\,\ref{sec:classification}.
\end{rem}

\begin{rem}
It can happen that the chiral symmetry is maximal, i.e $V=U$, as we will show in Section\,\ref{sec:can-edge}. We will also show in Section\,\ref{sec:general-edge} that $\langle \omega_U\rangle \subsetneq V\subsetneq U$ in general. 
\end{rem}

By the $V$-invariant boundary condition, $M_{x,y}$ are all $V$-modules in the usual sense \cite{fhl}. We denote the category of $V$-modules by $\Mod_V$. We obtain the following property of $Y_{A_x}$: 
\begin{itemize}
\item for $\phi\in V$, $v,w\in A_x$ and $w'\in A_x'$, the operator products 
$$
\langle w', \phi(z) Y_{A_x}(v, t) w\rangle, \quad\quad 
\langle w', Y_{A_x}(v, t)\phi(z)w\rangle, \quad\quad 
\langle w', Y_{A_x}(\phi(z)v, t-z)v, z)w\rangle
$$ 
converge absolutely in the domains $|z|>t>0$, $t>|z|>0$, $|z|>|t-z|>0$, respectively, and are analytic continuation of each other along any path in the $z$-plane (for example $\gamma_3\circ \gamma_i$ for $i=1,2$ as depicted in Figure\,\ref{fig:chiral-symmetry}). 
\end{itemize}
By the results in Section\,\ref{sec:osvoa}, this implies that $V$ lies in the meromorphic center of $A_x$. 
%It further implies that $A_x$ is a $V$-module, and, by 
By Theorem\,\ref{thm:intertwining}, $Y_{A_x}$ is an intertwining operator of $V$.

\medskip
This analysis can be generalized to more general vertex operators introduced in (\ref{eq:Y-xyz}), where $Y_{(z,y,x)}$ defines the OPE between defect fields in $M_{y,z}$ and those in $M_{x,y}$. Similarly, by the $V$-invariant boundary condition, we obtain the following property of $Y_{(z,y,x)}$:
\begin{itemize}
\item for $\phi\in V$, $v\in M_{y,z}$, $w\in M_{x,y}$ and $w'\in M_{x,z}'$, the operator products 
$$
\langle w', \phi(z) Y_{(z,y,x)}(v, t) w\rangle, \quad\quad 
\langle w', Y_{(z,y,x)}(v, t)\phi(z)w\rangle, \quad\quad 
\langle w', Y_{(z,y,x)}(\phi(z)v, t-z)v, z)w\rangle
$$ 
converge absolutely in the domains $t>|z|>0$, $|z|>t>0$, $|z|>|t-z|>0$, respectively, and are analytic continuation of each other along any paths in the $z$-plane. 
\end{itemize}
By Theorem\,\ref{thm:intertwining} again, $Y_{(z,y,x)}$ is an intertwining operator of $V$ \cite{fhl}. Note that $Y_{(x,x,x)}=Y_{A_x}$.

\medskip
Until the end of this section, we make the following assumption as a natural requirement in physics: 
the chiral symmetry $V$ is a unitary rational VOA such that the category of unitary $V$-modules, still denoted by $\Mod_V$, is a UMTC. 

\begin{rem}
There is a list of conditions on the VOA $V$ that guarantees $\Mod_V$ to be a modular tensor category by Huang's theorem \cite{huang-mtc,huang-mtc2}. In general, it is not clear if $\Mod_V$ is a UMTC when $V$ is unitary. See \cite{gui1,gui2,gui3} for discussion of the relation between the unitarity of a VOA $V$ and that of $\Mod_V$. 
\end{rem}

According to the tensor category theory of rational VOA developed by Huang and Lepowsky \cite{hl1,hl2,huang-JPAA}, the intertwining operator $Y_{(z,y,x)}$ of $V$ is equivalent to a morphism 
$$
\circledcirc: M_{y,z} \otimes_V M_{x,y} \to M_{x,z}
$$ 
in $\Mod_V$, where $\otimes_V$ is the tensor product in $\Mod_V$. Note that we have chosen the convention that the left factor of the relative tensor product $-\otimes_V -$ has a higher time coordinate. This convention is compatible with our OPE (recall (\ref{eq:Y-xyz})) and our convention in Eq.\,(\ref{eq:order-instantons}).

The associativity of OPE of intertwining operators of $V$ (recall the paragraph below Eq.\,(\ref{eq:Y-xyz})) 
%was rigorously proved in \cite{huang-JPAA,huang-ope-2}). Huang also proved that it implies that of $\circledcirc$, i.e. 
implies that the following diagram
\be \label{diag:M-wxyz}
\xymatrix{
M_{y,z} \otimes_V M_{x,y} \otimes_V M_{w,x} \ar[rr]^{1 \otimes_V \circledcirc} \ar[d]_{\circledcirc \otimes_V 1} & & M_{y,z} \otimes_V M_{w,y} \ar[d]^\circledcirc \\
M_{x,z} \otimes_V M_{w,x} \ar[rr]^\circledcirc & & M_{w,z}
}
\ee
is commutative \cite{huang-mtc}. When $w=x=y=z$, this commutative diagram together with the unit property $V\otimes_V A_x\simeq A_x$ imply that the triple $(A_x, \circledcirc, \iota_x)$ defines an algebra in $\Mod_V$ (see Definition\,\ref{def:algebra}) \cite[Theorem\,4.3]{osvoa}. 
Similarly, one can see that $M_{x,y}$ is an $A_y$-$A_x$-bimodule, which also induces the same $V$-action on $M_{x,y}$ because the $V$-action on $M_{x,y}$ is path independent. This leads to the following commutative diagrams: 
\be \label{diag:unit}
\xymatrix@!C=7ex{
  & M_{x,y}\otimes_V M_{x,x} \ar[rd]^\circledcirc \\
M_{x,y}\otimes_V V \ar[rr]^-\simeq \ar[ru]^{1_{M_{x,y}}\otimes\iota_x} &
% M_{x,y} \ar[r]^{1_{M_{x,y}}}  
& M_{x,y} \\
}
\quad\quad\quad
\xymatrix@!C=7ex{
  & M_{y,y}\otimes_V M_{x,y} \ar[rd]^\circledcirc \\
  V\otimes_VM_{x,y} \ar[rr]^-\simeq \ar[ru]^{\iota_y\otimes 1_{M_{x,y}}} & 
% M_{x,y} \ar[r]^{1_{M_{x,y}}} 
& M_{x,y}\, . \\
}
\ee
%where $\iota_x: V \hookrightarrow A_x$ is path independent. 

%\begin{rem} \label{rem:U-non-local} Since $U$ is a VOA-extension of $V$, by \cite{hkl}, $U$ is a commutative algebra in $\Mod_V$. It is clear that $M_{\one,x}$ is a right $U$-module. The braidings in $\Mod_V$ endow $M_{\one,x}$ with a structure of a $U$-$U$-bimodule. But $M_{\one,x}$ is not a local $U$-module \cite{KO} unless $V=U$.  \end{rem}

%\begin{rem}
%Although it seems natural to expect the formula $M_{y,z}\otimes_{M_{y,y}} M_{x,y} \simeq M_{x,z}$ to hold, it is not true for some topological edge excitations. We will show in Section\,\ref{sec:underlying-cat} that this formula holds for a subset of topological edge excitations.
%\end{rem}

In summary, all the observables on the 1+1D world sheet of a chiral gapless edge of a 2d topological order $(\CC,c)$ can be described by a pair $(V,\CXs)$, where $\CXs$ is a categorical structure: 
\begin{itemize}
\item objects in $\CXs$ are topological edge excitations: $x,y,z, \cdots$;  
\item for each pair $(x,y)$ of objects in $\CXs$, there is a space of morphisms: $\hom_{\CXs}(x,y):=M_{x,y}$, which is an object in $\Mod_V$;  
\item there is a map $\id_x: V = \one_{\Mod_V} \to M_{x,x}=A_x$ given by the canonical embedding $\iota_x: V\hookrightarrow A_x$, which is a morphism in $\Mod_V$;  
\item there is a map $\circledcirc: M_{y,z}\otimes_V M_{x,y} \to M_{x,z}$, which is also a morphism in $\Mod_V$,
\end{itemize}
satisfying the commutative diagrams in (\ref{diag:M-wxyz}) and (\ref{diag:unit}). By Definition\,\ref{defn:en-cat}, the categorical structure $\CXs$ is nothing but a category enriched in $\Mod_V$, or an $\Mod_V$-enriched category \cite{Ke}.

\medskip
There is a canonical $\Cb$-linear category $\CX$ associated to $\CX^\sharp$ defined as follows: 
\begin{itemize}
\item the objects are the objects in $\CX^\sharp$ (i.e. topological edge excitations);
\item the space of morphisms $\hom_\CX(x,y):=\hom_{\Mod_V}(\one_{\Mod_V}, M_{x,y})$; 
\item the identity morphism is given by $\id_x=\iota_x: V \hookrightarrow M_{x,x}$;
\item the composition morphism $\circ: \hom_\CX(y,z) \otimes \hom_\CX(x,y) \to \hom_\CX(x,z)$ is defined by the following composed morphism:
\begin{align} \label{eq:composition}
&\hom_{\Mod_V}(\one_{\Mod_V}, M_{y,z}) \otimes_\Cb \hom_{\Mod_V}(\one_{\Mod_V}, M_{x,y})
\to \hom_{\Mod_V} (\one_{\Mod_V}, M_{y,z}\otimes_V M_{x,y}) \nn
&\hspace{4cm} \xrightarrow{\circledcirc \circ -} 
\hom_{\Mod_V} (\one_{\Mod_V}, M_{x,z}). 
\end{align}
\end{itemize}
Note that $\id_x$ is indeed the identity morphism in the usual sense, i.e. $\id_y \circ f = f = f\circ \id_x$ for $f\in \hom_\CX(x,y)$, because of the commutative diagrams in (\ref{diag:unit}). 

Mathematically, this category $\CX$ is called {\it the underlying category of $\CX^\sharp$}. Physically, $\CX$ is nothing but {\it the category of topological edge excitations}. We will call the UMTC $\Mod_V$ {\it the background category of $\CXs$}.

\subsection{$\CX^\sharp$ is an enriched monoidal category}

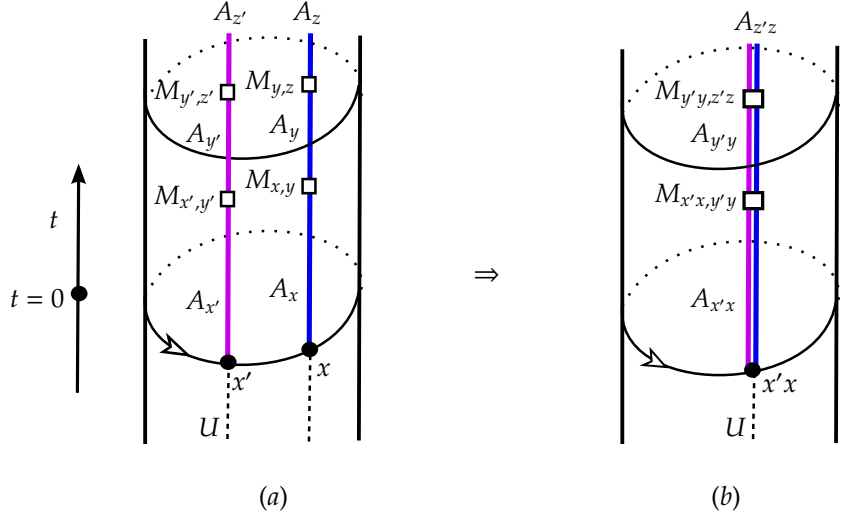
\begin{figure} 
$$
 \raisebox{-70pt}{
  \begin{picture}(80,165)
   \put(-30,8){\scalebox{0.6}{\includegraphics{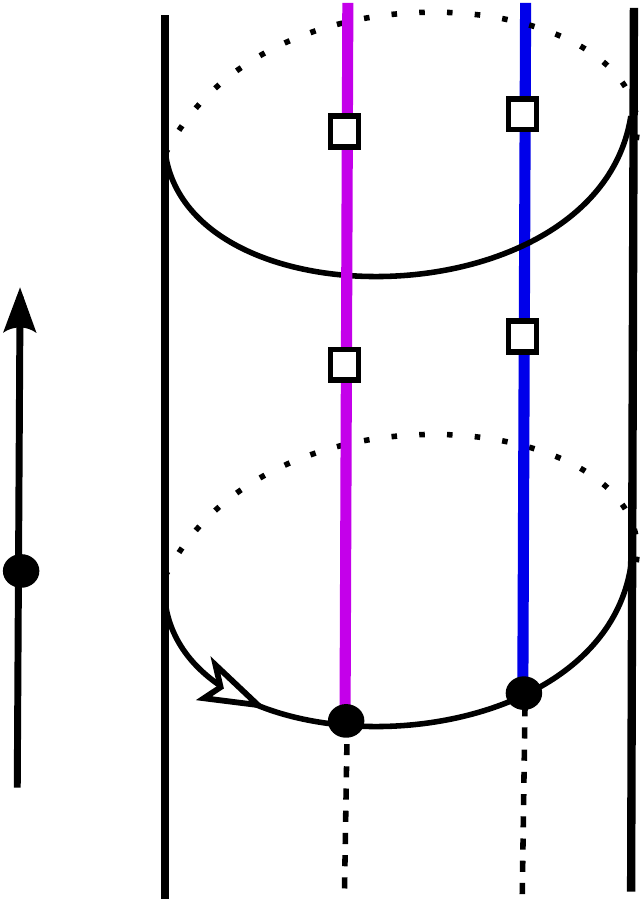}}}
   \put(-30,8){
     \setlength{\unitlength}{.75pt}\put(0,-70){
     \put(-30,140)  {$ t=0 $}
     \put(-10,180)  {$t$}
     %\put(65,120)  {$ \CB $}
     %\put(78,152)  {$ x \in \CB$}
     \put(101,145)  {$ A_x $}
     \put(101,228) {$A_y$}
     \put(89,200)  {$ M_{x,y} $}
     \put(89,250)  {$ M_{y,z} $}
     \put(113,285)  {$A_z$}
     \put(43,192)  {$M_{x',y'}$}
     \put(43,245)  {$M_{y',z'}$}
     \put(59,139)  {$A_{x'}$}
     \put(59,223)  {$A_{y'}$}
     \put(73,285)  {$A_{z'}$}
     \put(65,75) {$U$}
     \put(125, 105) {$x$}
     \put(82,98)  {$x'$}
     }\setlength{\unitlength}{1pt}}
  \end{picture}}
\quad\quad\quad\quad \Rightarrow \quad\quad\quad 
 \raisebox{-70pt}{
  \begin{picture}(100,165)
   \put(10,8){\scalebox{0.6}{\includegraphics{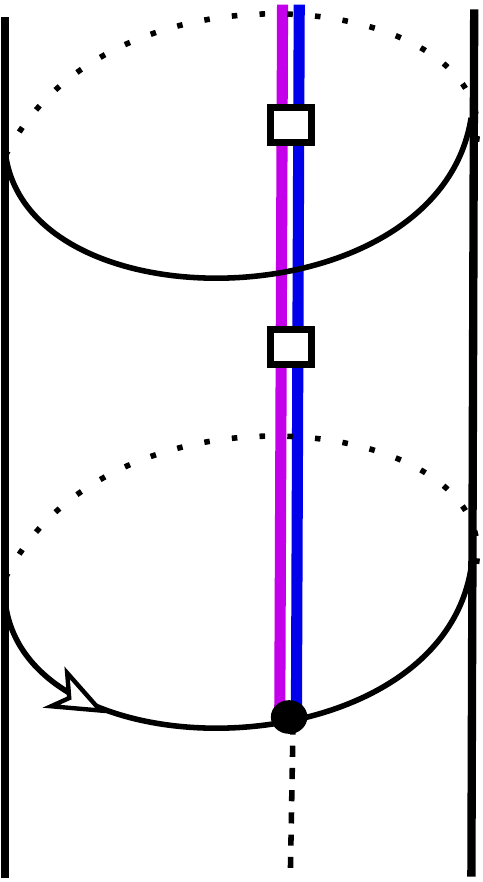}}}
   \put(10,8){
     \setlength{\unitlength}{.75pt}\put(-40,-75){
     %\put(-30,140)  {$ t=0 $}
     %\put(-10,180)  {$t$}
     %\put(65,120)  {$ \CB $}
     %\put(78,152)  {$ x \in \CB$}
     %\put(88,145)  {$ [x,x] $}
     %\put(85,235)  {$ [x,x']$}
     %\put(103,285)  {$[x',x']$}
     \put(58,198)  {$M_{x'x,y'y}$}
     \put(58,250)  {$M_{y'y,z'z}$}
     \put(76,228)  {$A_{y'y}$}
     \put(76,145)  {$A_{x'x}$}
     \put(98,285)  {$A_{z'z}$}
     \put(93,80) {$U$}
     %\put(125, 105) {$x$}
     \put(112,102)  {$x'x$}
     }\setlength{\unitlength}{1pt}}
  \end{picture}}  
$$
$$
(a) \quad\quad\quad\quad\quad\quad\quad\quad\quad\quad\quad\quad\quad\quad\quad\quad
(b)
$$
\caption{This picture depicts how to fuse spatially or horizontally two topological edge excitations $x$ and $x'$, together with boundary CFT's $A_x$, $A_y$, $A_{x'}$, $A_{y'}$ and walls $M_{x,y}$, $M_{x',y'}$.  
}
\label{fig:fusion}
\end{figure}

Two topological edge excitations $x'$ and $x$ can be fused along the edge (in the spatial dimension) to give a new edge excitation $x'\otimes x$ as depicted in Figure\,\ref{fig:fusion}. This spatial fusion automatically induces the spatial fusion of all observables on two world lines. If $M_{x',y'}$ and $M_{x,y}$ can both be viewed as defects in the spatial dimension, we should expect this spatial fusion to be given by $M_{x',y'}\otimes_U M_{x,y}$, where $M_{x',y'}$ and $M_{x,y}$ are viewed as $U$-$U$-bimodules and the relative tensor product $\otimes_U$ is well-defined in $\Mod_V$. However, it turns out to be wrong in general. The spatial fusion of observables (or instantons) in temporal dimension are very different from those in spatial dimensions. Recall that this is the recurring phenomena also happening on the gapped edges and when we move a bulk particle to the edge. The spatial fusion of $A_{x'}$ and $A_x$  is again a typical quantum quenching scenario. When we move $x'$ closer to $x$ at $t=0$, the Hamiltonian around $x$ changes suddenly. Renormalization flow will drive the world line theory $A_{x'}\otimes_U A_x$ supported on $x'\otimes x$ to a new fixed point theory, which is given by $A_{x'\otimes x}$. If the flow is non-trivial, then $A_{x'}\otimes_U A_x\nsimeq A_{x'\otimes x}$. But we expect that the RG flow gives a natural morphism from $A_{x'}\otimes_U A_x$ to $A_{x'\otimes x}$ because $A_{x'\otimes x}$ is not only physically universal but also universal in a mathematical sense, which automatically demands such a morphism (see Remark\,\ref{rem:univ} below). By generalizing the above argument to $M_{x',y'}$ and $M_{x,y}$, we expect that there is a natural morphism, for $x,y,x',y' \in \CXs$,
\be \label{eq:comp}
M_{x',y'} \otimes_U M_{x,y} \to M_{x'\otimes x, y'\otimes y},
\ee
which is not an isomorphism in general. Its failure of being an isomorphism is an indication of spatial fusion anomaly. 

\begin{rem} \label{rem:univ}
The universal property of $A_x$ and $M_{x,y}$ will be explained in details in Section\,\ref{sec:internal-hom} and Section\,\ref{sec:RG}. This is a recurring theme that will be summarized as a principle in Section\,\ref{sec:RG} for theories at RG fixed points. It is worthwhile to pointing out that, in some important situations, spatial fusion anomalies vanish (see Section\,\ref{sec:can-edge} and Remark\,\ref{rem:fusion-anomaly-free}). 
\end{rem}

Composing the morphism (\ref{eq:comp}) with the canonical morphism $M_{x',y'}\otimes_V M_{x,y} \to M_{x',y'}\otimes_U M_{x,y}$ defined by the universal property of the tensor product $\otimes_U$, we obtain a morphism
$$\otimes: M_{x',y'}\otimes_V M_{x,y} \to M_{x'\otimes x, y'\otimes y}.$$

The morphism $\otimes$ should satisfy some natural properties: 
\bnu
\item The chiral symmetry condition should be preserved under the spatial fusion. In other words, $V$ should be mapped into $M_{x\otimes y, x\otimes y}$ canonically (as $\iota_{x\otimes y}: V \to M_{x\otimes y, x\otimes y}$) and independent of the paths we choose. In particular, this implies that the following diagram
\be
\xymatrix{
& V\otimes_V V \simeq V \ar[dl]_{\iota_x \otimes_V \iota_y} \ar[dr]^{\iota_{x\otimes y}} & \\
M_{x,x} \otimes_V M_{y,y} \ar[rr]^\otimes & & M_{x\otimes y, x\otimes y} 
}
\ee
is commutative.

\item Consider the situation depicted in Figure\,\ref{fig:fusion}. If we fuse $M_{x,y}, M_{y,z}, M_{x',y'}, M_{y',z'}$ horizontally and vertically and let it flows to the fixed point theory $M_{x'x,z'z}$. For convenience, we abbreviated $x'\otimes x$ to $x'x$. This process should be independent of which fusion (horizontal or vertical) we do first. This implies that the following commutative diagram (called {\it the braided interchange property} \cite{MP}):  
\be \label{diag:comp-otimes}
\xymatrix{
M_{y',z'}\otimes_V M_{x',y'} \otimes_V M_{y,z} \otimes_V M_{x,y} \ar[dd]_{\circledcirc\, \otimes_V \, \circledcirc} \ar[rr]^{1\otimes_V c_{M_{x',y'}, M_{y,z}} \otimes_V 1} & &
M_{y',z'}\otimes_V M_{y,z}  \otimes_V M_{x',y'} \otimes_V M_{x,y} \ar[d]^{(\otimes) \otimes_V (\otimes)} \\
& & M_{y'y,z'z} \otimes_V M_{x'x,y'y} \ar[d]^{\circledcirc} \\
M_{x',z'} \otimes_V M_{x,z} \ar[rr]^\otimes & & M_{x'x,z'z}\, , 
}
\ee
where $c_{M_{x',y'}, M_{y,z}}: M_{x',y'} \otimes_V M_{y,z} \xrightarrow{\simeq} M_{y,z}\otimes_V M_{x',y'}$ is the braiding in $\Mod_V$. Our braiding convention is explained in Remark\,\ref{rem:braiding-convention} below. 
\enu
Mathematically, these two properties simply says that the spatial fusion $\otimes$ defines an enriched functor $\CX^\sharp \times \CX^\sharp \to \CX^{\sharp}$. 

\medskip
This spatial fusion is clearly associative and unital with respect to the tensor unit given by the trivial edge excitation $\one$. As a consequence, this spatial fusion upgrades $\CXs$ to an $\Mod_V$-enriched monoidal category (see Definition\,\ref{def:emc}) \cite{MP,kz2}. 

\begin{rem} \label{rem:braiding-convention}
Our convention of the braidings in $\Mod_V$ is that when $M_{x',y'}$ is moving around $M_{y,z}$ along a path sitting on the left side of the world line supported on $x$, i.e. along the world line supported on $x'$ in Figure\,\ref{fig:fusion}, then $M_{x',y'}$ will stay on the top during the braiding. 
More precisely, in this case, the initial (resp. final) time coordinate of $M_{x',y'}$ is higher (resp. lower) than that of $M_{y,z}$, the adiabatic move gives 
the braiding $c_{M_{x',y'}, M_{y,z}}: M_{x',y'} \otimes_V M_{y,z} \to M_{y,z}  \otimes_V M_{x',y'}$. We will use this convention throughout this work. 
\end{rem}

\begin{rem} \label{rem:MP-convention}
Note that the braiding $c$ used in the top horizontal arrow in Eq.\,(\ref{diag:comp-otimes}) is replaced by the anti-braiding in \cite[Definition\, 2.1]{MP}. Therefore, our definition of braided interchange property given in Eq.\,(\ref{diag:comp-otimes}) actually makes $\CXs$ an $\overline{\Mod_V}$-enriched monoidal category in the sense of Morrison and Penneys in \cite[Definition\, 2.1]{MP}. 
\end{rem}

We summarize the main result of this section as a physical theorem. 
\begin{thm}
The complete set of observables on a chiral gapless edge of a 2d topological order forms a pair $(V, \CXs)$, where 
\begin{itemize}
\item $V$ is the chiral symmetry, which is a unitary rational VOA such that $\Mod_V$ is a UMTC;  
\item $\CXs$ is an $\Mod_V$-enriched monoidal category, whose objects are topological edge excitations $x,y,z,\cdots$, and whose morphisms 
%$M_{x,x}$ are boundary CFT's, $M_{x,y}$ are domain walls consisting of boundary condition changing operators. 
$M_{x,y}$ are boundary CFT's (if $x\simeq y$) or domain walls consisting of boundary condition changing operators (if $x\nsimeq y$).
\end{itemize}
As a consequence, the underlying category $\CX$, i.e. the category of topological edge excitations, is a $\Cb$-linear monoidal category. 
\end{thm}

\section{Boundary-bulk CFT's} \label{sec:RCFT}

Further study of $\CX^\sharp$ needs the classification theory of rational CFT's (see Theorem${}^{\mathrm{ph}}$\,\ref{thm:bcft-3}  and Remark\,\ref{rem:internal-hom-include-all}). In this section, we review the mathematical theory of rational CFT's based on the representation theory of VOA's. 

%In Section\,\ref{sec:RCFT}, we review the mathematical results of boundary-bulk RCFT's. In Section \,\ref{sec:internal-hom}, we review the mathematical notion of an internal hom. 

\subsection{Definitions of boundary-bulk CFT's} \label{sec:def-RCFT}

The theory of conformal field theory (CFT) was developed by physicists at the end of 80's (see \cite{bpz,moore-seiberg} and \cite{CFTbook} for a lengthy review). It took, however, about 20 years for mathematicians to develop a successful mathematical foundation of CFT's. This foundation is not merely a rigorization of the existing physical theory in its old formalism, but a new mathematical theory written in a beautiful new language with powerful new tools, and it leads to many new results far beyond those in \cite{CFTbook}. These new results play a crucial role in this work, and are quite unfamiliar to most condensed matter physicists. Therefore, we briefly outline here where these results come from. See \cite{geometry} for a more detailed review.

\medskip
When physicists Belavin, Polyakov and Zamolodchikov published the first systematic study of 2D CFT's in 1984 \cite{bpz}, mathematicians Frenkel, Lepowsky and Meurman published their independent discovery of similar mathematical structures also in 1984 \cite{flm}. In 1987, the preprint of Segal's mathematical definition of a 2D CFT \cite{segal} came out and have made a big impact to mathematics community since then. We will give a sketchy presentation of an open-closed (or boundary-bulk) generalization of Segal's definition \cite{huang-cft,hukriz,geometry}.  

We define $\mathbf{Bord}_{\mathrm{op-cl}}^{\mathrm{cpx}}$ to be the category, in which (1) the objects are finite ordered set with two colors: ``c'' (for closed strings) and ``o'' (for open strings), e.g. $\{ o_1, c_2, o_3, c_4, c_5 \}$; (2) the spaces of morphisms are those of the conformal equivalence classes of open-closed complex bordisms, i.e. the moduli spaces of Riemann surfaces with boundary components being either completely parametrized or not parametrized but having none or some parametrized line segments, e.g. 
\begin{center}
\includegraphics[width=70mm]{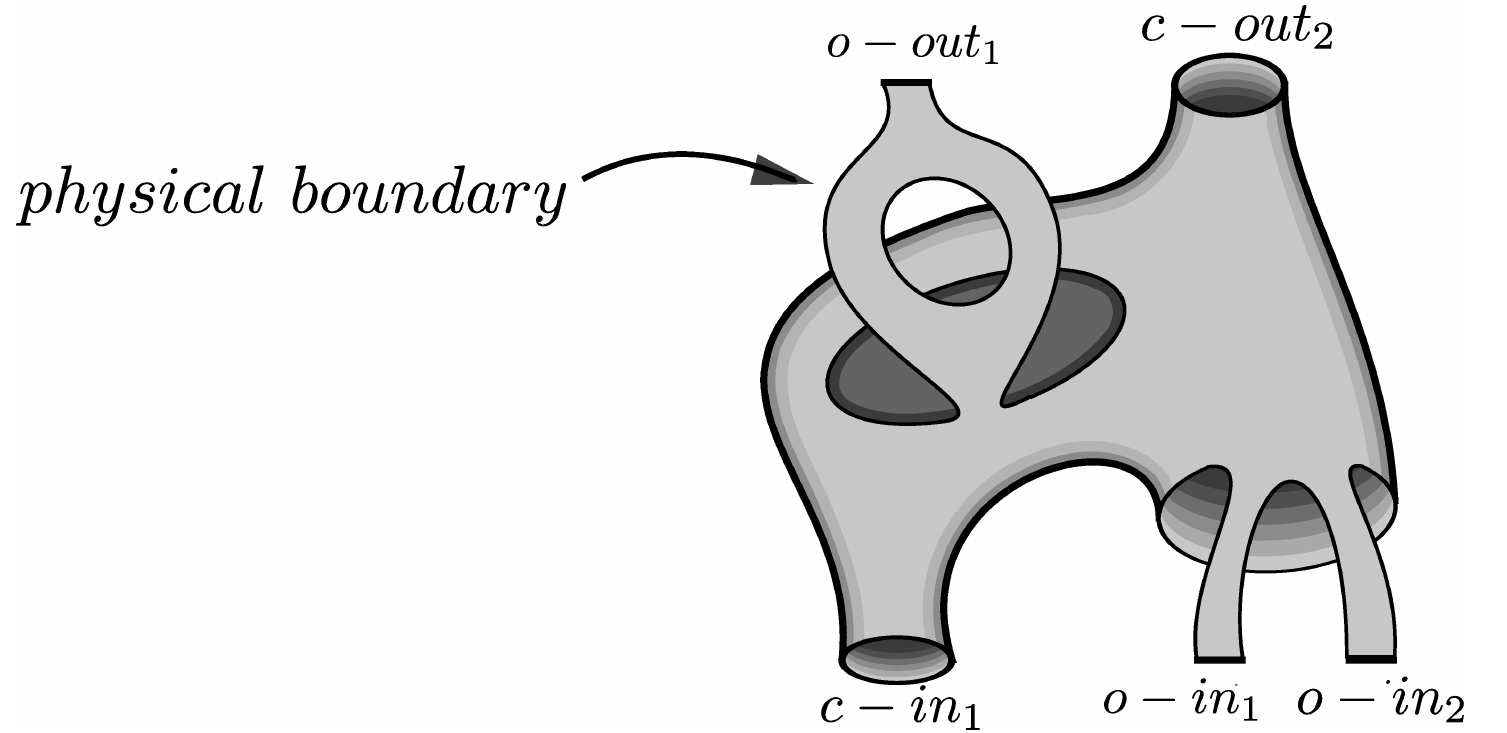}
\end{center}
where the label ``$in/out$'' is associated to domain/codomain of the morphism and is determined by the relation between the orientation of the boundary component induced from that of the surface and that induced from the parametrization. 

The category $\mathbf{Bord}_{\mathrm{op-cl}}^{\mathrm{cpx}}$ has the structure of a symmetric monoidal category with the tensor product defined by the disjoint union. Moreover, it has a $\ast$-structure defined by flipping the orientation of the Riemann surfaces. This flipping exchanges the domains with the codomains. Let $\mathbf{H}^\infty$ be the category of Hilbert spaces. It also has a $\ast$-structure $\hom_{\bh^\infty}(x,y) \xrightarrow[\simeq]{\dagger} \hom_{\bh^\infty}(y,x)$ defined by taking adjoint.

\begin{defn}[\cite{segal,huang-cft,hukriz}] \label{def:bcft-1}
A {\em boundary-bulk (or open-closed) CFT} is a real-analytic projective symmetric monoidal functor from $\mathbf{Bord}_{\mathrm{op-cl}}^{\mathrm{cpx}}$ to $\mathbf{H}^\infty$. It is called {\em unitary} if it is also a $\ast$-functor. 
\end{defn}

\begin{rem}
Since the spaces of morphisms in $\mathbf{Bord}_{\mathrm{op-cl}}^{\mathrm{cpx}}$ are moduli spaces naturally equipped with complex structures, ``real-analytic'' means that the functor, restricting on the spaces of morphisms, gives real-analytic functions. If the functor is complex analytic, then such CFT's are called holomorphic CFT's. 
\end{rem}

\begin{rem}
A CFT functor $F$ maps $\{ c_1 \}$, i.e. the set containing a single ``$c$''-colored element, to $H_\cl$, and maps $\{ o_1 \}$ to $H_\op$. The monoidalness of $F$ means that 
$$
F(\{ o_1, c_2, o_3, c_4, c_5 \}) \simeq H_\op \otimes_\Cb H_\cl \otimes_\Cb H_\op \otimes_\Cb H_\cl \otimes_\Cb H_\cl. 
$$
\end{rem}

This definition of CFT includes all consistence conditions, such as the modular invariant condition and the famous Cardy condition, etc. Unfortunately, this definition is not directly workable because chiral fields $\phi(z)$ in a CFT are associated to insertion at a point instead of a boundary component (or a line segment). It suggests to consider Riemann surfaces with parametrized punctures in the interior (closed strings stretched to infinity) or on the boundaries (open strings stretched to infinity) \cite{vafa,huang-book,geometry}. We will denote this surface-with-puncture variation of $\mathbf{Bord}_{\mathrm{op-cl}}^{\mathrm{cpx}}$ by $\mathbf{Bord}_{\mathrm{op-cl}}^{\mathrm{cpx}, \infty}$. Actually, $\mathbf{Bord}_{\mathrm{op-cl}}^{\mathrm{cpx}, \infty}$ is not a category because the composition of morphisms are only partially defined, thus is called a partial category. Similarly, we replace $\bh^\infty$ by a partial category of graded vector spaces (with a subtle definition of morphisms and their compositions), denoted by $\mathbf{GV}$ \cite{huang-book}. It is also possible to endow each of $\mathbf{Bord}_{\mathrm{op-cl}}^{\mathrm{cpx}, \infty}$ and $\mathbf{GV}$ with a $\ast$-structures \cite{gui1}. Then we obtain a working definition of a CFT. 

\begin{defn}[\cite{geometry}] \label{def:bcft-2}
A {\em boundary-bulk (or an open-closed) CFT} is a real-analytic projective symmetric monoidal functor $F: \mathbf{Bord}_{\mathrm{op-cl}}^{\mathrm{cpx},\infty} \to \mathbf{GV}$. It is {\em unitary} if it is also a $\ast$-functor. 
\end{defn}

To distinguish it from the first definition, we denote $F(\{c_1\})$ by $V_\cl$ and $F(\{o_1\})$ by $V_\op$.

\subsection{Classification theory of boundary-bulk CFT's} \label{sec:cf-RCFT}

If we restrict a CFT functor $F$ to the partial subcategory consisting of only bordisms of genus zero (Riemann spheres) from $\{c_1,\cdots, c_n\}_{n=0}^\infty$ to $\{ c_1\}$ and assume $F$ is complex analytic, it was proved by Huang that this restricted $F$ endows $V_\cl$ with the structure of a vertex operator algebra (VOA) \cite{huang-book}. Therefore, a VOA is a substructure of a CFT. In general, a VOA is not modular invariant except for holomorphic VOA's (such as the Monster Moonshine VOA \cite{flm}). The idea to find a modular invariant bulk (or closed) CFT $V_\cl$ is to realize it as certain extension of $V\otimes_\Cb \overline{V}$, where $V$ is a VOA and $\overline{V}$ is the same as the VOA $V$ in the formal variable but contains only the anti-chiral fields $\phi(\bar{z})=\sum_n \phi_n \bar{z}^{-n-1}$. Therefore, we need study the representation theory of VOA. Building on many earlier works by Huang and Lepowsky on tensor category theory for VOA's \cite{hl1,hl2,huang-JPAA}, Zhu's influential work on modular invariance \cite{zhu} and Huang's proof of Verlinde formula \cite{huang-mtc}, Huang proved that the category $\Mod_V$ of $V$-modules for $V$ satisfying certain rationality conditions is a modular tensor category \cite{huang-mtc2}. As a consequence, a bulk CFT $V_\cl$ is a commutative algebra in the category of $V\otimes_\Cb \overline{V}$-modules \cite{kong-ffa}, or equivalently, in the category $\Mod_V \boxtimes \overline{\Mod_V}$, which is also the Drinfeld center $\FZ(\Mod_V)$ of $\Mod_V$. Similarly, the boundary CFT $V_\op$ is a chiral extension of a VOA.

\medskip
Before we state the classification result, we need recall some basic notions in tensor categories. 
\begin{defn} \label{def:algebra}
An {\em algebra} in a monoidal category $\CA$ is a triple $A=(A,m,\eta)$ where $A$ is an object of $\CA$, $m$ (the multiplication) is a morphism $A \otimes A \rightarrow A$ and $\eta$ (the unit) is a morphism $\one_\CA \rightarrow A$ such that 
$$
m \circ (m \otimes 1_A) = m \circ (1_A \otimes m), \quad
m \circ (1_A \otimes \eta) = 1_A  = m \circ (\eta\otimes 1_A). 
$$ 
An algebra $A$ in a braided monoidal category is called {\em commutative} if $m_A \circ c_{A,A} = m_A$. 

Similarly, one can define a {\em coalgebra} $A = (A,\Delta,\epsilon)$, where $\Delta : A \rightarrow A \otimes A$ and $\epsilon : A \rightarrow \one_\CA$ obey the following coassociativity and counit conditions: 
$$
(\Delta \otimes 1_A) \circ \Delta = (1_A \otimes \Delta) \circ \Delta, \quad
(\epsilon \otimes 1_A) \circ \Delta = 1_A = (1_A \otimes \epsilon) \circ \Delta.
$$
\end{defn}

\begin{defn} {\rm 
A {\em Frobenius algebra} $A = (A,m,\eta,\Delta,\epsilon)$ is both an algebra and a coalgebra such that the coproduct $\Delta$ is an $A$-$A$-bimodules map, i.e.\ 
$$
(1_A \otimes m) \circ (\Delta \otimes 1_A) = \Delta \otimes m = (m \otimes 1_A) \circ (1_A \otimes \Delta).
$$ 
}
\end{defn}

We will use the following graphical representation:
$$
  m = \raisebox{-20pt}{
  \begin{picture}(30,45)
   \put(0,6){\scalebox{.75}{\includegraphics{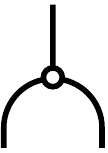}}}
   \put(0,6){
     \setlength{\unitlength}{.75pt}\put(-146,-155){
     \put(143,145)  {\scriptsize $ A $}
     \put(169,145)  {\scriptsize $ A $}
     \put(157,202)  {\scriptsize $ A $}
     }\setlength{\unitlength}{1pt}}
  \end{picture}}  
  ~~,\quad
  \eta = \raisebox{-15pt}{
  \begin{picture}(10,30)
   \put(0,6){\scalebox{.75}{\includegraphics{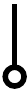}}}
   \put(0,6){
     \setlength{\unitlength}{.75pt}\put(-146,-155){
     \put(146,185)  {\scriptsize $ A $}
     }\setlength{\unitlength}{1pt}}
  \end{picture}}
  ~~,\quad
  \Delta = \raisebox{-20pt}{
  \begin{picture}(30,45)
   \put(0,6){\scalebox{.75}{\includegraphics{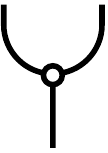}}}
   \put(0,6){
     \setlength{\unitlength}{.75pt}\put(-146,-155){
     \put(143,202)  {\scriptsize $ A $}
     \put(169,202)  {\scriptsize $ A $}
     \put(157,145)  {\scriptsize $ A $}
     }\setlength{\unitlength}{1pt}}
  \end{picture}}
  ~~,\quad
  \epsilon = \raisebox{-15pt}{
  \begin{picture}(10,30)
   \put(0,10){\scalebox{.75}{\includegraphics{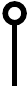}}}
   \put(0,10){
     \setlength{\unitlength}{.75pt}\put(-146,-155){
     \put(146,145)  {\scriptsize $ A $}
     }\setlength{\unitlength}{1pt}}
  \end{picture}}
  ~~.
$$

Let $\CB$ be a modular tensor category with the tensor product $\otimes$ and a tensor unit $\one_\CB$. The duality maps are expressed graphically (read from bottom to top) as follows: 
\be
\begin{array}{llll}
  \raisebox{-8pt}{
  \begin{picture}(26,22)
   \put(0,6){\scalebox{.75}{\includegraphics{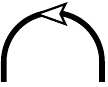}}}
   \put(0,6){
     \setlength{\unitlength}{.75pt}\put(-146,-155){
     \put(143,145)  {\scriptsize $ x^\ast $}
     \put(173,145)  {\scriptsize $ x $}
     }\setlength{\unitlength}{1pt}}
  \end{picture}}  
  = v_x : x^\ast \otimes x \rightarrow \one_\CB
  ~~,\qquad &
  \raisebox{-8pt}{
  \begin{picture}(26,22)
   \put(0,6){\scalebox{.75}{\includegraphics{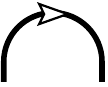}}}
   \put(0,6){
     \setlength{\unitlength}{.75pt}\put(-146,-155){
     \put(143,145)  {\scriptsize $ x $}
     \put(169,145)  {\scriptsize $ x^\ast $}
     }\setlength{\unitlength}{1pt}}
  \end{picture}}  
  = u_x^\dagger : x \otimes x^\ast \rightarrow \one_\CB
  ~~,
\\[2em]
  \raisebox{-8pt}{
  \begin{picture}(26,22)
   \put(0,0){\scalebox{.75}{\includegraphics{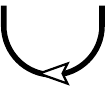}}}
   \put(0,0){
     \setlength{\unitlength}{.75pt}\put(-146,-155){
     \put(143,183)  {\scriptsize $ x $}
     \put(169,183)  {\scriptsize $ x^\ast $}
     }\setlength{\unitlength}{1pt}}
  \end{picture}}  
  = u_x : \one_\CB \rightarrow x \otimes x^\ast
  ~~,
  &
  \raisebox{-8pt}{
  \begin{picture}(26,22)
   \put(0,0){\scalebox{.75}{\includegraphics{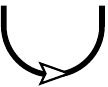}}}
   \put(0,0){
     \setlength{\unitlength}{.75pt}\put(-146,-155){
     \put(143,183)  {\scriptsize $ x^\ast $}
     \put(173,183)  {\scriptsize $ x $}
     }\setlength{\unitlength}{1pt}}
  \end{picture}}  
  = v_x^\dagger : \one_\CB \rightarrow x^\ast \otimes x~.
\end{array}
\ee

\void{
We fix a basis 
$\{ \lambda_{ij}^{k;\alpha} \}$ 
of $\hom_{\CC} (i\otimes j, k)$ and its dual basis 
$\{ y^{ij}_{k;\beta} \}$ in $\hom_{\CC} (k, i\otimes j)$, i.e. 
$$
y^{ij}_{k;\beta} = (\lambda_{ij}^{k;\beta})^\dagger, \quad\quad
\lambda_{ij}^{k;\alpha} \circ y^{ij}_{k;\beta}
 = \delta_{\alpha,\beta}\, \id_{k}, \quad\quad
 \sum_{k,\beta} y^{ij}_{k;\beta} \circ \lambda_{ij}^{k;\beta} = \id_{i\otimes j}. 
$$
We denote the basis vectors graphically as follows: 
\be
\lambda_{ij}^{k; \alpha} = 
  \raisebox{-23pt}{
  \begin{picture}(30,50)
   \put(0,8){\scalebox{.75}{\includegraphics{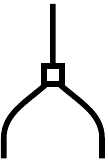}}}
   \put(0,8){
     \setlength{\unitlength}{.75pt}\put(-18,-11){
     \put(39,36)  {\scriptsize $ \alpha $}
     \put(32,61)  {\scriptsize $ k $}
     \put(17, 2)  {\scriptsize $ i $}
     \put(45, 2)  {\scriptsize $ j $}
     }\setlength{\unitlength}{1pt}}
  \end{picture}}
\quad , \qquad
y_{k; \alpha}^{ij} =
  \raisebox{-23pt}{
  \begin{picture}(30,50)
   \put(0,8){\scalebox{.75}{\includegraphics{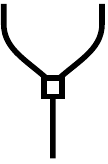}}}
   \put(0,8){
     \setlength{\unitlength}{.75pt}\put(-18,-11){
     \put(39,28)  {\scriptsize $ \alpha $}
     \put(32, 2)  {\scriptsize $ k $}
     \put(17,61)  {\scriptsize $ i $}
     \put(45,61)  {\scriptsize $ j $}
     }\setlength{\unitlength}{1pt}}
  \end{picture}} 
  ~~.
\ee
}

A Frobenius algebra $A$ in a modular tensor category $\CB$ is called {\em special} if $m\circ \Delta=\lambda_11_A$ and $\epsilon \circ \eta=\lambda_2\dim A$ for some $\lambda_1\lambda_2=1$. %, and is called {\em normalized-special} if $\lambda_1=\lambda_2=1$. 
It is called {\em symmetric} if it satisfies the following identity: 
\be
\raisebox{-35pt}{
  \begin{picture}(50,75)
   \put(0,8){\scalebox{.75}{\includegraphics{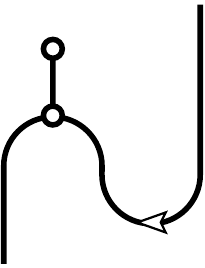}}}
   \put(0,8){
     \setlength{\unitlength}{.75pt}\put(-34,-37){
     \put(31, 28)  {\scriptsize $ A $}
     \put(87,117)  {\scriptsize $ A^\ast $}
     }\setlength{\unitlength}{1pt}}
  \end{picture}}  \quad
\,\, = \,\, \quad
  \raisebox{-35pt}{
  \begin{picture}(50,75)
   \put(0,8){\scalebox{.75}{\includegraphics{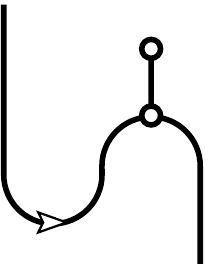}}}
   \put(0,8){
     \setlength{\unitlength}{.75pt}\put(-34,-37){
     \put(87, 28)  {\scriptsize $ A $}
     \put(31,117)  {\scriptsize $ A^\ast $}
     }\setlength{\unitlength}{1pt}}
  \end{picture}}
  ~~.
\label{eq:Frob-sym-cond}
\ee

Let $\CB=\Mod_V$ for a rational VOA $V$ such that it is a modular tensor category \cite{huang-mtc}. We have $\FZ(\CB)=\CB\boxtimes\overline{\CB}$ \cite{mueger2}. We denote the right adjoint functor to the tensor product functor $\otimes: \CB\boxtimes\overline{\CB} \to \CB$ by $\otimes^R$, which maps Frobenius algebras in $\CB$ to Frobenius algebras in $\CB\boxtimes\overline{\CB}$ \cite{kr2}. We denote the finite set of isomorphism classes of simple objects in $\CB$ by $\mathrm{Irr}(\CB)$. For $A \in \FZ(\CB)$, we also choose a basis $\{ b_{A}^{(i\boxtimes j;\alpha)} \}$ of $\hom_{\CC}(A, i\boxtimes j)$
and its dual basis $\{ b_{(i\boxtimes j;\beta)}^{A} \}$ of $\hom_{\CC}(i\boxtimes j, A)$ for $i,j\in \mathrm{Irr}(\CB)$, i.e. $b^{(i\boxtimes j;\alpha)}_{A} \circ b_{(i\boxtimes j;\beta)}^{A} = \delta_{\alpha\beta}\, 1_{i\boxtimes j}$, and use the following graphical notations
$$
b_A^{(i\boxtimes j; \alpha)} = 
  \raisebox{-23pt}{
  \begin{picture}(30,55)
   \put(0,8){\scalebox{.75}{\includegraphics{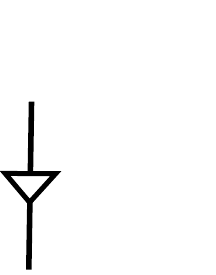}}}
   \put(0,8){
     \setlength{\unitlength}{.75pt}\put(-18,-11){
     \put(39,36)  {\scriptsize $ \alpha $}
     \put(25,65)  {\footnotesize $ i\boxtimes j $}
     \put(23, 2)  {\scriptsize $ A $}
     }\setlength{\unitlength}{1pt}}
  \end{picture}}
\quad  , \qquad
b_{(i\boxtimes j;\alpha)}^A =
  \raisebox{-23pt}{
  \begin{picture}(30,55)
   \put(0,8){\scalebox{.75}{\includegraphics{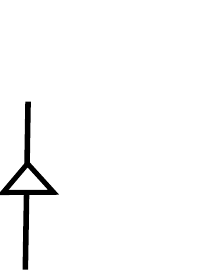}}}
   \put(0,8){
     \setlength{\unitlength}{.75pt}\put(-18,-11){
     \put(37,34)  {\scriptsize $ \alpha $}
     \put(23, 2)  {\footnotesize $ i\boxtimes j $}
     \put(22,65)  {\scriptsize $ A $}
     }\setlength{\unitlength}{1pt}}
  \end{picture}} 
  ~~.
$$
For an algebra $A$ in $\CB$, its full center $Z(A)$ in $\FZ(\CB)$ will be defined in Definition\,\ref{def:full-center}. If $A$ is Frobenius algebra,  $Z(A)$ can be defined by the subalgebra of $\otimes^R(A)$ given by the image of the morphism in Eq.\,(\ref{eq:cardy-CC}) for $A_\op=A$ \cite{fjfrs,kr2}.  

\medskip
Now we are ready to state a classification result for boundary-bulk CFT's. 
\begin{thm}[\cite{kong-cardy,kr2}] \label{thm:bcft-1}
A boundary-bulk CFT containing a single boundary condition, which preserves a rational chiral symmetry $V$, is necessary to be a triple $(A_\op |A_\cl, \iota)$, where
\bnu
\item $A_\op$ is a symmetric Frobenius algebra in $\CB=\Mod_V$; 
\item $A_\cl$ is a commutative symmetric Frobenius algebra in $\FZ(\CB)$; 
\item $\iota: A_\cl \rightarrow \otimes^R(A_\op)$ is an algebra homomorphism factoring through $Z(A_\op)\hookrightarrow \otimes^R(A_\op)$
\enu
satisfying the modular invariant condition, for all $i,j\in \mathrm{Irr}(\Mod_V)$, 
\be    \label{eq:mod-inv}
  \frac{\dim i \dim j}{\mathrm{dim} \CB} ~~
\raisebox{-55pt}{
  \begin{picture}(90, 110)
   \put(0,8){\scalebox{.6}{\includegraphics{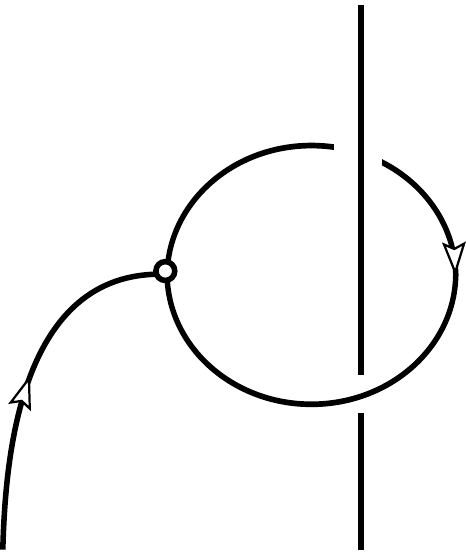}}}
   \put(0,8){
     \setlength{\unitlength}{.75pt}\put(-18,-19){
     \put(14, 8)     {\footnotesize $ A_\cl $}
     \put(126,80)  {\footnotesize $ A_\cl $}
     \put(95,8)    {$i \boxtimes j$ }
     \put(95,150){$i \boxtimes j $ }
     %\put(62, 83)  {\scriptsize $m$} 
     }\setlength{\unitlength}{1pt}}
  \end{picture}}
\quad = \quad  \, \sum_{\alpha}
 \raisebox{-55pt}{
  \begin{picture}(90,110)
   \put(8,8){\scalebox{.6}{\includegraphics{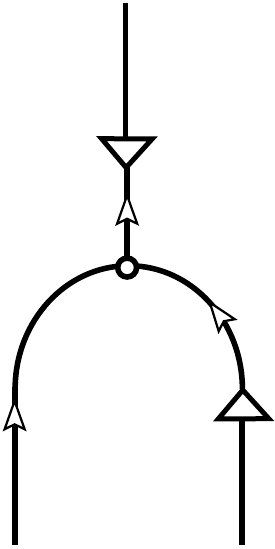}}}
   \put(0,8){
     \setlength{\unitlength}{.75pt}\put(-18,-19){
     \put(28, 8)  {\footnotesize $ A_\cl $}
     \put(50, 150)  {$ i \boxtimes j$}
     \put(80, 8)  {$ i \boxtimes j $}
     \put(27,95)  {\footnotesize $ A_\cl $}
     \put(85,75)    {\footnotesize $ A_\cl $}
     \put(67, 110)   {\scriptsize $\alpha$}
     \put(95, 50)     {\scriptsize $\alpha$}
     %\put(53, 75)  {\scriptsize $m$}
}\setlength{\unitlength}{1pt}}
  \end{picture}}
\ee  
and the Cardy condition 
\be   \label{eq:cardy-CC}
  \iota \circ \iota^\ast ~=~
  \raisebox{-40pt}{
  \begin{picture}(54,80)
   \put(0,8){\scalebox{.75}{\includegraphics{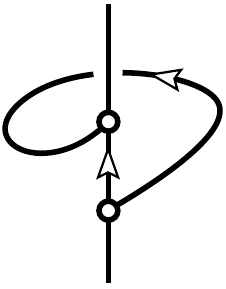}}}
   \put(0,8){
     \setlength{\unitlength}{.75pt}\put(-18,-19){
     \put(45, 10)  {\scriptsize $ \otimes^R(A_\op) $}
     \put(45,105)  {\scriptsize $ \otimes^R(A_\op) $}
     \put(80, 80)  {\scriptsize $ \otimes^R(A_\op) $}
     }\setlength{\unitlength}{1pt}}
  \end{picture}}
\ee
where $\iota^\ast$ is the right dual of $\iota$. The conjecture is that this triple is sufficient to give a boundary-bulk CFT defined in Definition\,\ref{def:bcft-2} \cite{lew}. 
\end{thm}

\begin{rem} \label{rem:vertex-tensor-cat}
The complete structure of a boundary-bulk CFT, including spectrums, correlation functions, OPE, partition functions, etc., is rather complicated. The drastic simplification achieved in Theorem\,\ref{thm:bcft-1} might looks surprising to physicists. This miracle happens because $\Mod_V$ has not only the structure of a modular tensor category but also a much richer structure called ``vertex tensor category'' \cite{hl1,huang-mtc2}, which makes the reduction possible. 
\end{rem}

\begin{defn}
An algebra $A$ in a fusion category is called {\it separable} if the multiplication morphism $m$ splits as an $A$-$A$-bimodule map. A separable algebra $A$ is called {\it simple} if $A$ is a simple $A$-$A$-bimodule. A commutative separable algebra $A$ in a braided fusion category $\CA$ is called {\it connected} if $\dim \hom_\CA(\one_\CA, A)=1$. A connected commutative separable algebra is also called a {\it condensable algebra} \cite{anyon}. A condensable algebra in a modular tensor category $\CA$ is called a {\it Lagrangian algebra} if $(\dim A)^2=\dim \CA$. 
\end{defn}

Using the results in \cite{frs1}, it is easy to show that a condensable algebra in a modular tensor category can be automatically upgraded to a commutative simple special symmetric Frobenius algebra (CSSSFA), which is unique up to a scalar factor in the definition of the counit (or equivalently, the comultiplication).

\begin{figure}
$$
\raisebox{-0pt}{
  \begin{picture}(95,80)
   \put(-20,10){\scalebox{1.2}{\includegraphics{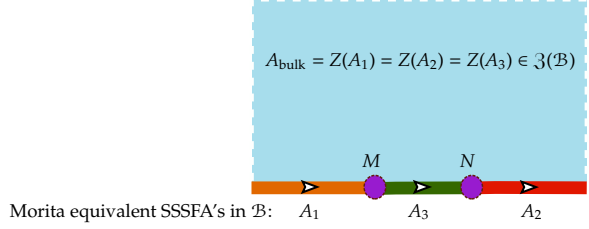}}}
   \put(-20,10){
     \setlength{\unitlength}{.75pt}\put(-18,-19){
     \put(-103,12) {\scriptsize Morita equivalent SSSFA's in $\CB$:}
     \put(43, 12)       {\scriptsize $A_1$}
     \put(155, 12)     {\scriptsize $ A_2 $}
     \put(98, 12)     {\scriptsize $ A_3 $}
     \put(75,37) {\scriptsize $M$}
     \put(124,37) {\scriptsize $N$}
     \put(26, 88)     {\scriptsize $ A_\cl=Z(A_1)=Z(A_2)=Z(A_3) \in \FZ(\CB)$}
          }\setlength{\unitlength}{1pt}}
  \end{picture}}
$$
\caption{This picture depicts a 1+1D world sheet, on which lives boundary CFT's $A_1,A_2,A_3\in \CB$ and a bulk CFT $A_\cl\in \FZ(\CB)$ given by the full center of the boundary CFT's. The domain wall $M$ is an invertible $A_1$-$A_2$-bimodule and $N$ is an invertible $A_2$-$A_3$-bimodule. 
}
\label{fig:bcft}
\end{figure}

\begin{thm} \label{thm:bcft-2}
For a given boundary-bulk CFT, if all its the boundary conditions preserve the same rational chiral symmetry $V$, and the bulk CFT has a unique vacuum, then 
\begin{itemize}
\item the bulk CFT $A_\cl$ is a Lagrangian algebra in $\FZ(\CB)$, which can be upgraded to a CSSSFA (unique up to a scalar factor for the counit) \cite{kr2}; 
\item boundary CFT's that share the same bulk $A_\cl$ are simple special symmetric Frobenius algebras (SSSFA) $A_\op$ in $\CB$ such that $A_\cl\simeq Z(A_\op)$ as algebras, where $Z(A_\op)$ is the full center of $A_\op$ \cite{frs1,fjfrs,kr2}; 
\item a domain wall between two different boundary CFT's $A_1$ and $A_2$ is given by a canonical invertible $A_1$-$A_2$-bimodule in $\Mod_V$ 
%(made more explicitly in Theorem\,\ref{thm:bcft-3}) 
\cite{ffrs,dkr2}. 
\end{itemize}
Two boundary CFT's are Morita equivalent if and only if their full centers are isomorphic as algebras \cite{kr1}. These results are illustrated in Figure\,\ref{fig:bcft}. 
\end{thm}

\begin{expl}  \label{expl:cardy-case}
In the so-called Cardy case \cite{fffs, frs1, kong-cardy} 
\begin{itemize}
\item the boundary CFT's are given by the SSSFA's in $\CB$ that are Morita equivalent to the trivial SSSFA $\one_\CB=V$. More precisely, these SSSFA's are precisely $x\otimes x^\ast$ for $x\in \CB$ equipped with the structure of a Frobenius algebra defined by 
\begin{align} \label{eq:frob-alg}
x\otimes x^\ast \otimes x \otimes x^\ast %\xrightarrow{m:=1v_x1} 
\xleftrightarrows[m:=1v_x1]{\Delta:=1v_x^\dagger1}
%\stackrel[\beta]{m:=1v_x1}{\leftrightarrows}
x\otimes x^\ast, &\quad  
\one_\CB \xleftrightarrows[\eta:=u_x]{\epsilon:=u_x^\dagger} x\otimes x^\ast; 
%\quad \Delta=m^\dagger, \quad \epsilon=\eta^\dagger. 
\end{align}

\item the bulk CFT is given by the full center $Z(x\otimes x^\ast)=Z(\one_\CB)=\oplus_{i\in \mathrm{Irr}(\CB)} i^\ast \boxtimes i$ in the category $\FZ(\CB)$ with the multiplication map defined by 
\be \label{eq:multiplication-Z(1)}
%\phi_2^R ~=
\bigoplus_{i,j,k\in \mathrm{Irr}(\CB)} \sum_{\alpha} ~~
  \raisebox{-25pt}{
  \begin{picture}(70,50)
   \put(0,8){\scalebox{.75}{\includegraphics{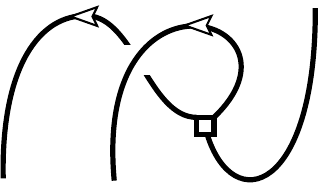}}}
   \put(0,8){
     \setlength{\unitlength}{.75pt}\put(-18,-11){
     \put(19, 2)        {\scriptsize $ i^\ast $}
     %\put(19, 120)    {\scriptsize $ x $}
     \put(48, 2)        {\scriptsize $ j^{\ast} $}
     \put(108, 65)  {\scriptsize $k^{\ast}$}
     \put(74, 35)      {\scriptsize $\alpha$}
     }\setlength{\unitlength}{1pt}}
  \end{picture}}
~~  \boxtimes ~~
\raisebox{-25pt}{
  \begin{picture}(20,50)
   \put(0,8){\scalebox{.75}{\includegraphics{pic-lambda-eps-converted-to.pdf}}}
   \put(0,8){
     \setlength{\unitlength}{.75pt}\put(-18,-11){
     \put(18, 2)     {\scriptsize $ i $}
     \put(46, 2)     {\scriptsize $ j $}
     \put(30, 26)   {\scriptsize $\alpha$} 
     \put(32, 62) {\scriptsize $ k $ }
     }\setlength{\unitlength}{1pt}}
  \end{picture}}
  \phantom{\frac{\dim U_i \,\dim U_j}{\dim U_k \,\dim\,\CC}} 
\ee
and the unit map defined by the canonical embedding $\one_\CB\boxtimes \one_\CB \hookrightarrow Z(\one_\CB)$ (see for example \cite{ffrs2,kr2}).  Note that $Z(\one_\CB)$ gives the famous charge conjugate modular invariant partition function;

\item the domain wall between two boundary CFT's $A=y\otimes y^\ast$ and $B=x\otimes x^\ast$ is given by the invertible $A$-$B$-bimodule $y\otimes x^\ast$ \cite{ffrs}. 

\end{itemize}
\end{expl}

%\begin{rem} \label{rem:xx-zz} Recall that CSSFA's $[x,x]$ for $x\in \CB$ are all Morita equivalent to the tensor unit $\one_\CB$, and are boundary CFT's of the same charge-conjugate modular invariant bulk CFT given by the full center of $\one_\CB$, i.e. $Z(\one_\CB)=\oplus_i i\boxtimes i^\ast$, in $\CB\boxtimes\overline{\CB}$. Except in a few cases (such as the Ising and Fibonacci UMTC's), in general, there are more SSSFA's in $\CB$ that are not Morita equivalent to $\one_\CB$. % (see Remark\,\ref{rem:cl-cft}). \end{rem}

\subsection{Internal homs} \label{sec:internal-hom}

It turns out that the object $y\otimes x^\ast$, appeared in Example\,\ref{expl:cardy-case}, is the simplest example of a universal construction called the internal hom. Since the notion of an internal hom plays a crucial role in this work, we review its definition.

\begin{defn}
Let $\CA$ be a monoidal category.  A {\em left $\CA$-module} $\CM$ is a category equipped with an action functor $\odot: \CA \times \CM \to \CM$, $(a,x) \mapsto a\odot x$ and natural isomorphisms: 
\be
\one_\CA \odot x \xrightarrow{\simeq} x \quad\quad \mbox{and} \quad\quad (a\otimes b) \odot x \xrightarrow{\simeq} a \odot (b\odot x) \quad\quad \forall a,b\in \CA,x\in\CM, \label{eq:oplax-action}
\ee
satisfying the following commutative diagrams:
$$
{\small 
\xymatrix@C=1em{
((a\otimes b) \otimes c) \odot x \ar[r] \ar[d] & (a \otimes (b\otimes c)) \odot x \ar[r] & a \odot ((b\otimes c) \odot x) \ar[d] \\
(a\otimes b) \odot (c\odot x) \ar[rr] & & a \odot (b\odot (c\odot x))
}
\quad \mbox{and} \quad 
\xymatrix@C=1em{
(a \otimes \one_\CA) \odot x \ar[r] \ar[dr] & a \odot (\one_\CA \odot x) \ar[d] \\
& a\odot x\, .
}
}
$$  
\end{defn}

\begin{expl}
$\CA$ itself is a left $\CA$-module with the action functor defined by the tensor product functor $\otimes:\CA\times \CA \to \CA$. 
\end{expl}

\begin{defn}[Internal hom by adjunction]
The {\em internal hom} $[x,y]_\CA$, or $[x,y]$ for simplicity, is an object in $\CA$ uniquely determined (up to unique isomorphism) by a family of isomorphisms: 
\be
\hom_\CM(a\odot x, y) \xrightarrow{\simeq} \hom_\CA(a, [x,y])   \label{eq:adjunction}
\ee
which are natural in all variables $a, x, y$. 
\end{defn}

If $f: a\odot x \to y$ is mapped to $g: a \to [x,y]$ under the isomorphism (\ref{eq:adjunction}), then $f$ and $g$ are called the mates of each other. Taking $a=[x,y]$, we denote the mate of the identity map $1_{[x,y]}: [x,y] \to [x,y]$ by $\ev: [x,y]\odot x \to y$. 

\begin{expl}
When $\CA$ is a UFC and $\CM=\CA$, we have $[x,y]\simeq y\otimes x^\ast$ and $\ev: [x,y]\odot x \to y$ is precisely given by $1_y\otimes v_x: y\otimes x^\ast \otimes x\to y$.
\end{expl}

We can equivalently define the notion of an internal hom by its universal property. 
\begin{defn}[Internal hom by the universal property] \label{def:int-hom-univ}
The {\em internal hom} is a pair 
$$ %\label{eq:[xy]-ev-1}
([x,y], \,\, [x,y]\otimes x \xrightarrow{\ev} y),
$$
which is universal among all such pairs. It means that for any such a pair $(Q, Q\otimes x\xrightarrow{f} y)$ there exists a unique morphism $\underline{f}: Q \to [x,y]$ rendering the following diagram commutative:
\be \label{eq:[xy]-ev-2}
\xymatrix{
& [x,y] \odot x \ar[rd]^\ev & \\
Q\otimes x \ar[rr]^f \ar[ur]^{\exists ! \, \underline{f} \otimes 1} & & y\, . 
}
\ee
\end{defn}

\begin{rem}
The standard mathematical symbol $\exists !$ represents the phrase ``there exists a unique''. It is an important and enjoyable exercise to show that this universal property determines $[x,y]$ up to unique isomorphism. Intuitively, the internal hom $[x,y]$ is in some sense the "maximal" one among all possible $(Q,f)$. 
\end{rem}

There are two canonical morphisms defined by their mates shown as follows:
\begin{align}
[y,z]\otimes [x,y]\rightarrow [x,z] \quad\quad &\longleftrightarrow \quad\quad  [y,z]\otimes [x,y] \otimes x \xrightarrow{\ev \circ (1\otimes \ev)} z \\
\one_\CA \to [x,x] \quad\quad &\longleftrightarrow \quad\quad \one_\CA \otimes x \xrightarrow{\simeq} x. 
\end{align}
When $x=y=z$, these morphisms endow $[x,x]$ with the structure of an algebra. 

\begin{expl}
When $\CA$ is a fusion category and $\CM=\CA$, the algebra structure on $[x,x]$ coincides with that of $x\otimes x^\ast$ defined in Example\,\ref{expl:cardy-case}. 
\end{expl}

\begin{expl} \label{expl:int-hom}
Let $\CA$ be a fusion category with the tensor product $\otimes$ and $A$ a simple separable algebra in $\CA$. The category $\CA_A$ of right $A$-modules is naturally an indecomposable semisimple left $\CA$-module category with the left $\CA$-action $\odot: \CA \times \CA_A \to \CA_A$ defined by $(a, x) \mapsto a\otimes x$. For $x,y\in \CA_A$, $[x,y]_\CA$ exists and we have
\be \label{eq:[xy]}
[x,y]_\CA \simeq (x\otimes_A y^*)^*. 
\ee
In particular, $[A,-]: \CA_A \to \CA$ is the precisely the forgetful functor. 
\end{expl}

%\medskip All separable algebras in a fusion category can be realized by internal homs. Indeed, when $\CA$ is a fusion category and $\CM$ is an indecomposable semisimple left $\CA$-module, $[x,x]$ is a simple separable algebra in $\CA$ for every non-zero $x\in \CM$. Moreover, let $\CA_{[x,x]}$ be the category of right $[x,x]$-modules in $\CA$. Then the functor $\CM \to \CA_{[x,x]}$ defined by $y\mapsto [x,y]$ is an equivalence of left $\CA$-modules \cite{ostrik}.  
%The category $\CA_{[x,x]|[x,x]}$ of $[x,x]$-$[x,x]$-bimodules in $\CA$ is again a fusion category and is monoidal equivalent to the category $\fun_\CA(\CM,\CM)$ of $\CA$-module functors, and is Morita equivalent to $\CA$ . 

%\medskip
The notion of full center can also be defined as an internal hom \cite{davydov}. Let $\CA$ be a fusion category and $\FZ(\CA)$ be its Drinfeld center. Let $A$ be a simple separable algebra in $\CA$. Then 
%$\CA_A$ is an indecomposable semisimple left $\CA$-module, and 
the category $\CA_{A|A}$ of $A$-$A$-bimodules in $\CA$ is a fusion category \cite{ostrik}. There is an action functor $\FZ(\CA) \times \CA_{A|A} \to \CA_{A|A}$ defined by $(z, x) \to z\otimes x$. 
\begin{defn} \label{def:full-center}
The {\em full center} $Z(A)$ of $A$ is the internal hom $[A,A]_{\FZ(\CA)}$. 
\end{defn}

%be an indecomposable left semisimple $\CA$-module and $\id_\CM$ the tensor unit of $\fun_\CA(\CM,\CM)$. There is a unital action functor $\FZ(\CA) \times \fun_\CA(\CM,\CM) \to \fun_\CA(\CM,\CM)$ defined by $(z, F) \to z\odot F(-)$. It defines a left $\FZ(\CA)$-module structure on $\fun_\CA(\CM,\CM)$. 

%For example, the bulk CFT in Example\, \ref{expl:cardy-case} is an internal hom. Indeed, when $\CA=\FZ(\Mod_V)$ and $\CM=\Mod_V$, there is an action functor $\odot: \FZ(\Mod_V) \times \Mod_V \to \Mod_V$ defined by $$\FZ(\Mod_V) \times \Mod_V \xrightarrow{\forget \times 1} \Mod_V \times \Mod_V \xrightarrow{\otimes} \Mod_V, $$ where $\forget$ is the forgetful functor. Then the full center $Z(\one_\CM)$ is nothing but $[\one_\CM,\one_\CM]_\CA$. 

%Recall that a condensable algebra in a UMTC has a unique structure of a simple normalized-special symmetric Frobenius algebra. %As a consequence, this Frobenius algebra is automatically $\dagger$-Frobenius. 

\subsection{Unitary boundary-bulk CFT's}  \label{sec:unitary-RCFT}

We have discussed the classification result of boundary-bulk rational CFT's. What about unitary rational CFT's? The representation theory of a unitary VOA has not been seriously studied until recently \cite{gui1,gui2,gui3}. As far as we know, the classification theory of unitary rational CFT's derived directly from Definition\,\ref{def:bcft-1} or \ref{def:bcft-2}, is not yet available. There is, however, another formulation of a unitary rational CFT based on the representation theory of rational conformal nets, and its classification theory was known \cite{longo-rehren,rehren1,rehren2}. It is essentially the same as that of boundary-bulk CFT's given in Theorem\,\ref{thm:bcft-1} but with a rational VOA replaced by a rational conformal net, whose module category is a UMTC, and the two types of Frobenius algebras replaced by two types of $\dagger$-Frobenius algebras (see \cite{kr3} for a review). A {\em $\dagger$-Frobenius algebra} in a UMTC is a Frobenius algebra satisfying the conditions $\Delta=m^\dagger$ and $\epsilon=\eta^\dagger$. For example, the Frobenius algebra constructed in (\ref{eq:frob-alg}) is a $\dagger$-SSSFA. We will assume that this classification also works for the unitary boundary-bulk CFT's defined in Definition\,\ref{def:bcft-1} or Definition\,\ref{def:bcft-2}. We make the classification result more precise below. 

\medskip
From now on, we assume that $V$ is a unitary rational VOA of central charge $c$ such that the category of unitary $V$-modules (see \cite{gui1}), still denoted by $\Mod_V$, is a UMTC. %We set $\CB=\Mod_V$ for a simpler notation. 

\medskip
If $\CB$ is a UMTC and $\CM$ is an indecomposable unitary left $\CB$-module, then the category $\fun_\CB(\CM,\CM)$ of $\CB$-module functors from $\CM$ to $\CM$
%, which is monoidal equivalent to the category of $\CB$-module $\ast$-functors, 
can be upgraded to a UFC \cite{ghr}. Therefore, for non-zero $x\in\CM$, the category $\CB_{[x,x]|[x,x]} \simeq \fun_\CB(\CM,\CM)$ can be upgraded to a UFC. In this case, the simple separable algebra $[x,x]$ is automatically a Frobenius algebra with $\Delta:=m^\dagger$ and $\epsilon:=\eta^\dagger$. Moreover, it is a $\dagger$-SSSFA because $m^\dagger$ is automatically a bimodule map
%\footnote{Without loss of generality, the Frobenius algebra can be chosen to be normalized-special, because the rescaling $m\mapsto \lambda^{-1}m$ and $\eta\mapsto \lambda\eta$ defines an isomorphic algebra.} 
by the unitarity of $\CB_{[x,x]|[x,x]}$.

%In this case, for a right $A$-module $(M,\mu_M: M\otimes A \to M)$, $\mu_M^\dagger: M \to M \otimes A$ is also a right $A$-module map and can be explicitly expressed as follows: 

Conversely, all $\dagger$-SSSFA's in $\CB$ can be realized by internal homs. Indeed, for a $\dagger$-SSSFA $A$, 
\void{
a right $A$-module $M$ in $\CB$ is called a {\em $\dagger$-module} if the following equation 
\be \label{eq:mu-dagger}
\mu_M^\dagger = (\mu_M \otimes 1_A) \circ (1_M \otimes (\Delta \circ \eta)) 
\ee
holds. Then 
}
the category $\CB_A$ of right $A$-modules is an indecomposable unitary left $\CB$-module. If $x$ is a right $A$-module, then $x^\ast$ is automatically a left $A$-module. For $x,y\in \CB_A$, we have \cite{ostrik}
$$ %\label{eq:[xy]}
[x,y] = (x\otimes_A y^\ast)^\ast \simeq y\otimes_A x^\ast. 
$$
In particular, $A \simeq [A,A]$ as Frobenius algebras.

If $B$ is a $\dagger$-SSSFA in $\CB$ that is Morita equivalent to $A$, we can also realize $B$ as an internal hom in the following way. Suppose that an invertible $B$-$A$-bimodule $x$ defines the Morita equivalence. 
%We have $$\Cb \simeq \hom_{B|B}(B, B) \simeq \hom_{B|B}(y\otimes_A x, B) \simeq \hom_{B|A}(y, B\otimes_B x^\ast) \simeq \hom_{B|A}(y,x^\ast), $$ where, in the second ``$\simeq$'', we have used the fact that the equalizer is automatically a coequalizer in a unitary category. As a consequence, $y\simeq x^\ast$ as $B$-$A$-bimodules. Moreover, since 
Since $-\otimes_B x: \CB_B \to \CB_A$ defines an equivalence between the two left $\CB$-modules and maps $B$ to $x$, we obtain $B\simeq[B,B]\simeq [x,x]$ as $\dagger$-SSSFA's.

\void{
\begin{conj} \label{conj:bcft}
A unitary boundary-bulk CFT (defined in Definition\,\ref{def:bcft-1} or Definition\,\ref{def:bcft-2}) with all its boundary conditions preserving the same chiral symmetry $V$ and a unique vacuum in the bulk can be mathematically described as follows: 
\begin{itemize}
\item the bulk CFT $A_\cl$ in $\FZ(\Mod_V)$ is a Lagrangian algebra such that its canonical upgrading to a CSSSFA is $\dagger$-Frobenius (i.e. $\dagger$-CSSSFA); 
\item all boundary CFT's that share the same bulk $A_\cl$ are simple normalized-special symmetric $\dagger$-Frobenius algebras ($\dagger$-SSSFA's) $A_\op$ in $\Mod_V$ such that $A_\cl\simeq Z(A_\op)$, where $Z(A_\op)$ is the full center of $A_\op$; 
\item a domain wall between two different boundary CFT's $A_1$ and $A_2$ is given by an invertible $A_1$-$A_2$-bimodule in $\Mod_V$
\end{itemize} 
\end{conj}

\begin{expl}
By our assumption on $V$ in Assumption\,\ref{ass:unitary-rational}, the Cardy case boundary-bulk CFT defined in Example \ref{expl:cardy-case} is actually unitary.  
\end{expl}
}

\medskip
For applications in unitary boundary-bulk CFT's with only $V$-invariant boundary conditions, we set $\CB=\Mod_V$. 
For a given unitary boundary CFT $A$, i.e. a $\dagger$-SSSFA in $\CB$, all unitary boundary CFT's that share the same bulk with $A$ are those $\dagger$-SSSFA's in $\CB$ that are Morita equivalent to $A$ \cite{kr1}. Therefore, all of them can be recovered as internal homs $[x,x]$ for $x\in \CB_A$. The category $\CB_A$ is an indecomposable unitary left $\CB$-module, which is uniquely determined by the unitary bulk CFT $Z(A)$ up to equivalence. Moreover, there is a bijection from the set of the equivalence classes of indecomposable unitary left $\CB$-modules to that of the equivalence classes of Lagrangian algebras in $\FZ(\CB)$ defined by $\CM \mapsto [\id_\CM, \id_\CM]_{\FZ(\CB)}$ \cite{kr1,dmno}, where $\id_\CM$ is the tensor unit of the UFC $\fun_\CB(\CM,\CM)$ and the action functor $\FZ(\CB) \times \fun_\CB(\CM,\CM) \to \fun_\CB(\CM,\CM)$ is defined by $(z, F) \to z\odot F(-)$. When $\CM=\CB_A$, we have $\fun_\CB(\CM,\CM)=\CB_{A|A}$ and $\id_\CM=A$. 
%The inverse map is defined by $L \mapsto \FZ(\CB)_L^\dagger$ for a Lagrangian algebra in $\FZ(\CB)$. 

\medskip
Combining the above discussion with the assumption that unitary boundary-bulk CFT's based on VOA's is equivalent to those based on conformal nets \cite{longo-rehren,rehren1,rehren2}, we obtain the following physical ``theorem''. 
\begin{pthm} \label{thm:bcft-3}
Let $\CB=\Mod_V$ be a UMTC. For a given unitary bulk CFT $A_\cl$ in $\FZ(\CB)$ with a unique vacuum, we have the following assertions. 
\begin{itemize}
\item The category of boundary conditions of $A_\cl$ is given by an indecomposable unitary left $\CB$-module $\CM$ which is canonically associated to $A_\cl$. We have $\CM\simeq \CB_A$ for a unitary boundary CFT $A$, i.e. a $\dagger$-SSSFA in $\CB$ such that $Z(A)\simeq A_\cl$ as algebras. An object $x\in\CM$ is called a boundary condition. 
\item %Let $\id_\CM$ be the tensor unit in $\fun_\CB(\CM,\CM)$. Then 
$A_\cl=[\id_\CM,\id_\CM]_{\FZ(\CB)}$ is a Lagrangian algebra in $\FZ(\CB)$.   
\item For $x\in \CM$, the unitary boundary CFT associated to the boundary condition $x$ is given by $[x,x]_\CB$, which is a $\dagger$-SSSFA in $\CB$ such that $Z([x,x]_\CB)\simeq A_\cl$; 
\item For $x,y\in \CM$, the domain wall between the two boundary CFT's $[x,x]_\CB$ and $[y,y]_\CB$ is precisely given by the invertible bimodule $[x,y]_\CB$. 
\end{itemize}
\end{pthm}

\begin{rem}
What we are saying above is that upgrading a rational CFT (Theorem\,\ref{thm:bcft-2}) to a unitary rational CFT (Theorem$^{\mathrm{ph}}$\,\ref{thm:bcft-3}) amounts to replacing the adjective ``Frobenius'' by ``$\dagger$-Frobenius''. % and ``modules'' by ``$\dagger$-modules''.
\end{rem}

\begin{rem} \label{rem:stable-bcft}
Strictly speaking, we should require not only the bulk CFT to have a unique vacuum, but each boundary CFT also to have a unique vacuum. Usually, a QFT (defined on $S^n$ or $\Rb^n$) with multiple vacuums is not stable. It will flow to a stable one (with a unique vacuum) under perturbation. This requirement amounts to consider only simple boundary conditions in $\CM$ and boundary CFT's $[x,x]_\CB$ for simple $x\in \CM$, or equivalently, haploid $\dagger$-SSSFA's $A$ (i.e. $\dim\hom_\CB(\one_\CB,A)=1$). For this reason, we sometimes discuss only simple boundary conditions and simple topological edge excitations in many physical discussions. 
\end{rem}

\begin{rem} \label{rem:bcft-enrich-cat}
For the category of boundary condition $\CM$ of a given bulk CFT $A_\cl$, we have a $\CB$-enriched category ${}^\CB\CM$ via the canonical construction (see Example\, \ref{exam:SC}). For a condensable algebra $A$ in $\CB$, the category $\CM:=\overline{\CB}_A$ is a UFC and $-\otimes A: \overline{\CB} \to \CM$ is a central functor. We obtain a $\CB$-enriched monoidal category ${}^\CB\CM$ via the canonical construction given in Theorem\,\ref{thm:SC}. 
\end{rem}

\begin{expl}
The unique unitary rational VOA of central charge $c=0$ is $V=\Cb$. In this case, Theorem${}^{\mathrm{ph}}$ gives a classification of 1+1D unitary wall-boundary-bulk TQFT's. In particular, in this case, $\CB=\bh$, the only Lagrangian algebra in $\bh$ is $\Cb$. The category of boundary conditions associated to the algebra $\Cb$ is $\bh$, which is the unique indecomposable semisimple $\bh$-module. For $x,y\in\bh$, the associated boundary TQFT's are matrix algebras $[x,x]=x\otimes_\Cb x^\ast=\hom_\bh(x,x)$ and $[y,y]$, respectively, and the domain wall between them is $[x,y]=y\otimes x^\ast=\hom_\bh(x,y)$. Note that the full center of $[x,x]$ is just the usual center $Z([x,x])=\Cb$. 
\end{expl}

\begin{rem} \label{rem:internal-hom-include-all}
Theorem${}^{\mathrm{ph}}$\,\ref{thm:bcft-3} plays a crucial in this work. We want physics readers to keep in mind that this radical simplification of wall-boundary-bulk CFT's to internal homs does not lose any physical information. More precisely, a single internal hom $[x,y]_\CB$ or $[\id_\CM,\id_\CM]_{\FZ(\CB)}$ (for $\CB=\Mod_V$) automatically includes the information of the spectrum, chiral (or non-chiral) fields, correlation functions, OPE, structure constants and partition functions on torus or any other higher genus surfaces, etc. The key point of this miracle is stated in Remark\,\ref{rem:vertex-tensor-cat}. 

\end{rem}

\section{Constructions of gapless edges} \label{sec:examples}
In Section\,\ref{sec:can-edge}, we give an explicit construction of the so-called canonical gapless edge of a 2d chiral topological order, and show that it is a special case of the so-called canonical construction. In Section\,\ref{sec:general-edge}, we construct more gapless edges by fusing canonical gapless edges with gapped domain walls. 

\subsection{A natural construction of chiral gapless edge of $(\CB,c)$} \label{sec:can-edge}

We are ready to give an explicit construction of a chiral gapless edge for a 2d chiral topological order defined by a pair $(\CB,c)$, where $\CB$ is a UMTC realized by a unitary rational VOA $V$, i.e. $\CB=\Mod_V$ and $c$ is the central charge of $V$. We denote the tensor product in $\CB$ by $\otimes=\otimes_V$

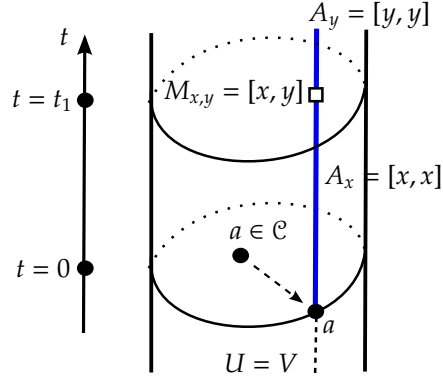
\begin{figure} 
$$
 \raisebox{-70pt}{
  \begin{picture}(130,130)
   \put(-20,8){\scalebox{0.6}{\includegraphics{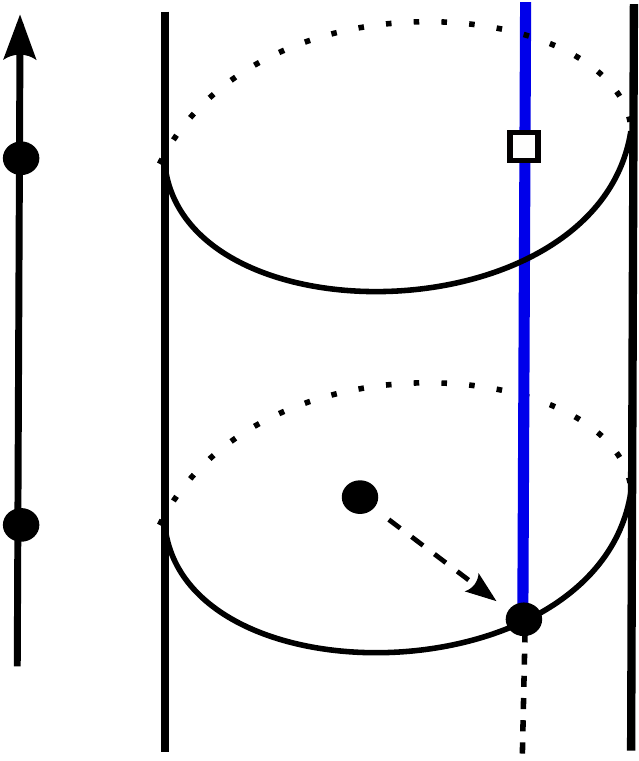}}}
   \put(-20,8){
     \setlength{\unitlength}{.75pt}\put(0,-83){
     \put(-30,133)  {$ t=0 $}
     \put(-32,219)  {$ t=t_1$}
     \put(-8, 250)  {$t$}
     %\put(65,120)  {$ \CB $}
     \put(78,152)  {$ a \in \CC$}
     \put(126,180)  {$ A_x = [x,x] $}
     \put(118,262)  {$A_y=[y,y]$}
     \put(45,223)   {$M_{x,y}=[x,y]$}
     \put(75,85) {$U=V$}
     \put(125, 105) {$a$}
     }\setlength{\unitlength}{1pt}}
  \end{picture}}
\void{
\quad\quad\quad \rightsquigarrow \quad\quad\quad 
\raisebox{-70pt}{
  \begin{picture}(100,75)
   \put(0,40){\scalebox{0.6}{\includegraphics{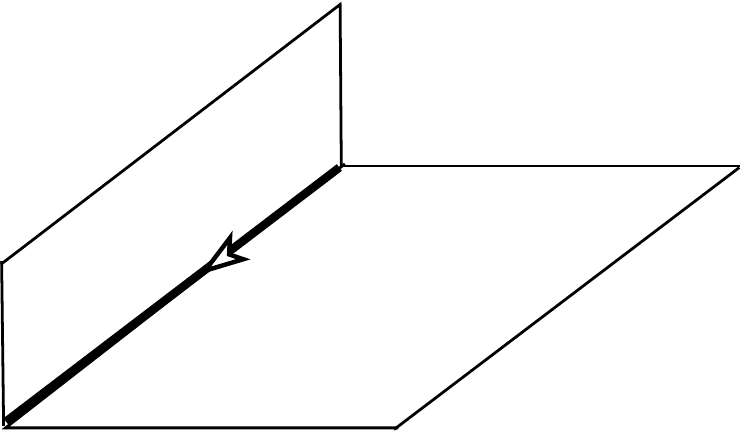}}}
   \put(0,40){
     \setlength{\unitlength}{.75pt}\put(0,0){
     \put(105,40)  {\scriptsize $ (\CB,c) $}
     \put(38,50) {\scriptsize $\CB$}
     \put(38, 77)  {\scriptsize $V$}
     \put(38,21)   {\scriptsize $\CB$}
     \put(86,82)  {\scriptsize $(V,{}^\CB\CB)$}
     %\put(110, 80) {$x$}
     }\setlength{\unitlength}{1pt}}
  \end{picture}}
  }
$$
%$$(a) \quad\quad\quad\quad\quad\quad\quad\quad\quad \quad\quad\quad\quad\quad\quad\quad\quad\quad (b)$$
\caption{This picture depicts a 2d topological order $(\CB,c)$, together with the canonical gapless edge $(V, {}^\CB\CB)$, in which $\hom_{{}^\CB\CB}(x,y):=[x,y]=y\otimes x^\ast \in \CB$ for $x,y\in\CB$. %The picture (b) provide a simple graphical way to record the gapless edge $(V, {}^\CB\CB)$. Two $\CB$'s in the pair $(\CB,\CB)$, which is used to define ${}^\CB\CB$ via the canonical construction, are given independent meanings. 
}
\label{fig:cylinder-2}
\end{figure}

\medskip
%We will call the following gapless edge as the {\it canonical gapless edge} of $(\CB,c)$. 
In this case, we give a very natural construction of a chiral gapless edge of the 2d topological order $(\CB,c)$. 
We describe its ingredients explicitly as follows (also illustrated in Figure\,\ref{fig:cylinder-2}): 
\begin{itemize}
\item Topological edge excitations are all obtained from moving topological bulk excitations to the edge. They are labeled by objects in $\CB$. 
%\item $\CB=\Mod_V$ for a rational VOA $V$. 
\item $U=M_{\one_\CB,\one_\CB}=[\one_\CB,\one_\CB]=\one_\CB=V$, namely, all the boundary CFT's $A_x$ and walls between them $M_{x,y}$ are required to preserve the largest chiral symmetry $V$. As a consequence,  $M_{x,y}$ are objects in $\CB$. 

%\item $M_x=x$ as a $V$-module. For each $x\in \CB$, the boundary CFT $A_x$ is given by the internal hom $[x,x]:=x\otimes x^\ast$, which is a CSSFA in $\CB$ \cite{fffs,frs1,kong-cardy}. Its algebra structures are defined in Eq.\,(\ref{eq:composition}) when $x=y=z$. When $x=\one_\CB$, we have $A_{\one_\CB}=\one_\CB=V$. 

\item $M_{x,y}=[x,y]=y\otimes x^\ast \in \CB$. In particular, $M_{\one,x}=[\one,x]=x$. 

\item Defect fields in $M_{x,y}$ can be fused with those in $M_{y,z}$ to give defect fields in $M_{x,z}$. This amounts to a morphism $[y,z] \otimes [x,y] \to [x,z]$ in $\CB$, which is defined as follows: 
\be \label{eq:composition-2}
[y,z] \otimes [x,y] = z\otimes y^\ast \otimes y \otimes x^\ast \xrightarrow{1_z \otimes v_y \otimes 1_{x^\ast}} z\otimes x^\ast = [x,z]. 
\ee
In this case, $[x,y]$ is automatically an invertible $[y,y]$-$[x,x]$-bimodule.
\item The spatial fusion of topological edge excitations $M_{x',y'}\otimes M_{x,y} \to M_{x'\otimes x, y'\otimes y}$ (depicted in Figure\,\ref{fig:fusion}) is a morphism in $\CB$ defined as follows: 
\be \label{eq:tensor-product}
[x',y']\otimes [x,y]  = y' \otimes x'^\ast \otimes y \otimes x^\ast \xrightarrow{1_{y'} \otimes c_{x'^\ast, y\otimes x^\ast}} (y' \otimes y) \otimes (x'\otimes x)^\ast = [x'\otimes x, y'\otimes y],
\ee
where we have used our braiding convention: the braiding $c_{x', y\otimes x^\ast}: x' \otimes (y\otimes x^\ast) \to (y\otimes x^\ast) \otimes x'$ is defined by moving $x'$ from $t_{x'}>t_{yx^\ast}$ to $t_{x'}<t_{yx^\ast}$ along a path lying entirely to the left of the world line supported on $x$ in Figure\,\ref{fig:fusion}. 

\end{itemize}

\begin{rem}
This chiral gapless edge of $(\CB,c)$ is the most studied edge in physics. But our description of the complete set of observables and their fusions in (\ref{eq:composition}) and (\ref{eq:tensor-product}) is new. 
\end{rem}

\begin{rem} \label{rem:gannon}
For a generic $(\CB,c)$, it is not known if there exists any unitary rational VOA $V$ of central charge $c$ such that $\CB\simeq\Mod_V$. If it indeed exists, it is not known how many are there? In a special case, when $c=0$, the only unitary VOA of central charge zero is $V=\Cb$ \cite{gannon2}. It means that, in this case, a 2d topological order $(\CB,0)$ can have gapped edges ($V=\Cb$) but does not have any non-trivial chiral gapless edges. It may have non-chiral gapless edges \cite{cjkyz,kz4}. 
\end{rem}

These observables on the above chiral gapless edge of $(\CB,c)$ can be summarized by a pair $(V,\CBs)$, where $\CBs$ is a $\CB$-enriched monoidal category defined as follows:
\begin{itemize}

\item an object in $\CBs$ is a topological edge excitation, i.e. an object in $\CB$; 

\item $\hom_\CBs(x,y)=[x,y]=y\otimes x^\ast$;  

\item the identity morphism $\id_x: \one_\CB \to [x,x]=x\otimes x^\ast$ is defined by $u_x: \one_\CB \to x\otimes x^\ast$; %In terms of chiral fields, it is also the OSVOA homomorphism $\iota: V \to A_x=[x,x]$; 

\item the composition morphism $\circledcirc:[y,z] \otimes [x,y] \to [x,z]$ is defined by Eq.\,(\ref{eq:composition-2}); 

\item the morphism $\otimes: [x',y']  \otimes [x,y]  \to [x'\otimes x, y'\otimes y]$ is defined by Eq.\,(\ref{eq:tensor-product}). 

\end{itemize}
It was proved by Morrison and Penneys in \cite{MP} that this categorical structure $\CBs$ is indeed a $\CB$-enriched monoidal category. Moreover, it is just a special case of the so-called canonical construction.

\begin{thm}[{\bf Canonical Construction} \cite{MP}] \label{thm:SC}
Let $\CD$ be a braided monoidal category and $\CY$ a monoidal category equipped with a braided oplax monoidal functor $F_\CY: \overline{\CD} \to \FZ(\CY)$ such that $F_\CY(\one)=\one$. It endows $\CY$ with structure of an action functor $\odot: \CD \times \CY \to \CY$ defined by $(c,y)\mapsto c\odot y:=F_\CY(c)\otimes y$. We assume that internal hom $[x,y]$ in $\CD$ exists for all $x,y\in \CY$. Then we obtain from the triple $(\CD,\CY, F_\CY)$ a $\CD$-enriched monoidal category, denoted by ${}^\CD\CY$, as follows: 
\bnu
\item objects in ${}^\CD\CY$ are precisely the objects in $\CY$; 
\item for $x,y\in \CY$, $\hom_{{}^\CD\CY}(x,y):=[x,y]$ in $\CD$;  
\item the identity morphism $\id_x: \one_\CD \to [x,x]$ is the morphism in $\CD$ given by the mate of the canonical isomorphism $\one_\CD\odot x \simeq x$ in $\CY$; 

\item the composition morphism $\circ: [y,z] \otimes [x,y] \to [x,z]$ is the mate of the following composed morphism:
$$
([y,z]\otimes [x,y])\odot x \to [y,z] \odot ([x,y] \odot x) \xrightarrow{1_{[y,z]}\odot \ev} [y,z]\odot y \xrightarrow{\ev} z;
$$ 
\item $\otimes: [x',y']\otimes [x,y] \to [x'\otimes x, y'\otimes y]$ is the mate of the following composed morphism:
\begin{align}
([x',y']\otimes [x,y]) \odot (x' \otimes x) &= F_\CY([x',y'] \otimes [x,y]) \otimes x' \otimes x \nn
&\rightarrow F_\CY([x',y']) \otimes F_\CY([x,y]) \otimes x' \otimes x \nn
&\hspace{-2cm} \xrightarrow{1 \otimes \beta_{F_\CY([x,y]), x'} \otimes 1_x}  F_\CY([x',y']) \otimes x' \otimes F_\CY([x,y]) \otimes  x \to y' \otimes y, \nonumber
\end{align}
where $\beta_{F_\CY([x,y]), x'}: F_\CY([x,y]) \otimes x' \to x' \otimes F_\CY([x,y])$ is the half-braiding. 
\enu
The underlying category of ${}^\CD\CY$ is $\CY$, and the background category of ${}^\CD\CY$ is $\CD$. 
\end{thm}

\medskip
Using the notation of the canonical construction, we see that $\CBs={}^{\CB}\CB$. From now on, we denote $(V,\CBs)$ by the pair $(V,{}^\CB\CB)$. As we will show later, only physically relevant $F_\CY$ are braided monoidal functors (NOT oplax). Hence, among all possible $\CY$ (viewed as a finite left $\CB$-module), $\CY=\CB$ is the canonical choice. For this reason, from now on, we will refer to $(V,{}^\CB\CB)$ as the {\it canonical chiral gapless edge of $(\CB,c)$}. 

\begin{rem}
Strictly speaking, for a given 2d topological order $(\CB,c)$, there is no canonical chiral gapless edge even if we assume the existence of a VOA $V$ such that $\CB\simeq\Mod_V$ because such VOA's are often not unique (see \cite{cmbcn} and \cite{dm,ems,lamlin}). Once $V$ is fixed, however, the categorical data ${}^{\CB}\CB$ is indeed canonical with respect to $\CB$. 
\end{rem}

We would like to provide a graphical notation for this canonical gapless edge as shown below.  
\be \label{graph:can-edge}
(V, {}^\CB\CB) \quad\quad  = \quad\quad
\raisebox{-30pt}{
  \begin{picture}(100,65)
   \put(0,10){\scalebox{0.5}{\includegraphics{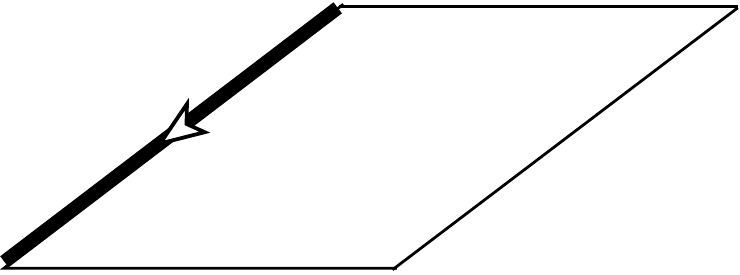}}}
   \put(0,10){
     \setlength{\unitlength}{.75pt}\put(0,0){
     \put(90,35)  {\scriptsize $ (\CB,c) $}
     %\put(38,47) {\scriptsize $\CB$}
     %\put(38, 71)  {\scriptsize $V$}
     %\put(38,21)   {\scriptsize $\CB$}
     \put(0,33)  {\scriptsize $(V,{}^\CB\CB)$}
     %\put(110, 80) {$x$}
     }\setlength{\unitlength}{1pt}}
  \end{picture}}
\quad\quad = \quad\quad
\raisebox{-27pt}{
  \begin{picture}(100,65)
   \put(0,0){\scalebox{0.5}{\includegraphics{pic-edge-eps-converted-to.pdf}}}
   \put(0,0){
     \setlength{\unitlength}{.75pt}\put(0,0){
     \put(95,35)  {\scriptsize $ (\CB,c) $}
     \put(32,47) {\scriptsize $(\CB,c)$}
     %\put(38, 71)  {\scriptsize $V$}
     \put(38,21)   {\scriptsize $\CB$}
     \put(70,67)  {\scriptsize $(V,{}^\CB\CB)$}
     %\put(110, 80) {$x$}
     }\setlength{\unitlength}{1pt}}
  \end{picture}}
\ee
In the second picture, we try to provide different ``physical meanings'' to the two ``$\CB$''s in ${}^\CB\CB$. 
\begin{itemize}
\item The background category ``$\CB$'', as a UMTC, can be interpreted as a 2d topological order. Therefore, we use this ``$\CB$'' to label a ``fictional 2d phase'' in the time direction (depicted as the vertical plane). Note that the fictional vertical plane also remind us the 1+1D world sheet of this chiral gapless edge. 

\item The underlying category ``$\CB$'', viewed as a UFC (by forgetting its braiding), is the category of topological edge excitations. It can be interpreted as a ``fictional gapped domain wall'' between the fictional 2d phase in the time direction and the physical 2d phase $(\CB,c)$ in the spatial dimensions. Notice that $\CB$ is indeed a legitimate gapped domain wall (actually the trivial one) between two 2d phases $(\CB,c)$ and $(\CB,c)$. 
\end{itemize}

Note that the whole fictional vertical plane, including the fictional ``wall'' labeled by $\CB$, should be viewed as a single gapless edge $(V,{}^\CB\CB)$. This graphical notation of canonical gapless edge in (\ref{graph:can-edge}) has an immediate advantage. It seems to suggest that this gapless edge can be obtained, as illustrated below, by starting from a horizontal plane in the spatial dimensions then folding the left half of the plane to a vertical spacetime plane. 
\be \label{pic:wick-1}
\raisebox{-30pt}{
  \begin{picture}(120,75)
   \put(0,15){\scalebox{0.5}{\includegraphics{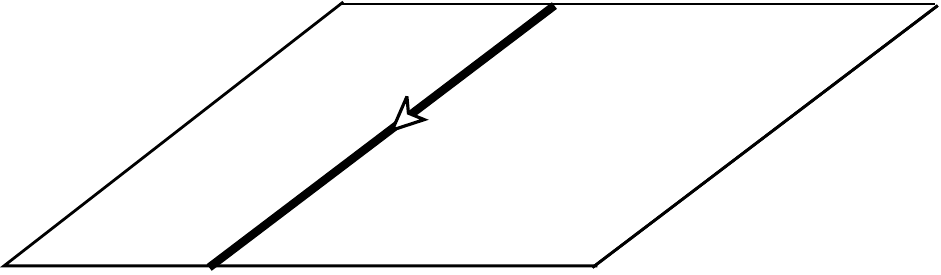}}}
   \put(0,15){
     \setlength{\unitlength}{.75pt}\put(0,0){
     \put(140,40)  {\scriptsize $ (\CB,c) $}
     \put(15,7) {\scriptsize $(\CB,c)$}
     %\put(38,21)   {\scriptsize $\CB$}
     \put(85,27)   {\scriptsize $\CB$}
     %\put(110, 80) {$x$}
     }\setlength{\unitlength}{1pt}}
  \end{picture}} 
\quad \xrightarrow{\mbox{\footnotesize topological Wick rotation}} \quad 
\raisebox{-30pt}{
  \begin{picture}(100,75)
   \put(0,10){\scalebox{0.5}{\includegraphics{pic-edge-eps-converted-to.pdf}}}
   \put(0,10){
     \setlength{\unitlength}{.75pt}\put(0,0){
     \put(100,40)  {\scriptsize $ (\CB,c) $}
     \put(25,43) {\scriptsize $(\CB,c)$}
     \put(70, 70)  {\scriptsize $(V,{}^\CB\CB)$}
     \put(30,15)   {\scriptsize $\CB$}
     %\put(110, 80) {$x$}
     }\setlength{\unitlength}{1pt}}
  \end{picture}}
\ee
Of course, such a folding process is physically impossible. A picture might be closer to the physical reality is the following one. Imagine that we cut out the left half of a 2d topological order $(\CB,c)$ defined on $\Rb^2$ by brutal force. It takes the system in the neighborhood of the edge away from a RG fixed point. As time goes by, the edge will undergo a self-healing process by flowing to a new RG fixed point. One of the possible RG fixed point is the canonical chiral gapless edge $(V,{}^\CB\CB)$. We will show later that, in general, there are other RG fixed points corresponding to different chiral gapless edges. What makes this picture so amazing is its holographical nature. That is, the information lost through the brutal force cutting in the spatial dimensions can be completely restored in the temporal dimension!

Therefore, we believe that this fictional folding process can not be just an ad hoc bookkeeping trick. It should have some yet-to-be-clarified deep physical meanings, and will be proved to be very useful later. %For example, in Section\,\ref{sec:underlying-cat}, we will show that the category of topological edge excitations is precisely the so-called underlying category of ${}^\CB\CB$. 
Therefore, we would like to refer to this fictional folding trick as a ``topological Wick rotation''.

\begin{rem} \label{rem:chiral-cft}
It is generally accepted that a chiral gapless edge of a 2d topological order is given by a so-called ``chiral CFT'', the precise meaning of which has never been clarified. Our result in this subsection gives a precise meaning to a ``chiral CFT'' as a pair $(V,{}^\CB\CB)$. Note that the spaces of conformal blocks do not live on the 1+1D world sheet of the gapless edge directly. It can, however, be recovered by the underlying category of ${}^\CB\CB$.
\end{rem}

%The conclusion of this section is that the complete mathematical description of the canonical gapless edge of $(\CB,c)$ is given by a pair $(V, \CBs)$, where $V$ is a VOA such that $\Mod_V=\CB$, and $\CBs$ is the $\CB$-enriched monoidal category obtained from the pair $(\CB,\CB)$ via the canonical construction given in Remark\,\ref{rem:canonical-construction}. 

\begin{rem}
The unitary boundary-bulk CFT obtained by applying the dimensional reduction depicted in Figure\,\ref{fig:cylinder-1} on the canonical gapless edge is precisely the Cardy case presented in Example\,\ref{expl:cardy-case}.
\end{rem}

It turns out that the canonical construction of enriched monoidal categories also naturally includes the mathematical description of a gapped edge of a 2d topological order, i.e. a UFC, as a special case. Indeed, for an ordinary UFC $\CM$, there is a unitary braided monoidal functor $\bh \hookrightarrow \FZ(\CM)$ defined by $\Cb\mapsto \one_{\FZ(\CM)}$. The UFC $\CM$ can be viewed as the $\bh$-enriched monoidal category obtain from the triple $(\bh, \CM, \bh\hookrightarrow \FZ(\CM))$ via the canonical construction, i.e. $\CM={}^\bh\CM$. Moreover, $\bh$ can be viewed as the UMTC $\Mod_V$ for $V=\Cb$, which should be viewed as the trivial unitary rational VOA of central charge $c=0$. %Actually, $\Cb$ is the unique unitary rational VOA of central charge $c=0$. 
As a consequence, a gapped edge $\CM$ of a 2d topological order is described by the pair $(\Cb,{}^\bh\CM)$, where the underlying category is again the category of topological edge excitations $\CM$ and the background category is $\bh$. In this case, $[x,x]_\bh=\hom_\CM(x,x)$, as a direct sum of matrix algebras, should be viewed as a boundary TQFT of the trivial bulk TQFT given by its full center $Z([x,x]_\bh)=\Cb$.

%Note that the usual graphic way of presenting a gapped edge as in the first picture in Eq.\,(\ref{pic:fusing-edge-wall}) is the same as the second one because the fictional ``vertical bulk phase'' is the trivial topological order $\bh$ thus can be ignored. 

\subsection{General gapless edges}  \label{sec:general-edge} 

Let $\CB=\Mod_V$. Let $(\CC,c)$ be a 2d topological order that is Witt equivalent to $(\CB,c)$. In other words, the 2d topological orders $(\CB,c)$ and $(\CC,c)$ can be connected by a gapped domain wall, which is described by 
\begin{itemize}
\item a UFC $\CM$ equipped with a unitary braided monoidal equivalence $\phi_\CM: \overline{\CB}\boxtimes \CC \xrightarrow{\simeq} \FZ(\CM)$ (i.e. a closed fusion $\CB$-$\CC$-bimodule \cite[Definition\,2.6.1]{kz1}).
\end{itemize}
In this case, by fusing this gapped domain wall $\CM$ with the canonical gapless edge $(V,{}^\CB\CB)$ of $(\CB,c)$, we obtain a gapless edge of $(\CC,c)$. We illustrate this fusion process below: 
\be \label{pic:fusing-edge-wall}
\raisebox{-30pt}{
  \begin{picture}(120,75)
   \put(0,10){\scalebox{0.5}{\includegraphics{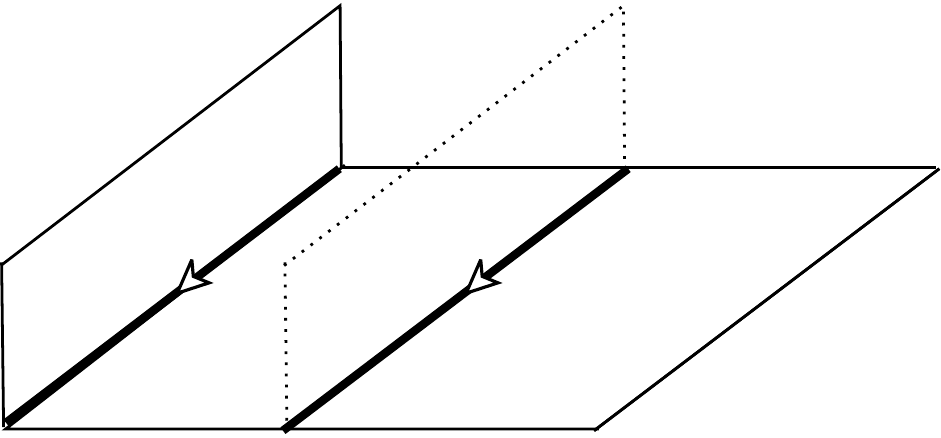}}}
   \put(0,10){
     \setlength{\unitlength}{.75pt}\put(0,0){
     \put(140,40)  {\scriptsize $ (\CC,c) $}
     \put(38,47) {\scriptsize $\CB$}
     \put(28,7) {\scriptsize $(\CB,c)$}
     \put(70, 75)  {\scriptsize $(V,{}^\CB\CB)$}
     \put(38,21)   {\scriptsize $\CB$}
     \put(105,30)   {\scriptsize $\CM$}
     \put(105,60)  {\scriptsize $\bh$}
     \put(122,75) {\scriptsize $(\Cb,{}^\bh\CM)$}
     %\put(110, 80) {$x$}
     }\setlength{\unitlength}{1pt}}
  \end{picture}} 
\quad\quad\quad \rightsquigarrow \quad\quad\quad 
\raisebox{-30pt}{
  \begin{picture}(100,75)
   \put(0,10){\scalebox{0.5}{\includegraphics{pic-edge-eps-converted-to.pdf}}}
   \put(0,10){
     \setlength{\unitlength}{.75pt}\put(0,0){
     \put(100,40)  {\scriptsize $ (\CC,c) $}
     \put(23,42) {\scriptsize $\CB \boxtimes \bh$}
     \put(70, 70)  {\scriptsize $(V,{}^\CB\CM)$}
     \put(23,9)   {\scriptsize $\CB\boxtimes_\CB\CM=\CM$}
     %\put(110, 80) {$x$}
     }\setlength{\unitlength}{1pt}}
  \end{picture}}
\ee
The superscripts for ${}^\CB\CB$ and ${}^\bh\CM$ can be viewed as fictional topological orders in the time direction and the UFC's $\CB$ and $\CM$ can be viewed as fictional gapped domain walls.
If we take this for granted, then it suggests immediately the following fusion formula: 
\be \label{eq:edge-wall-2}
(V, {}^\CB\CB) \boxtimes_{(\CB,c)} (\Cb,{}^\bh\CM) = (V\otimes_\Cb \Cb, {}^{\CB\boxtimes \bh}(\CB\boxtimes_\CB\CM)) = (V, {}^\CB\CM). 
\ee
Note that the unitary braided monoidal equivalence $\phi_\CM: \overline{\CB} \boxtimes \CC \to \FZ(\CM)$ provides a composed braided monoidal functor $\overline{\CB} \hookrightarrow \overline{\CB} \boxtimes \CC \xrightarrow{\phi_\CM} \FZ(\CM)$. Therefore, ${}^\CB\CM$ is a well-defined enriched monoidal category via the canonical construction. In this case, $\CB$ is the background category, and $\CM$ is the underlying category and the category of topological edge excitations.

\medskip
The fusion formula (\ref{eq:edge-wall-2}) is a little bit mysterious. Another perhaps equally mysterious way to understand this chiral gapless edge $(V, {}^\CB\CM)$ is by the topological Wick rotation illustrated below:  
\be \label{pic:wick-2}
\raisebox{-30pt}{
  \begin{picture}(120,75)
   \put(0,15){\scalebox{0.5}{\includegraphics{pic-edge-4-eps-converted-to.pdf}}}
   \put(0,15){
     \setlength{\unitlength}{.75pt}\put(0,0){
     \put(140,40)  {\scriptsize $ (\CC,c) $}
     \put(15,7) {\scriptsize $(\CB,c)$}
     %\put(38,21)   {\scriptsize $\CB$}
     \put(85,27)   {\scriptsize $\CM$}
     %\put(110, 80) {$x$}
     }\setlength{\unitlength}{1pt}}
  \end{picture}} 
\quad \xrightarrow{\mbox{\footnotesize topological Wick rotation}} \quad 
\raisebox{-30pt}{
  \begin{picture}(100,75)
   \put(0,10){\scalebox{0.5}{\includegraphics{pic-edge-eps-converted-to.pdf}}}
   \put(0,10){
     \setlength{\unitlength}{.75pt}\put(0,0){
     \put(100,40)  {\scriptsize $ (\CC,c) $}
     \put(25,43) {\scriptsize $(\CB,c)$}
     \put(70, 70)  {\scriptsize $(V,{}^\CB\CM)$}
     \put(30,15)   {\scriptsize $\CM$}
     %\put(110, 80) {$x$}
     }\setlength{\unitlength}{1pt}}
  \end{picture}}
\ee

\begin{rem} \label{rem:left-right-convention}
Note that our left-right convention in $\phi_\CM: \overline{\CB}\boxtimes \CC \xrightarrow{\simeq} \FZ(\CM)$ is that if the orientation of the wall is the same (resp. the opposite) as the induced orientation of a bulk, then this bulk phase acts on the wall from right (resp. left). We will use this convention throughout this work. 
\end{rem}

We will postpone the proof of \eqref{eq:edge-wall-2} until Section\,\ref{sec:RG}. 
Let us take \eqref{eq:edge-wall-2} for granted for now. We list a few basic ingredients of the gapless edge $(V,{}^\CB\CM)$.
\bnu
\item Topological edge excitations $x,y,z,\cdots$ are objects in $\CM$, and the trivial topological edge excitation $\one$ is given by the tensor unit $\one_\CM$ of $\CM$. 
\item The VOA $U=M_{\one,\one}$ that can live on the entire 1+1D world sheet is given by $[\one_\CM, \one_\CM]_\CB$, which a condensable algebra in $\CB$ \cite{dmno}. According to \cite{hkl}, $U$ is indeed a VOA extension of $V$.  
More generally, we have $M_{x,y} = [x,y]_\CB$ for $x,y\in \CM$. % will be explained later.
%\item Boundary CFT's and domain walls $M_{x,y}$ are given by $[x,y]_\CB$ for $x,y\in \CM$. We have explained in Section\,\ref{sec:unitary-RCFT} why these internal homs are indeed boundary CFT's and domain walls of the same bulk CFT $Z(U)$.  
\enu

From the explicit construction of the gapless edge $(V,{}^\CB\CM)$, we can see some general features of gapless edges. 

\begin{itemize}
\item In general, $V\subsetneq U$. For example, when $(\CC,c)$ is obtained by condensing a non-trivial condensable algebra $U$ in $(\CB,c)$ (i.e. $\dim U>1$), we have $\CC=\CB_U^0$, where $\CB_U^0$ is the category of local $U$-modules in $\CB$ \cite{dmno,anyon}. This condensation also produces a gapped domain wall between $(\CB,c)$ and $(\CC,c)$ described by the UFC $\CB_U$ of right $U$-modules in $\CB$. In this case, $\CM=\CB_U$, and $U=[\one_\CM,\one_\CM]_\CB\neq V$ is a non-trivial VOA extension of $V$. By results in \cite{hkl}, we have $\CB_U^0=\Mod_U$. Therefore, $\CC\nsimeq \Mod_V$ in this case. 

\item In general, $\Mod_V\nsimeq \CC\nsimeq\Mod_U$. For example, when $\CB=\Mod_V$ is obtained by condensing a non-trivial condensable algebra $A$ in $\CC$, i.e. $\CB=\CC_A^0$,  the condensation produces a gapped domain wall $\CM=\CC_A$. In this case, the central functor $\overline{\CB} \to \CM$ is the canonical embedding $\overline{\CC_A^0} \hookrightarrow \CC_A$. Therefore, $U=[\one_\CM,\one_\CM]_\CB=V$. Note that $\dim \CC > \dim\CC/\dim A = \dim \CB$ because $\dim A>1$. Therefore, we have $\CC\nsimeq \Mod_V=\Mod_U=\CB$ in this case. 
\end{itemize}

Recall that the category of boundary conditions of the unitary bulk CFT $Z(U)$ is given by the category $\CB_U$ of right $U$-modules in $\CB$ (Theorem$^{\mathrm{ph}}$\,\ref{thm:bcft-3}). It is illuminating to study the relation between the category $\CB_U$ of boundary conditions and that of topological edge excitations $\CM$ on the gapless edge $(V,{}^\CB\CM)$. 
\begin{itemize}
\item The composed functor $L: \overline{\CB} \to \overline{\CB} \boxtimes \CC \xrightarrow{\phi_\CM} \FZ(\CM) \xrightarrow{\forget} \CM$ has a right adjoint functor $L^R: \CM \to \overline{\CB}$, which factors as $\CM \xrightarrow{R} \overline{\CB}_U \xrightarrow{\forget} \overline{\CB}$, where $U=L^R(\one_\CM)=[\one_\CM,\one_\CM]_\CB$ is a condensable algebra in $\overline{\CB}$ \cite[Lemma\,3.5]{dmno}. Moreover, the left adjoint functor $R^L: \overline{\CB}_U \to \CM$ of $R$ is monoidal and fully faithful \cite[Lemma\,3.5]{dmno}, i.e. a monoidal embedding
\be \label{map:BU-M}
R^L: \overline{\CB}_U \hookrightarrow \CM.
\ee
Namely, the category of boundary conditions $\overline{\CB}_U$ is a fusion subcategory of that of topological edge excitations. 
We illustrate these functors in the following diagram: 
\be \label{eq:L-R}
\xymatrix@R=3em@C=3em{ \overline{\CB} \ar@/^/[rr]^{- \otimes U} \ar@/^/[rd]^L & & \overline{\CB}_U \ar@/^/[ll]^{\forget}_\perp \ar@/_/[ld]_{R^L} \\
& \CM \ar@/_/[ur]_{R}^\perp  \ar@/^/[ul]^{L^R}_\perp & 
}
\ee
where left (resp. right) adjoints form a commutative diagram. The functor $L: \overline{\CB} \to \CM$ is a central functor. It endows $\CM$ with a structure of a unitary left $\overline{\CB}$-module with the action functor $\odot: \overline{\CB} \times \CM \to \CM$ defined by the composed functor $\overline{\CB} \times \CM\xrightarrow{L \times \id_\CM} \CM \times \CM \xrightarrow{\otimes} \CM$. 

\item In general, we have $\overline{\CB}_U \subsetneq \CM$. For example, let $A$ be a non-trivial condensable algebra $A \in\CC$. When $\Mod_V=\CB=\CC_A^0$ and $\CM=\CC_A$, on the gapless edge $(V,{}^\CB\CM)$, we have $U=A=V$ and $\overline{\CB}_U \subsetneq \CM$. In this case, $\CM$ splits as a direct sum of indecomposable unitary left $\overline{\CB}$-modules, i.e. 
$$
\CM \simeq \overline{\CB}_U \oplus \CM_1\oplus \cdots \oplus \CM_N.
$$ 
For two simple objects $x\in \overline{\CB}_U$ and $y\in \CM_i$ for $i=1,\cdots, N$, we have $R(y)=0$ and $M_{x,y}=0$ because $\hom_\CB(b,M_{x,y}) \simeq \hom_\CM(b\odot x,y)=0$ for all $b\in \CB$. But $M_{\one,y\otimes y^\ast} \neq 0$. Moreover, the boundary CFT $M_{y,y}=[y,y]\neq 0$ because $\one_\CB \hookrightarrow [y,y]$ is non-zero. In general, the boundary CFT's $M_{y,y}$ and $M_{x,x}$ are not compatible in the sense that they have different bulk CFT's. By Theorem\,\ref{thm:bcft-3}, each unitary left $\overline{\CB}$-modules is associated to a unique bulk CFT. The bulk CFT's associated to $\overline{\CB}_U$ and $\CM_i$ are different in general. In other words, the dimensional reduction process depicted in Figure\,\ref{fig:cylinder-1} might produce different bulk CFT's in general. Also note that both $L$ and $R^L$ are monoidal functors, but $L^R$ and $R$ are not monoidal in general.  

%\item For a simple object $x\in\CM$ but $x\notin \overline{\CB}_U$, we have $M_{\one,x}=0$. Such an edge excitation cannot appear on the edge alone.  But it might appear in pairs, i.e. $M_{\one,x\otimes x^\ast} \neq 0$. As we will show that these ghost-like excitations, all coming from the gapped domain wall $\CM$, play a mysterious but indispensable role in defining the edge as they are demanded by the boundary-bulk relation (see Section\,\ref{sec:classification}) and the physical construction given here. 

\end{itemize}

\section{Chiral gapless edges II}  \label{sec:gapless-edge-II}

In this section, we continue the analysis started in Section\,\ref{sec:gapless-edge-1} on the structures and properties of observables on the 1+1D world sheet of a chiral gapless edge. This analysis leads us to a classification theory of all chiral gapless edges of any 2d topological orders. %This section is more mathematically technical than other sections. 

%\subsection{Relations between $\Mod_V$ and $\CX$} \label{sec:underlying-cat}

\subsection{$\CXs$ is an enriched unitary fusion category} \label{sec:MP}

At the end of Section\,\ref{sec:gapless-edge-1}, we conclude that the observables on the 1+1D world sheet of a chiral gapless edge of the 2d topological order $(\CC,c)$ can be summarized by a pair $(V,\CXs)$, where $V$ is the chiral symmetry and $\CXs$ is an $\Mod_V$-enriched monoidal category. The objects in $\CXs$ are topological edge excitations, and the hom spaces $\hom_{\CXs}(x,y):=M_{x,y}$ are boundary CFT's (if $x\simeq y$) or domain walls between boundary CFT's (if $x\nsimeq y$). In order to gain a better understanding of the hidden structure of $\CXs$, it is crucial to understand the mysterious relation between the objects in $\CX^\sharp$ and the morphisms in $\CX^\sharp$. In terms of the underlying category $\CX$ of $\CXs$, it amounts to ask for the relation between $\CX$ and $\CB=\Mod_V$.

\begin{figure} 
$$
 \raisebox{-30pt}{
  \begin{picture}(70,130)
   \put(0,5){\scalebox{0.6}{\includegraphics{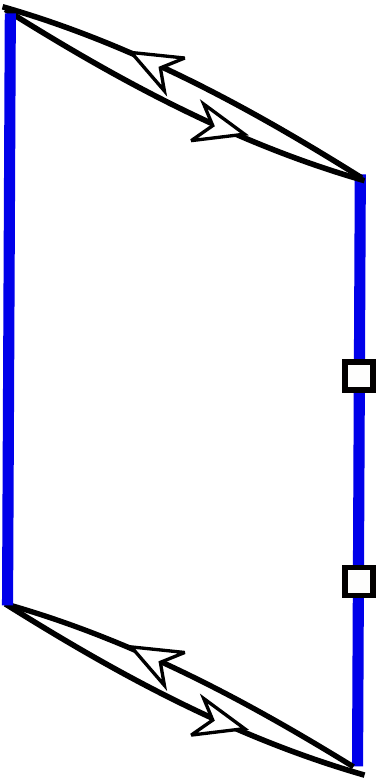}}}
   \put(0,5){
     \setlength{\unitlength}{.75pt}\put(0,-83){
     \put(88,149)  {$ A_{y_0} $}
     \put(88,205)  {$A_{z_0}$}
     \put(93,173)   {$M_{y_0,z_0}$}
     \put(88,97) {$A_{x_0}$}
     \put(92, 124) {$M_{x_0,y_0}$}
     \put(-44,183) {$U=A_{\one}$}
     \put(25,163) {$Z(U)$}
     }\setlength{\unitlength}{1pt}}
  \end{picture}}
  \quad\quad\quad\quad\quad\quad\quad\quad\quad\quad\quad\quad\quad
  %\rightsquigarrow\quad\quad\quad \quad
 \raisebox{-30pt}{
  \begin{picture}(70,130)
   \put(0,5){\scalebox{0.6}{\includegraphics{pic-band-1-eps-converted-to.pdf}}}
   \put(0,5){
     \setlength{\unitlength}{.75pt}\put(0,-83){
     \put(88,142)  {$ A_{y_i} $}
     \put(88,205)  {$A_{z_i}$}
     \put(90,172)   {$M_{y_i,z_i}$}
     \put(88,93) {$A_{x_i}$}
     \put(89, 115) {$M_{x_i,y_i}$}
      \put(-21,183) {$A_{x_i^\ast}$}
      \put(25,163) {$Z(A_{x_i})$}
     }\setlength{\unitlength}{1pt}}
  \end{picture}}
$$
$$
(a) \quad\quad\quad\quad\quad\quad\quad\quad\quad\quad\quad\quad\quad\quad\quad (b)
$$
\caption{These two pictures depicts two possible 1+1D world sheets obtained by applying dimensional reduction to Figure\,\ref{fig:cylinder-1} (a). The associated two sets of boundary CFT's $\{ A_\one, A_{x_0}, A_{y_0}\}$ and $\{ A_{x_i}, A_{y_i}, A_{z_i}\}$ are potentially incompatible and $M_{x_0,x_i}=0$ for $x_0\in \CX_0, x_i\in \CX_i$}
\label{fig:bulk-CFT}
\end{figure}

\medskip
%It turns out that this mysterious relation is clarified by the classification theory of boundary-bulk CFT's presented in Section \ref{sec:RCFT}. 
Consider the 1+1D world sheets depicted in Figure\,\ref{fig:bulk-CFT} (a), which is essentially the same as Figure\,\ref{fig:cylinder-1} (b). The boundary CFT's $A_{x_0}, A_{y_0}, A_{z_0}$ are compatible with $U$ in the sense that they share the same bulk CFT $Z(U)$. As we have shown at the end of Section\,\ref{sec:examples}, it is possible that there are $x\in \CX$ such that $M_{\one, x}=0$, i.e. there is no boundary condition changing operator that can change the boundary condition associated to $\one$ to that associated to $x$. As a consequence, the whole enriched category $\CX^\sharp$ splits into a direct sum of connected components: 
$$
\CX^\sharp \simeq \CX_0^\sharp \oplus \CX_1^\sharp \oplus \cdots \oplus \CX_N^\sharp,
$$ 
where $\one \in \CX_0^\sharp$ and $M_{x_i, x_j}=0$ for $x_i\in \CX_i^\sharp$, $x_j\in \CX_j^\sharp$, $i,j=0, \cdots, N$ and $i\neq j$. Within each connected component $\CX_i^\sharp$, all boundary CFT's and domain walls $M_{x_i,y_i}$ (for $x_i,y_i\in\CX_i$) are necessarily compatible. Figure\,\ref{fig:bulk-CFT} (b) depicts three compatible boundary CFT's $A_{x_i}, A_{y_i}, A_{z_i}$ for $x_i,y_i,z_i\in \CX_i^\sharp$ and their domain walls and the associated bulk CFT.

By the Naturality Principle in physics, if $A_{x_i}$ appears as a boundary CFT on the 1+1D world sheet of the gapless edge $(V, \CX^\sharp)$, then all boundary CFT's and their domain walls that are compatible with $A_{x_i}$ should appear on this world sheet\footnote{unless there are additional unknown symmetry principles that forbid certain boundary CFT's or walls to appear.}. As a consequence, by Theorem\,\ref{thm:bcft-3} and Remark\,\ref{rem:bcft-enrich-cat}, each connected component $\CX_i^\sharp$ of $\CX^\sharp$, is precisely the enriched category ${}^\CB\CX_i$ via the canonical construction (see Example\,\ref{exam:SC}), where $\CX_i$ is an indecomposable unitary left $\CB$-module and, at the same time, the category of boundary conditions associated to the bulk CFT $Z(A_{x_i})$. Therefore, we obtain 
$$
\CX^\sharp \simeq {}^\CB\CX_0 \oplus {}^\CB\CX_1 \oplus \cdots \oplus {}^\CB\CX_N
$$
as $\CB$-enriched categories and 
$$
\CX \simeq \CX_0 \oplus \CX_1 \oplus \cdots \oplus \CX_N 
$$
as unitary left $\CB$-modules. We denote the left $\CB$-action on $\CX$ by $\odot: \CB \times \CX \to \CX$. In other words, $\CX^\sharp \simeq {}^\CB\CX$ as $\CB$-enriched categories with $[x,y]_\CB \simeq M_{x,y}$ for $x,y\in \CX$. We want to show that it is also an equivalence of enriched monoidal categories. 

\begin{rem} \label{rem:gapped-split}
When $V=\Cb$, $\CB=\bh$ and the edge $(\Cb, {}^\bh\CX)$ is gapped. In this case, $\bh$ has a unique indecomposable unitary left $\bh$-module $\bh$. Therefore, $\CX \simeq \bh \oplus \cdots \oplus \bh$ as unitary left $\bh$-modules, i.e. $\CX_i=\bh$. In this case, $M_{x_i,x_j}=0$ for $x_i\in\CX_i,x_j\in\CX_j$ and $i\neq j$. For $x_i\in \CX_i=\bh$, the matrix algebra $[x_i,x_i]_\bh=\hom_\CX(x_i,x_i)$ is actually a 0+1D boundary TQFT, which has a trivial bulk TQFT given by its center $Z([x_i,x_i])=\Cb$. The ordinary center is also the full center in this case. 
\end{rem}

\begin{rem}
By Theorem\,\ref{thm:bcft-3} and \cite[Proposition 7.3]{dkr2}, the result $\CX^\sharp \simeq {}^\CB\CX$ implies that 
$$
M_{y_i,z_i}\otimes_{M_{y_i,y_i}} M_{x_i, y_i} \simeq M_{x_i, z_i}, \quad\quad\quad \mbox{for non-zero objects $x_i,y_i,z_i\in \CX_i$}.
$$ 
\end{rem}

Morrison-Penneys' classification of $\CB$-enriched monoidal categories \cite{MP} indicates that $\CX^\sharp$ is necessarily obtained from a triple $(\CB,\CX,F_\CX)$ via the canonical construction, where $F_\CX:\overline{\CB}\to\FZ(\CX)$ is a braided oplax monoidal functor such that the composed functor $L: \overline{\CB} \xrightarrow{F_\CX} \FZ(\CX) \xrightarrow\forget \CX$ is left adjoint to a functor $M_{\one,-}:\CX\to\CB$, which will be defined later, i.e.
\be \label{diag:central-functor-2}
\xymatrix{
\overline{\CB} \ar[r]^-{F_\CX} \ar[dr]_L &  \FZ(\CX) \ar[d]^\forget \\
& \CX\, .
}
\ee
In our case, since $\CX$ is already a unitary left $\CB$-module with $[x,y]_\CB=M_{x,y}$, the result turns out to be even stronger: $F_\CX$ is a unitary braided monoidal functor (NOT oplax), or equivalently, $L$ is a central functor (recall Diagram\,(\ref{diag:central-functor})). To see this, let us go through some parts of Morrison-Penneys' proof that are adapted and simplified to suit our case.

%We want to show that this is also an equivalence of $\CB$-enriched monoidal categories, where ${}^\CB\CX$ is given by the canonical construction in Theorem\,\ref{thm:SC} from a triple $(\CB,\CX,F_\CX)$, where $F_\CX: \overline{\CB} \to \FZ(\CX)$ is a yet-to-be-constructed unitary braided monoidal functor (NOT oplax). We denote the left $\CB$-action on $\CX$ by $\odot: \CB \times \CX \to \CX$. It is enough to show that $\odot$ is induced from the composed functor $L:=(\overline{\CB} \xrightarrow{F_\CX} \FZ(\CX) \xrightarrow{\forget} \CX)$, i.e.

%The idea of proof is to first construct the functor $L: \overline{\CB} \to \CX$, then show that $L$ is a central functor, which means that it is a monoidal functor that can be lifted to a unitary braided monoidal functor $F_\CX: \overline{\CB} \to \FZ(\CX)$. 

\bnu
\item There is a well-defined $\Cb$-linear functor $M_{\one,-}: \CX \to \CB$ defined by $y\mapsto M_{\one,y}$ and %for $f:y\to y'$ in $\CX$, 
$$
f \mapsto (M_{\one,y} \xrightarrow{\simeq} V\otimes_V M_{\one,y}  \xrightarrow{f \otimes_V 1} M_{y,y'}\otimes_V M_{\one,y} \xrightarrow{\circledcirc} M_{\one,y'}) \quad \mbox{for $f: y\to y'$ in $\CX$}.
$$ 
It is clear that $M_{\one,-}(\CX_i)=0$ for $i>0$. 

\item Let $L: \CB \to \CX$ be the left adjoint functor of $M_{\one,-}$. The category of topological edge excitations $\CX$ is monoidal. The natural physical requirement is that each topological edge excitation should have its anti-particle. This requires $\CX$ to be rigid thus a UFC. We want to show that $L$ is a central functor. 

\item By definition (see \cite[Sec.\,2.7]{MP}), $\CX^\sharp$ is rigid if and only if $\CX$ is rigid. The rigidity of $\CXs$ implies the Frobenius reciprocity: $M_{x,y\otimes z^\ast} \simeq M_{x\otimes z,y} \simeq M_{z, x^\ast \otimes y}$ \cite[Sec.\,2.7]{MP}. In particular, we have 
\be \label{eqn:fro-rec}
M_{x,y} \simeq M_{\one, x^\ast \otimes y} \simeq M_{\one, y\otimes x^\ast} \simeq M_{y^\ast, x^\ast}.
\ee

\item We have $L(\one_\CB)\simeq \one$ because $\hom_\CX(L(\one_\CB), x) \simeq \hom_\CB(\one_\CB, M_{\one,x}) \simeq \hom_\CX(\one, x)$ for $x\in\CX$. Moreover, we have 
$$\hom_\CX(L(a)\otimes x,y) 
\simeq \hom_\CX(L(a),y\otimes x^\ast) 
\simeq \hom_\CB(a,M_{\one,y\otimes x^\ast}) 
\simeq \hom_\CB(a,M_{x,y}) 
\simeq \hom_\CX(a\odot x,y). 
$$
Thus $L(a)\otimes x\simeq a\odot x$ for $a\in\CB$ and $x,y\in\CX$. Then $L(a)\simeq a\odot\one$. This further implies the following isomorphisms: 
$$
L(a)\otimes L(b) \simeq a\odot(b\odot\one) \simeq L(a\otimes b).
$$
Therefore, $L$ is monoidal. 

\item The monoidal functor $L: \overline{\CB} \to \CX$ is a central functor. This amounts to show that each $L(a)$ can be endowed with the structure of a half-braiding \cite{dmno}. Indeed, we have
$$\hom_\CX(L(a)\otimes x,y)
%\simeq \hom_\CB(a,M_{\one,y\otimes x^\ast}) 
\simeq \hom_\CB(a,M_{x,y}) 
\simeq \hom_\CB(a,M_{\one,x^\ast\otimes y}) 
\simeq \hom_\CX(L(a),x^\ast\otimes y)
\simeq \hom_\CX(x\otimes L(a),y). 
$$
Therefore, we obtain a half-braiding isomorphism $L(a)\otimes x\simeq x\otimes L(a)$ as desired by Yoneda Lemma. Further details of the proof could be found in \cite{MP}.

\item Moreover, by the physical requirement, the natural isomorphisms in Eq.\,\eqref{eqn:fro-rec} should be unitary\footnote{An isomorphism $f$ in a $\ast$-category is unitary if $f^\dagger=f^{-1}$}. Consequently, the natural isomorphisms $L(a)\otimes L(b)\simeq L(a\otimes b)$ and $L(a)\otimes x\simeq x\otimes L(a)$ are all unitary. Namely, $F_\CX:\overline{\CB}\to\FZ(\CX)$ is a unitary braided monoidal functor.

\enu

The physical meaning of above proof is not very explicit. In the rest of this subsection, we would like to explain some of its hidden structures in more physical terms and some physical consequences. %This discussion also provides an independent proof of the monoidalness of $L$. 

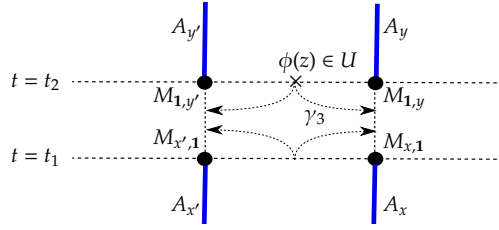
\begin{figure} 
$$
 \raisebox{-70pt}{
  \begin{picture}(130,110)
   \put(-20,8){\scalebox{0.6}{\includegraphics{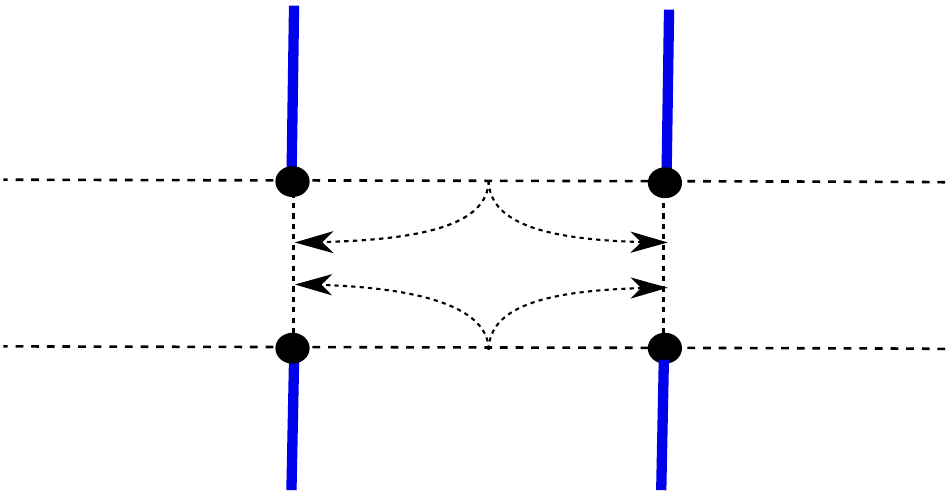}}}
   \put(-20,8){
     \setlength{\unitlength}{.75pt}\put(0,-83){
     \put(-30,153)  {\footnotesize $ t=t_2 $}
     \put(-30,115)  {\footnotesize $ t=t_1 $}
     \put(51,180)  {\footnotesize $ A_{y'}$}
     \put(51,90) {\footnotesize $ A_{x'}$}
     \put(158,180)  {\footnotesize $A_y$}
     \put(158,90)  {\footnotesize $A_x$}
     %\put(60,220)   {\footnotesize $M_{y,y}$}
     %\put(140,220)   {\footnotesize $M_{x,x}$}
     %\put(72,135) {\footnotesize $U$}
     \put(42,145) {\footnotesize $M_{\one,y'}$}
     \put(42,124) {\footnotesize $M_{x',\one}$}
     \put(158,145) {\footnotesize $M_{\one,y}$}
     \put(158,122) {\footnotesize $M_{x,\one}$}
     \put(118, 136) {\footnotesize $\gamma_3$}
     \put(105,163) {\footnotesize $\phi(z)\in U$}
     \put(110,153)  {\footnotesize $\times$}
     }\setlength{\unitlength}{1pt}}
  \end{picture}}
$$
%$$(a) \quad\quad\quad\quad\quad\quad\quad\quad\quad \quad\quad\quad\quad\quad\quad\quad\quad\quad (b)$$
\caption{This picture depicts a special situation on the 1+1D world sheet of the chiral gapless edge $(V,\CXs)$ for $x,y,x',y'\in\CX_0$. It shows that the horizontal $U$-actions on $M_{\one,y'}, M_{x,\one}, M_{\one,y}, M_{x,\one}$ are determined by the vertical $U$-actions via the braidings long the fours paths, respectively. 
%there is a physically meaningful left $U$-action on $M_{\one,y}$ defined by the right $U$-action and a braiding along the path $\gamma_3$. 
}
\label{fig:left-U-action}
\end{figure}

\bnu
\item Since $\one\in\CX_0$ and $A_\one=U$, by Theorem\,\ref{thm:bcft-3}, we must have $\CX_0\simeq \CB_U$ as left $\CB$-modules. We want to construct a canonical equivalence $\CX_0\simeq \CB_U$ explicitly. Note that, for $x_0\in \CX_0$, there is a natural right $U$-action on $M_{\one,x_0}$ (see Figure\,\ref{fig:left-U-action}): 
$$
\circledcirc: M_{\one,x_0} \otimes_V M_{\one,\one} \to M_{\one,x_0},
$$
which endows $M_{\one,x_0}$ with the structure of a right $U$-module. Therefore, the functor $M_{\one,-}: \CX \to \CB$ factors through the forgetful functor $\forget: \CB_U \to \CB$, i.e. $M_{\one,-} = \forget \circ R$, where $R: \CX \to \CB_U$ is defined by $x \mapsto M_{\one,x}$. 
We want to show that the functor $R: \CX_0 \to \CB_U$ is the canonical equivalence we are looking for. 

\item It is clear that $R(\CX_i)=0$ for $i>0$. By Theorem\,\ref{thm:bcft-3}, each boundary condition $X\in\CB_U$ can be realized as the domain wall $[U,X]_\CB=X$ between the boundary CFT's $U$ and $[X,X]_\CB$ because $[X,-]_\CB: \CB_U \to \CB$ is precisely the forgetful functor (recall Example\,\ref{expl:int-hom}). Therefore, $M_{\one, x_0}$ for $x_0\in \CX_0$ recovers all boundary conditions in $\CB_U$, or equivalently, $R: \CX_0 \to \CB_U$ is essentially surjective. Moreover, for $x\in\CX_0$, the domain wall $R(x)=M_{\one,x}$ as an invertible domain wall uniquely determines the boundary CFT $M_{x,x}$, which has to be $[R(x),R(x)]_\CB$ by Theorem\,\ref{thm:bcft-3}. Then, for $x,y\in\CX_0$, we must have $M_{x,y}\simeq [R(x),R(y)]_\CB$ and 
$$ 
\hom_\CX(x,y) = \hom_\CB(\one_\CB, M_{x,y}) \simeq \hom_\CB(\one_\CB, [R(x),R(y)]_\CB) \simeq
\hom_{\CB_U}(R(x), R(y)). 
$$
It is implies that $R$ is fully faithful. Therefore, $R: \CX_0\xrightarrow{\simeq} \CB$ is an equivalence of finite unitary categories. 

\item Let $R^L$ be the left adjoint functors of $R$. The left adjoint functor of $\forget$ is $-\otimes_V U$. we obtain the following diagram (similar to the diagram (\ref{eq:L-R})): 
$$
\xymatrix@R=4em@C=6em{ \overline{\CB} \ar@/^/[rr]^{- \otimes_V U} \ar@/^/[rd]^{L} & & \overline{\CB}_U \ar@/^/[ll]^{\forget}_\perp \ar@/_/[ld]_{R^L} \\
& \CX \ar@/_/[ur]_{R}^\perp  \ar@/^/[ul]^{M_{\one,-}}_\perp & 
}
$$ 
where the right and left adjoint functors form two commutative diagrams, respectively. Since $R(\CX_i)=0$ for $i>0$,  $R^L$ factors as $\overline{\CB}_U \to \CX_0 \hookrightarrow \CX$.

\item We explain that $\overline{\CB}_U$ is also a UFC. Mathematically, for $y\in \CX_0$, the right $U$-module $M_{\one,y}$ may have more than one possible left $U$-actions. But there is only one that is physically meaningful. It is determined\footnote{Mathematically, it means that the OPE between the fields in $U$ and those in $M_{\one,y}$ along the $t=t_2$-line is determined by that along the vertical line via an analytic continuation along the path $\gamma_3^{-1}$ \cite{huang-mtc2}.} by the adiabatic move along the path $\gamma_3$ depicted in Figure\,\ref{fig:left-U-action}. According to our braiding convention (recall Remark\,\ref{rem:braiding-convention}), this left $U$-action is defined by
$$
U \otimes_V M_{\one,y} \xrightarrow{c_{U, M_{\one,y}}} M_{\one,y} \otimes_V U \to M_{\one,y},
$$
where $c_{-,-}$ is the braiding in $\CB$. This left $U$-action allows us to fuse $M_{\one,y'}$ with $M_{\one,y}$ horizontally (in Figure\,\ref{fig:left-U-action}) to give $M_{\one,y'} \otimes_U M_{\one,y}$. As a consequence, $\overline{\CB}_U$ is a UFC with the tensor product $\otimes_U$ and the tensor unit $U$. Moreover, the functor $-\otimes_V U$ is monoidal. 

\item Since $L: \overline{\CB} \to \CX$ is a central functor, by \cite[Lemma\,3.5]{dmno}, the functor $R^L$ defineds a monoidal equivalence from $\overline{\CB}_U$ to its image in $\CX$. Since the left $\overline{\CB}$-module structure on $\CX$ is induced from the monoidal functor $L$, the monoidal functor $R^L$ is also a left $\overline{\CB}$-module functor. Since $\CX_0\simeq \overline{\CB}_U$ as left $\overline{\CB}_U$-modules, $R^L: \overline{\CB}_U \to \CX_0$ must be an equivalence of left $\overline{\CB}_U$-modules. Therefore, $R: \CX_0\to \overline{\CB}_U$ is an equivalence of left $\overline{\CB}_U$-modules and a monoidal equivalence. In particular, we have a canonical isomorphism: 
\be \label{eq:Mxy}
M_{\one,y'\otimes y} \simeq M_{\one,y'}\otimes_U M_{\one,y}, \quad\quad\quad \mbox{for $y,y'\in\CX_0$}. 
\ee
%On the other hand, this formula follows automatically from (\ref{eq:[xy]}), and implies $R:\CX_0 \to \CB_U$ is a monoidal equivalence.  This provides an independent proof of the fact that the functor $L$ is monoidal. 

\item Note that (\ref{eq:Mxy}) does not hold for general $x,y \in \CX$. For example, if $x\in \CX_i$ for $i>0$, then $M_{\one,x}=0$ but $M_{\one,x\otimes x^\ast}\neq 0$. For a gapped edge $(\Cb,{}^\bh\CX)$, (\ref{eq:Mxy}) is almost never true as we have seen in (\ref{map-otimes}). On the other hand, (\ref{eq:Mxy}) is always true for either $y\simeq\one$ or $y'\simeq\one$. 

\enu

\begin{rem} \label{rem:fusion-anomaly-free}
Using $R: \CX_0\xrightarrow{\simeq} \overline{\CB}_U$, we can identify $\CX_0$ with $\overline{\CB}_U$. For $x,y,x',y'\in\CX_0$, by the properties of internal homs, we obtain $M_{x,y} \simeq M_{\one,y} \otimes_U M_{x,\one}$. It means that $M_{x,y}$ can be split into $M_{\one,y}$ and $M_{x,\one}$ without altering the physics (depicted in the right half of Figure\,\ref{fig:left-U-action}). It is also clear that the dual formula of (\ref{eq:Mxy}): 
$$
M_{x',\one}\otimes_U M_{x,\one} \simeq M_{x'\otimes x, \one}, \quad\quad\quad \mbox{for $x,x'\in\CX_0$}. 
$$ 
also holds. As a consequence, the spatial fusion of $M_{x',y'}$ and $M_{x,y}$ can be achieved by fusing $M_{\one,y'}, M_{x',\one}, M_{\one,y}, M_{x,\one}$ as shown in Figure\,\ref{fig:left-U-action}. The braided interchange property (recall Diagram\,\ref{diag:comp-otimes}) implies that the morphism (\ref{eq:comp}) must be an isomorphism: 
\be \label{eq:Mxyxy}
M_{x',y'} \otimes_U M_{x,y} \simeq M_{x'\otimes x, y'\otimes y}, \quad\quad\quad \mbox{for $x,x',y,y'\in\CX_0$}.
\ee
This formula does not hold for general $x,x',y,y'\in\CX$. This also explain why (\ref{eq:tensor-product}) is an isomorphism because $\CX=\CX_0$ for the canonical chiral gapless edge. When $V=0$ (i.e. edge is gapped), $\CX_0=\bh$ and (\ref{eq:Mxyxy}) simply says that the tensor product of two matrix algebras is again a matrix algebra. More formula of this type will be given and studied in \cite{kyz}. 
\end{rem}

\medskip
The results in this subsection motivates us to introduce the following definition. 
\begin{defn} \label{def:en-UFC}
An enriched monoidal category ${}^\CB\CX$ obtained by the canonical construction from a triple $(\CB,\CX,F_\CX)$ is called a {\em $\CB$-enriched (unitary) multi-fusion category} if \bnu
\item $\CB$ is a (unitary) braided multi-fusion category; 
\item $\CX$ is a (unitary) multi-fusion category; 
\item $F_\CX: \overline{\CB} \to \FZ(\CX)$ is a (unitary) braided monoidal functor.  
\enu
It is called a {\em $\CB$-enriched (unitary) fusion category} if both $\CB$ and $\CX$ are also fusion categories. 
\end{defn}

Then the results in this subsection can be summarized as follows:
\begin{itemize}
\item A chiral gapless edge $(V, \CXs)$ of a 2d topological order $(\CC,c)$ is precisely given by $(V, {}^\CB\CX)$, where $\CB=\Mod_V$ is a UMTC and ${}^\CB\CX$ is a $\CB$-enriched unitary fusion category. %obtained from a triple $(\CB,\CX, F_\CX)$ via the canonical construction. 
\end{itemize} 
%defined by the canonical construction from the triple $(\CB,\CX,F_\CX: \overline{\CB} \to \FZ(\CX))$ via  (see Theorem\,\ref{thm:XBX}). 

\subsection{Classification theory of chiral gapless edges} \label{sec:classification}

%We have shown that a chiral gapless edge of a 2d topological order $(\CC,c)$ is given by $(V, {}^\CB\CX)$, where $\CB=\Mod_V$ is UMTC and ${}^\CB\CX$ is a $\CB$-enriched unitary fusion category. It also has to satisfy the boundary-bulk relation, i.e. $\FZ({}^\CB\CX)\simeq \CC$. 

The last constraint on the mathematical description of a chiral gapless edge is that it must satisfy the so-called boundary-bulk relation, which says that the center of a gapped/gapless edge should coincide with the UMTC describing the bulk phase \cite{kong-wen-zheng-2}. Motivated by the physical meaning of the boundary-bulk relation for 2d topological orders, we introduced the notion of the center of an enriched monoidal category in \cite{kz2}. We will not give the precise definition of this notion in this work. Instead, we will do that in Part II \cite{kz4}. %Moreover, we have proved the following theorem. 

%\medskip Let $\CA$ be a MTC and $\CY$ a fusion category. Let ${}^\CA\CY$ be the $\CA$-enriched fusion category obtained from the triple $(\CA,\CY,F_\CY)$, where $F_\CY: \overline{\CA} \to \FZ(\CY)$ is a braided monoidal functor, via the canonical construction. 
\begin{thm}[\cite{kz2}] \label{thm:kz}
For a $\CB$-enriched fusion category ${}^\CB\CX$ obtained from a triple $(\CB,\CX,F_\CX)$, its center $\FZ({}^\CB\CX)$ is given by the centralizer of the image of $\overline{\CB}$ in $\FZ(\CX)$, denoted by $F_\CX(\overline{\CB})'|_{\FZ(\CX)}$. 
\end{thm}

Therefore, the requirement of boundary-bulk relation, i.e. $\FZ({}^\CB\CX) \simeq \CC$, is equivalent to the condition: $F_\CX(\overline{\CB})'|_{\FZ(\CX)} \simeq \CC$ as UMTC's. Since $\overline{\CB}$ is a UMTC, the functor $F_\CX$ is necessarily a braided monoidal embedding. Therefore, we obtain $\overline{\CB} \boxtimes \CC \simeq \FZ(\CX)$ as UMTC's \cite{dmno}. Similar to gapped edges, to uniquely determine the edge, we need specify how the topological excitations in the bulk are mapped to those on the edge \cite{anyon}. This is given by a central functor $\CC \to \CX$ (recall Diagram (\ref{diag:central-functor})). Together with $F_\CX$, we obtain a unitary braided monoidal equivalence: 
\be \label{eq:phi_X}
\phi_\CX: \overline{\CB} \boxtimes \CC \xrightarrow{\simeq} \FZ(\CX),
\ee
which should be viewed as a defining data of the edge. This completes our analysis. 

\begin{rem} \label{rem:af}
The existence of a braided monoidal equivalence $\phi_\CX$ in (\ref{eq:phi_X}) should be viewed as an anomaly-free condition. Namely, the bulk $(\CC,c)$ should resolve all the anomalies of the edge $(V,{}^\CB\CX)$ viewed as an anomalous 1d phase. 
\end{rem}

%What does this relation means? Since both $\overline{\CB}$ and $\FZ(\CX)$ are UMTC's, the existence of the braided monoidal functor $F_\CX: \overline{\CB} \to \FZ(\CX)$ implies that $\FZ(\CX) \simeq \overline{\CB} \boxtimes \overline{\CB}'$, where $\overline{\CB}'$ denotes the centralizer of $\overline{\CB}$ in $\FZ(\CX)$ and is also a UMTC. By Theorem\,\ref{thm:KZ}, we have $\FZ({}^\CB\CX)\simeq \overline{\CB}'$. The boundary-bulk relation implies that $\CD\simeq \CC$. 

How do we know whether we have found all the physical requirements of ${}^\CB\CX$? We have indeed found all of them because the pairs $(V, {}^\CB\CX)$ are precisely those chiral gapless edges constructed in Section\,\ref{sec:general-edge} via the physical process of fusing canonical gapless edges with gapped domain walls (with a loophole to be fixed in Section\,\ref{sec:RG}). In other words, topological Wick rotations realize all chiral gapless edges.
% of a given 2d topological order $(\CC,c)$. 

\medskip
We have found the precise mathematical description of the observables on the 1+1D world sheet of a chiral gapless edge as a pair $(V,{}^\CB\CX)$. We need to answer a few physical questions before we claims that this mathematical description actually classify all chiral gapless edges. 
\bnu

\item What do we mean by a chiral gapless edge physically? In reality, if a chiral gapless edge of a 2d topological order is realized in lab, it might not have any topological edge excitations on it at all before we introduce topological defects (or impurities) onto the edge. In this context, by a chiral gapless edge described by $(V,{}^\CB\CX)$, we mean the maximal way of inserting topological defects onto the edge without breaking the chiral symmetry $V$.

\item How do we know different pairs describes different edges? We provide three answers. First, two different edges can be obtained by fusing canonical edges with two different gapped domain walls along two potentially different 2d topological orders. Therefore, a transition between two chiral gapless edges can be understood as a gap-closing topological phase transition between two 2d topological orders, together with a gap-closing transition between two gapped domain walls. Secondly, for two different ${}^\CB\CX$ and ${}^{\CB'}\CX'$, the categories of topological edge excitations are different; mathematically, ${}^\CB\CX$ and ${}^{\CB'}\CX'$, as two $E_1$-algebras, can be viewed as different topological invariants defined on open 1-disks. They lead to different global topological invariant by integration or factorization homology \cite{ai}. Changing of topological invariants is associated with a phase transition in general. Secondly, for a given chiral gapless edge $(V,{}^\CB\CX)$, when we introduce more defects that break the chiral symmetry $V$ to a smaller one (see Remark\,\ref{rem:breaking-cs-2}), we believe that this process of breaking of chiral symmetry should cause a purely edge phase transition.

\item How to define a phase transition between two 1+1D gapless phase? A phase transition between two gapped phases are defined by closing the gap. But a phase transition between two gapless phases is not so clear from this perspective. As far as we know, there is no model-independent definition of such a phase transition. Interestingly, the physical intuition of chiral gapless edges actually provides us three possible model-independent definitions of a phase transition between two potentially anomalous 1+1D gapless phases.  
\bnu 

\item We can define such a phase transition as a process of changing or breaking local quantum symmetries (i.e. chiral symmetries in this case). 

\item Since all chiral gapless edges can be obtained by fusing the canonical chiral gapless edges with gapped domain walls along two 2d bulk phases, we can define the phase transition between two chiral gapless edges by a gap-closing topological phase transition between two bulk phases and that between two gapped domain walls. 

\begin{figure} 
$$ 
\raisebox{-30pt}{
  \begin{picture}(140,75)
   \put(0,15){\scalebox{0.7}{\includegraphics{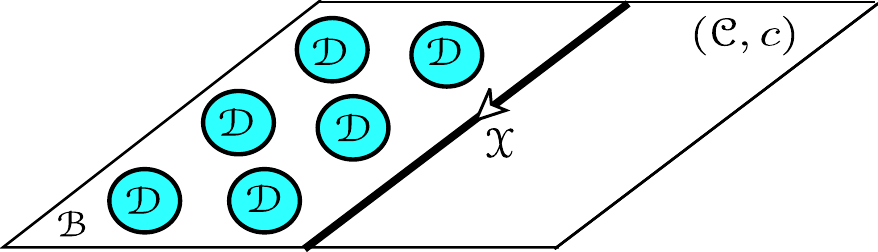}}}
   \put(0,15){
     \setlength{\unitlength}{.75pt}\put(0,0){
     %\put(140,40)  {\scriptsize $ \CB $}
     %\put(12,3) {\scriptsize $\CB$}
     %\put(38,21)   {\scriptsize $\CB$}
     %\put(85,27)   {\scriptsize $\CB$}
     %\put(110, 80) {$x$}
     }\setlength{\unitlength}{1pt}}
  \end{picture}} 
\quad \xrightarrow{\mbox{\footnotesize topological Wick rotation}} \quad 
\raisebox{-30pt}{
  \begin{picture}(100,75)
   \put(0,10){\scalebox{0.7}{\includegraphics{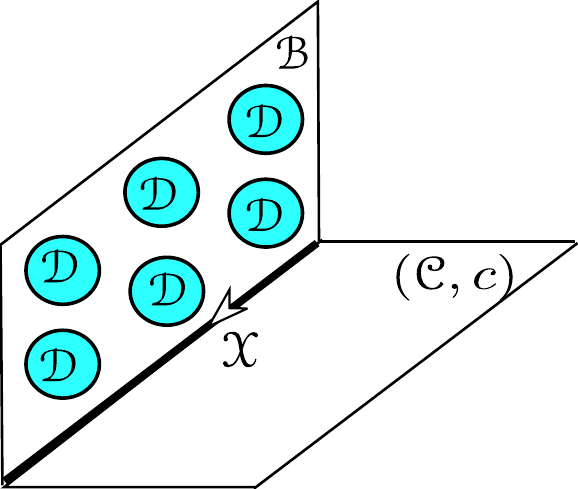}}}
   \put(0,10){
     \setlength{\unitlength}{.75pt}\put(0,0){
     %\put(100,40)  {\scriptsize $ (\CC,c) $}
     %\put(25,43) {\scriptsize $(\CB,c)$}
     %\put(70, 70)  {\scriptsize $(V,{}^\CB\CB)$}
     %\put(30,15)   {\scriptsize $\CB$}
     %\put(110, 80) {$x$}
     }\setlength{\unitlength}{1pt}}
  \end{picture}}
$$
\caption{These two pictures depict a physical description of pure edge phase transition via a topological Wick rotation. 
}
\label{fig:phase-transition}
\end{figure}
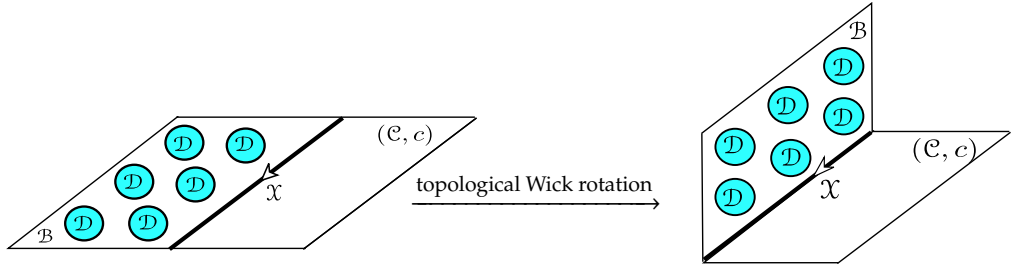

\item We can define such a 1+1D phase transition by a 2d topological phase transition via a topological Wick rotation. More precisely, a 2d topological phase transition from a 2d topological order $(\CB,c)$ to a new one $(\CD,c)$ can be achieved by first introducing some islands of $(\CD,c)$-phases into the 2d topological order $(\CB,c)$ as depicted in the first picture in Figure\,\ref{fig:phase-transition}. When these islands proliferated, a topological phase transition from $(\CB,c)$ to $(\CD,c)$ occurs. By Wick rotating this process, we obtain a description (or a definition) of a purely edge 1+1D phase transition as depicted in the second picture in Figure\,\ref{fig:phase-transition}. This definition automatically makes sense in any dimensions. We will return to this point in Section\,\ref{sec:outlooks}. 

\enu
As a consequence of these definitions, any chiral gapless edge $(V',{}^{\CB'}\CX')$ can be obtained from another one $(V,{}^\CB\CX)$ via a purely edge phase transition. 

\enu

We conclude that the mathematical description of observables on a 1+1D world sheet of a chiral gapless edge as a pair $(V, {}^\CB\CX)$ provides a classification of all chiral gapless edges.  We summarize this result as a physical theorem. 
\begin{pthm} \label{thm:main}
Gapped and chiral gapless edges of a 2d topological order $(\CC,c)$ are precisely described and classified by pairs $(V, {}^\CB\CX)$, where 
\begin{itemize}
\item $V$ is a unitary rational VOA of central charge $c$ such that $\CB:=\Mod_V$ is a UMTC;  
\item $\CX$ is a UFC equipped with a unitary braided monoidal equivalence $\phi_\CX: \overline{\CB} \boxtimes \CC \to \FZ(\CX)$;  
\item ${}^\CB\CX$ is the $\CB$-enriched UFC obtained via the canonical construction from the triple $(\CB,\CX,F_\CX)$, where $F_\CX$ is the unitary braided monoidal functor defined as follows: 
$$
F_\CX: \overline{\CB} \hookrightarrow \overline{\CB} \boxtimes \CC \xrightarrow{\phi_\CX} \FZ(\CX). 
$$ 
\end{itemize}
Equivalently, for the convenience of numerical computation, 
\begin{itemize}
\item all gapped and chiral gapless edges of $(\CC,c)$ are classified by pairs $(V, A)$, where $A$ is a Lagrangian algebra in $\overline{\Mod}_V \boxtimes \CC$. 
\end{itemize}
When $V=\Cb$, the edge is gapped. In this case,  $\Mod_V=\bh$. 
\end{pthm}

\begin{rem}
Unstable edges naturally occur if we fuse a chiral gapless edge with a gapped domain wall. If we allow unstable edges in our mathematical description, we can simply replace the condition that $\CX$ is UFC by a weaker condition that $\CX$ is a unitary multi-fusion category (see also Remark\,\ref{rem:unstable}).
\end{rem}

\begin{rem}
There is a nice classification of gapless edges for abelian 2d topological orders given in \cite{ccmnpy}. It will be very interesting to compare their results with ours. Constructing new gapless edges via anyon condensations of a non-abelian bulk phase $(\CC,c)$ was considered in some special cases in \cite{bn}. Our constructions of gapless edges are more general than those obtained by condensing the bulk \cite{dmno,anyon}. 
\end{rem}

\begin{expl}
Consider a conformal embedding $V \subset A$ of unitary rational VOA's of central charge $c$, e.g.
\begin{align*}
&su(m)_n \times su(n)_m \subset su(mn)_1, \quad 
sp(2m)_n \times sp(2n)_m  \subset so(4mn)_1,  \\
&so(m)_n \times so(n)_m \subset so(mn)_1, \quad 
so(m)_4 \times su(2)_m \subset sp(2m)_1, \cdots . 
\end{align*} 
Then $A$ can be viewed as a condensable algebra in $\CB=\Mod_V$ \cite{hkl,dmno} and we have $\Mod_A\simeq \CB_A^0$. Therefore, the two topological orders $(\CB,c)$ and $(\Mod_A,c)$ can be connected by a gapped domain wall given by $\CB_A$. By topological Wick rotations, we obtain 
\begin{itemize}
\item a chiral gapless edge of $(\CB,c)$ defined by 
$(A, {}^{\Mod_A}(\CB_A))$, in which $M_{\one,\one}=A$; 
\item a chiral gapless edge of $(\Mod_A,c)$ defined by 
$(V, {}^{\CB}(\CB_A^\rev))$, in which $M_{\one,\one}=A$. 
\end{itemize}
\end{expl}

%\subsection{Remarks to the classifications of chiral gapless edges} \label{sec:breaking-cs}

\begin{rem} \label{rem:breaking-cs-2}
Recall Remark\,\ref{rem:breaking-cs-1}, for a given chiral gapless edge, the chiral symmetry is chosen to be the maximal one that is transparent on the entire 1+1D world sheet of the edge. If we choose a smaller one, i.e. $V' \subsetneq V$, then the category of boundary conditions for the same bulk CFT $Z(U)$ will be enlarged to $(\Mod_{V'})_U$. Note that $(\Mod_V)_U \simeq ((\Mod_{V'})_V^0)_U \subsetneq (\Mod_{V'})_U$. The category of topological edge excitations will be enlarged to $(\Mod_{V'})_V \boxtimes_{\Mod_V} \CX$. 
\end{rem}

It is not yet possible to list explicitly all chiral gapless edges for a given chiral 2d topological order because, for a fixed UMTC $\CB$, how many unitary rational VOA's $V$ satisfy $\Mod_V \simeq \CB$ as UMTC's is still an open question. It was conjectured that every UMTC $\CC$ (without fixing $c$) can be realized by the category of unitary modules over a unitary VOA (see for example \cite{gannon1,tew}). It is not true if we also fix $c$. More precisely, for a pair $(\CC,c)$, in general, it is not possible to find a VOA $V$ of central charge $c$ such that $\CC\simeq\Mod_V$. For example, when $\CC$ is non-chiral and $\CC\nsimeq \bh$, such $V$ does not exist because the only unitary VOA of central charge $0$ is the trivial one $V=\Cb$. In other words, the non-chiral 2d topological order $(\CC,0)$ does not have any chiral gapless edges. It has only gapped and non-chiral gappable gapless edges (recall Remark\,\ref{rem:gannon}). It is, however, possible to find $V$ of central charges of $8\Zb$ such that $\CC\simeq\Mod_V$ \cite{gannon1}. It is interesting to note that our classification result actually supports a different conjecture: 
\begin{conj}
For a UMTC $\CC$ and a central charge $c$ such that $c_\CC^{\mathrm{top}}=c\, (\mathrm{mod}\,\, 8)$, there is at least one unitary VOA $V$ of central charge $c$ such that $\Mod_V$ is a UMTC Witt equivalent to $\CC$.  
\end{conj}

\begin{rem}
Interestingly, if there are only finitely many chiral gapless edges of a given bulk phase $(\CC,c)$, an assumption which is not totally unreasonable, then our theory suggests that there are only finitely many unitary rational VOA's with central charge $c$ such that their module categories are UMTC's that are Witt equivalent to $\CC$.
\end{rem}

\void{
\begin{rem} \label{rem:gauge-free}
For example, let $(V,{}^\CB\CM)$ be a chiral gapless edge in our classification. Assume that $V_0$ is a unitary rational VOA such that there is a conformal embedding $V_0\hookrightarrow V$. In this case, $V$ can be viewed as a condensable algebra in $\CB_0:=\Mod_{V_0}$. We have $\CB\simeq (\CB_0)_V^0$ as UMTC's. We denote the left adjoint functor of the canonical embedding $\CB \simeq (\CB_0)_V^0 \hookrightarrow (\CB_0)_V$ by $G$. Then the following composed functor 
$$
\CB_0 \xrightarrow{-\otimes_{V_0} V} (\CB_0)_V \xrightarrow{G} \CB
$$
is braided oplax monoidal. Therefore, we construct a new enriched monoidal category ${}^{\CB_0}\CM$. Interesting, the pair $(V_0, {}^{\CB_0}\CM)$ describes the same set of physical observables on the world sheet of a wall as those for the pair $(V,{}^\CB\CM)$ except that the chiral symmetry is taken to be a smaller one $V_0 \subsetneq V$. Therefore, the pair $(V_0, {}^{\CB_0}\CM)$ can be viewed as a ``gauge equivalent'' way of describing the same chiral gapless edge. Our classification theory has already gotten rid of this kind of ``gauge freedoms". Besides this example, we feel that it is still possible for $F_\CX$ being only a braided oplax monoidal functor to describe some real physical situations, which should go beyond the current setup. We suspect that they might be related to gapless 2d bulk phases.  
\end{rem}
}

\begin{rem} \label{rem:wall}
A domain wall between two 2d topological orders can be realized as an edge by the folding trick. Therefore, the classification theory of the gapped and chiral gapless edges also provides that of the gapped and chiral gapless domain walls. %If $c_1=c_2$, there are only gapped walls; if $c_1\neq c_2$, then all the walls are necessarily gapless. 
\end{rem}

\subsection{Universality at RG fixed points} \label{sec:RG}

The reasoning behind our classification theory of chiral gapless edges given in Section\,\ref{sec:classification} will not be complete unless we fill the one last loophole. Note that we have not yet provided any explanation of the mysterious formula (\ref{eq:edge-wall-2}). We will do that in this subsection.

\void{
\begin{figure} 
$$
 \raisebox{-50pt}{
  \begin{picture}(100,100)
   \put(-120,0){\scalebox{0.8}{\includegraphics{pic-fusing-wall-eps-converted-to.pdf}}}
   \put(-120,0){
     \setlength{\unitlength}{.75pt}\put(0,0){
     \put(30,10)  {$ (\CC,c_1) $}
     \put(170,10) {$(\CD,c_1+c_2)$}
     \put(280,10) {$(\CE,c_1+c_2+c_3)$}
     \put(210, 110) {$(V_\CM,{}^\CA\CM)$}
     \put(160, 80)  {$(\CA,c_2)$}
     \put(170,40)   {$\CM$}
     \put(313,40)   {$\CN$}
     \put(353, 110) {$(V_\CN,{}^\CB\CN)$}
     \put(302, 80)  {$(\CB,c_3)$}
     %\put(110, 80) {$x$}
     }\setlength{\unitlength}{1pt}}
  \end{picture}}
$$
\caption{This picture illustrate the fusion of a wall $(U,\CA,\CM)$ between $(\CC,c_1)$ and $(\CD,c_1+c_2)$ with a wall $(V,\CB,\CN)$ between $(\CD,c_1+c_2)$ and $(\CE,c_1+c_2+c_3)$.  
}
\label{fig:fusing-walls}
\end{figure}
}

\begin{figure} 
$$
 \raisebox{-50pt}{
  \begin{picture}(170,150)
   \put(-50,0){\scalebox{0.65}{\includegraphics{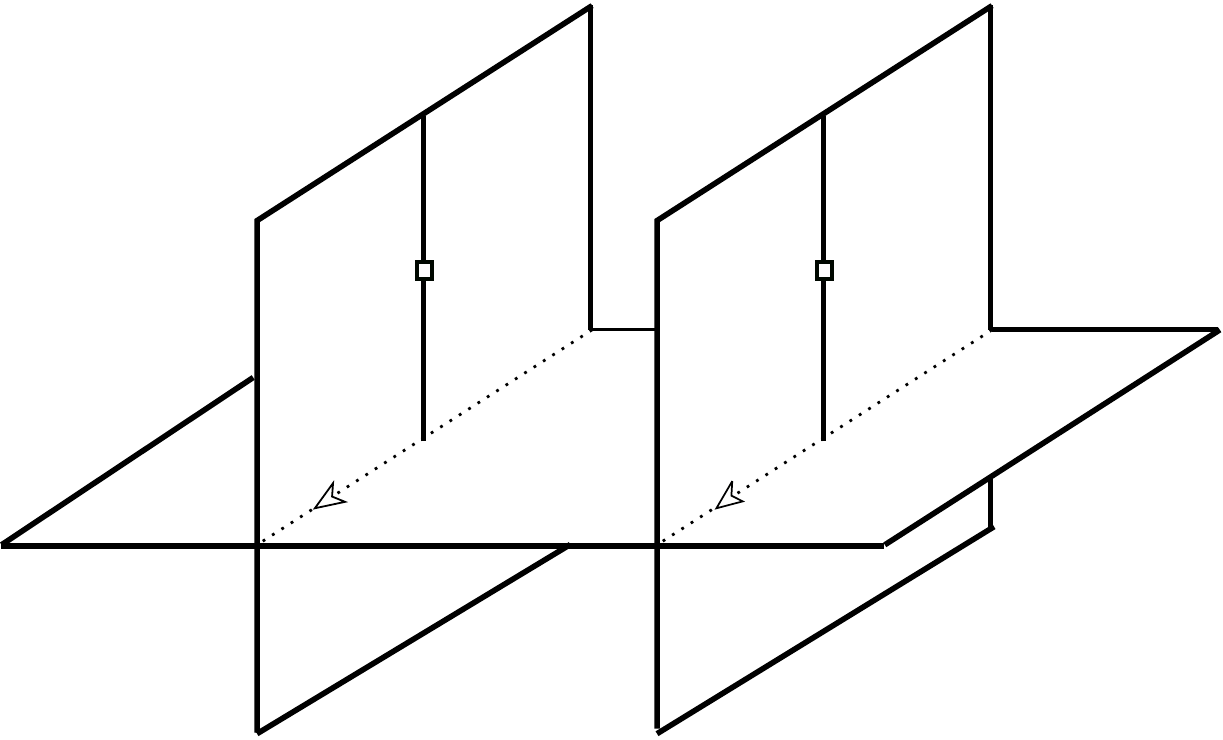}}}
   \put(-50,0){
     \setlength{\unitlength}{.75pt}\put(0,0){
   
     \put(253,165)  {\scriptsize $(V_\CB,{}^\CB\CN)$}
     \put(153,165)    {\scriptsize $(V_\CA,{}^\CA\CM)$}
     
     \put(125,90) {\scriptsize $\CM$}
     \put(115,130) {\scriptsize $(\CA,c_2)$}
     \put(227,90) {\scriptsize $\CN$}
     \put(215,130) {\scriptsize $(\CB,c_3)$}
     
     \put(110,72)  {\scriptsize $x\in \CM$}
     \put(210,72)  {\scriptsize $p\in \CN$}
     \put(110,152) {\scriptsize $y\in \CM$}
     \put(210,152) {\scriptsize $q \in \CN$}
     
     \put(26,55)  {\scriptsize $(\CC,c_1)$}
     \put(100,55) {\scriptsize $(\CD,c_1+c_2)$}
     \put(270,108) {\scriptsize $(\CE,c_1+c_2+c_3)$}
     %\put(72,190)    {$V$}
     \put(73,115)   {\scriptsize $[x,y]_\CA$}
     \put(175,115)   {\scriptsize $[p,q]_\CB$}
     
     %\put(110, 80) {$x$}
     }\setlength{\unitlength}{1pt}}
  \end{picture}}
 \quad\quad
   \raisebox{-50pt}{
  \begin{picture}(100,150)
   \put(0,0){\scalebox{0.65}{\includegraphics{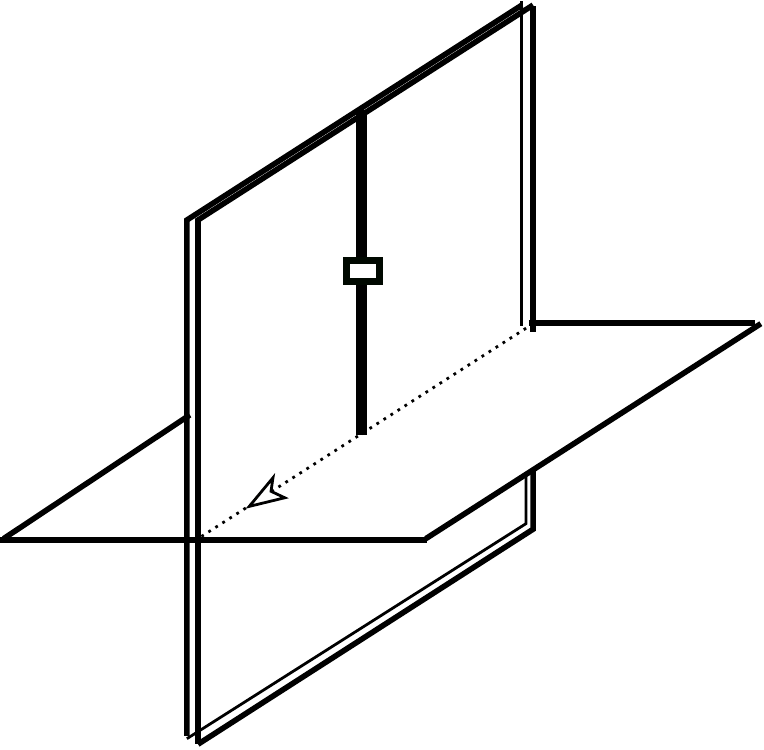}}}
   \put(0,0){
     \setlength{\unitlength}{.75pt}\put(0,0){
    
    \put(20,58)  {\scriptsize $(\CC,c_1)$}
     \put(150,113) {\scriptsize $(\CE,c_1+c_2+c_3)$}
     
     \put(100,95) {\scriptsize $\CM\boxtimes_\CD\CN$}
     \put(100,130) {\scriptsize $\CA\boxtimes\CB$}
     
     \put(95,75) {\scriptsize $xp$}
     \put(95,153) {\scriptsize $yq$}
     
     \put(53,118)   {\scriptsize $[xp,yq]$}
     \put(-5,177) {\scriptsize $(V_\CA\otimes_\Cb V_\CB, {}^{\CA\boxtimes\CB}(\CM\boxtimes_\CD\CN)$}
     
     %\put(110, 80) {$x$}
     }\setlength{\unitlength}{1pt}}
  \end{picture}}  
$$
$$
(a) \quad\quad\quad\quad\quad\quad\quad\quad\quad\quad\quad\quad\quad\quad\quad\quad \quad\quad\quad\quad
(b)
$$
\caption{The picture (a) depicts two chiral gapless domain walls $(V_\CA, {}^\CA\CM)$ and $(V_\CB, {}^\CB\CN)$. The vertical direction is the direction of time. The picture (b) depicts the new wall obtained after the fusion, where $xp:=x\boxtimes_\CD p, yp:=y\boxtimes_\CD q \in \CM\boxtimes_\CD\CN$. The arrows on the dotted lines are the orientation of the wall. It determines the order of the fusion product of wall excitations. 
}
\label{fig:fusion-walls}
\end{figure}

\medskip
Instead of explaining the mysterious formula (\ref{eq:edge-wall-2}), we would like to explain a more general formula, which describes the fusion of two chiral gapless domain walls (recall Remark\,\ref{rem:wall}). We illustrate two chiral gapless walls before the fusion in Figure\,\ref{fig:fusion-walls} (a) and after the fusion in Figure\,\ref{fig:fusion-walls} (b). More precisely, $\CA,\CB,\CC,\CD,\CE$ are UMTC's, and $\CM$ and $\CN$ are UFC's equipped with unitary braided monoidal equivalences $\phi_\CM: \overline{\CC}\boxtimes \overline{\CA} \boxtimes \CD \xrightarrow{\simeq} \FZ(\CM)$ and $\phi_\CN: \overline{\CD}\boxtimes \overline{\CB} \boxtimes \CE \xrightarrow{\simeq} \FZ(\CN)$.\footnote{Our convention is that the fictional bulk phase $\CA$ (or $\CB$) sits on the left side of the oriented wall (recall Remark\,\ref{rem:left-right-convention}).} 
The vertical direction is the direction of time. Two vertical planes depict the 1+1D world sheets (or fictional bulk phases) of two chiral gapless domain walls $(V_\CA, {}^\CA\CM)$ and $(V_\CB, {}^\CB\CN)$. Note that $V_\CA$ and $V_\CB$ have central charge $c_2$ and $c_3$, respectively. They precisely make up the difference of the chiral central charges of the two sides of the wall. Now we propose a formula for the spatial fusion of these two walls: 
\be \label{eq:fusing-walls-1}
(V_\CA, {}^\CA\CM) \boxtimes_{(\CD, c_1+c_2)} (V_\CB, {}^\CB\CN) 
= (V_\CA\otimes_\Cb V_\CB, {}^{\CA\boxtimes \CB} (\CM\boxtimes_\CD\CN)).  
\ee
Note that this formula covers the formula (\ref{eq:edge-wall-2}) as a special case.

\begin{rem}  \label{rem:unstable}
Note that $\CM\boxtimes_\CD\CN$ is in general a multi-fusion category even if both $\CM$ are $\CN$ are fusion ones. In this case, it describes an unstable domain wall, which can flow to a stable one under RG flow \cite{ai}. Since the multi-fusion categories naturally appear in such a fusion process, it is also natural to use unitary multi-fusion categories to describe the category of topological wall excitations.  
\end{rem}

Before we explain the formula (\ref{eq:fusing-walls-1}), we first recall that $M_{x,y} = [x,y]_\CB$. By Definition\,\ref{def:int-hom-univ}, $M_{x,y}$ is universal in the mathematical sense. More precisely, $[x,y]_\CB$, as a space of boundary condition changing operators, is equipped with a map $\ev: [x,y]_\CB \odot x \to y$, which specifies how the boundary condition changing is done. For any space of boundary condition changing operators, i.e. $(Q, f: Q\odot x \to y)$, where $Q$ is the space and $f$ specifies how the boundary condition changing is done, there exists a unique morphism $\underline{f}: Q \to [x,y]_\CB$ exhibiting the following diagram
\be \label{diag:univ-Mxy}
\xymatrix{
& [x,y]_\CB \odot x \ar[dr]^\ev & \\
Q \odot x \ar[rr]^f \ar[ur]^{\exists ! \underline{f} \odot 1_x} & & x
}
\ee
commutative. We believe that it is just a special case of a more general principle for physics at a RG fixed point. More precisely, we propose
\begin{quote}
{\bf Principle of Universality at RG fixed points}: A physical theory at a RG fixed point always satisfies a proper universal property in the mathematical sense. 
\end{quote}

Another example of the Principle of Universality is the boundary-bulk relation, i.e. bulk = the center of a boundary  \cite{kong-wen-zheng-2}. More precisely, the unique gapped $n+$1d bulk $B(\CX)$ of a (gapped or gapless) $n$d boundary $\CX$ satisfies the following universal property: 
\be \label{diag:univ-property-center}
\raisebox{2em}{\xymatrix{
& B(\CX) \boxtimes \CX \ar[rd]^m & \\
\CQ \boxtimes \CX \ar[rr]^g \ar[ur]^{\exists ! \, \underline{g} \boxtimes \id_\CX} & & \CX.
}}
\ee
where $B(\CX)$ is viewed as an $n$d topological order by forgetting additional structures, $\boxtimes$ is the stacking operation of two phases, and both $m$ and $g$ are morphisms between potentially anomalous $n$d (gapped or gapless) phases (introduced in \cite{kong-wen-zheng-2}) and define a unital $B(\CX)$-action and a unital $\CQ$-action on $\CX$, respectively. When $\CX$ is a gapped boundary of a 2d topological order, the 2d bulk $B(\CX)$ is precisely given by the Drinfeld center $\FZ(\CX)$ of $\CX$, which is viewed as a unitary fusion category. Mathematically, in this case, $\boxtimes$ is the Deligne tensor product; $m$ is just the tensor product functor $\otimes$ of $\CX$, and is unital and monoidal; and the unital action $g$ is also monoidal; and (\ref{diag:univ-property-center}) is nothing but the mathematical universal property of the Drinfeld center.

More examples will appear below. Note that the condition ``at RG fixed point'' is crucial. This principle 
can not be true for general QFT's or many-body condensed matter systems that are not at RG fixed point. However, for general QFT's, we believe that observables (such as those in the bulks, boundaries, walls or other defects, instantons, etc.) should factor through the universal ones.

\medskip
Now we return to the physical situation depicted in Figure\,\ref{fig:fusion-walls}. A typical example of physical observables on the 1+1D world sheet of each gapless wall before the fusion are $[x,y]_\CA$ for $x,y\in\CM$ and $[p,q]_\CB$ for $p,q\in\CN$. If we treat two vertical planes as two real bulk phases in space, then, when we fuse them, the observables $[x,y]_\CA$ and $[p,q]_\CB$ should also fuse and give $[x,y]_\CA \boxtimes [p,q]_\CB\in \CA\boxtimes \CB$ (see Figure\,\ref{fig:fusion-walls} (b)). This naive picture contradicts to both the formula (\ref{eq:fusing-walls-1}) and the Principle of Universality.

It is again a typical quantum quenching scenario. When we fuse two walls in the spatial dimensions, only thing we should expect is that the categories of wall excitations $\CM$ and $\CN$ are fused together to give $\CM\boxtimes_\CD \CN$. Indeed, the topological excitations in $\CD$ acts on those in $\CM$ from right according to $(m, d) \mapsto m\odot d$ and on those in $\CN$ from left according to $(d, n) \mapsto d\odot n$. Assume that this fusion produces a functor $F$ from $\CM \boxtimes \CN$ to the not-yet-known category of topological excitations in the final wall. We should expect that $F(m\odot d, n)\simeq F(m, d\odot n)$ for $m\in \CM,n\in\CN$. Such $F$ is called a $\CD$-balanced functor. The relative tensor product $\boxtimes_\CD: \CM \times \CN \to \CM\boxtimes_\CD \CN$, defined by $(m,n)\mapsto m\boxtimes_\CD n$, is an example of such $\CD$-balanced functors, i.e. $(m\odot d)\boxtimes_\CD n \simeq m\boxtimes_\CD (d\odot n)$, and is universal among all $\CD$-balanced functors. That is, for any $\CD$-balanced functor $G: \CM\boxtimes \CN \to \CY$, there exists a unique $\underline{G}: \CM\boxtimes_\CD\CN \to \CY$ such that the following diagram
$$
\xymatrix{
\CM \boxtimes \CN \ar[r]^{\boxtimes_\CD} \ar[rd]_G & \CM\boxtimes_\CD \CN \ar[d]^{\exists ! \, \underline{G}} \\
& \CY
}
$$
is commutative. By the Principle of Universality, we conclude that the category of topological excitations on the wall after the fusion is given by $\CM\boxtimes_\CD \CN$. It is worthwhile to remind the readers that this is precisely how two gapped walls are fused \cite{kk,fsv,anyon}. As a consequence, by \cite[Theorem\,3.3.6]{kz1} and the boundary-bulk relation (recall Theorem\,\ref{thm:kz}), the background category of the wall after the fusion has to be $\CA\boxtimes\CB$. %Therefore, the naive formula (\ref{eq:fusing-walls-1}) should be correct by (\ref{diag:univ-Mxy}) or the Principle of Universality!

%This fusion process of two walls is again a typical quantum quenching process. 
Assume that this fusion happens at $t=0$. Then two wall excitations $x\in \CM$ and $p\in\CN$ are fused into a new wall excitation $xp:=x\boxtimes_\CD p\in \CM\boxtimes_\CD \CN$. According to the discussion in Section\,\ref{sec:osvoa}, observables living on the world line support on it will be changed to the boundary CFT $M_{xp, xp}$, which is precisely given by $[xp,xp]_{\CA\boxtimes\CB}$. Similarly, topological excitations $y\in \CM$ and $q\in \CN$ is fused to $yq:=y\boxtimes_\CD q$. Therefore, the domain wall between the two boundary conditions $xp$ and $yq$ should be given by $M_{xp,yq}=[xp,yq]_{\CA\boxtimes\CB}$.

In general, the spatial fusion of instantons $[x,y]_\CA \boxtimes [p,q]_\CB$ is not invariant under RG flow\footnote{For example, in the case discussed in (\ref{pic:fusing-edge-wall}), take $x=y=\one_\CB$ and $p=q=\one_\CM$. On the one hand, we have $[\one_\CB,\one_\CB]_\CB\boxtimes [\one_\CM,\one_\CM]_\bh \simeq \one_\CB=V$. On the other hand, we have $[\one_\CM,\one_\CM]_\CB=U\neq V$ in general (see Section\,\ref{sec:general-edge}).}, which will drive it to a new RG fixed point given by $[xp,yq]_{\CA\boxtimes\CB}$. Indeed, $[x,y]_\CA \boxtimes [p,q]_\CB$ is a space of boundary condition changing operators because it is equipped with a natural way to do the ``boundary condition changing'': 
$$
([x,y]_\CA \boxtimes [p,q]_\CB) \odot (x\boxtimes_\CD p) := 
([x,y]_\CA \odot x) \boxtimes_\CD ([p,q]_\CB \odot p) \xrightarrow{\ev \boxtimes_\CD \ev}
y\boxtimes_\CD q. 
$$
Therefore, there exists a canonical morphism $g:[x,y]_\CA \boxtimes [p,q]_\CB \to [xp, yq]$ by the universal property of the internal hom (\ref{diag:univ-Mxy}). When $g$ is not an isomorphism, it means that additional operators such as excitations tunneling between two 1d walls becoming local as they getting close, and it is a manifestation of the spatial fusion anomaly and, at the same time, a mathematical description of the RG flow. The appearance of spatial fusion anomaly is due to the fact that both domain walls are anomalous 1d phases. This fact allows additional operators to become local as two 1d walls getting close. Indeed, when both of the 1d phases are anomaly-free, what live on two world sheets are modular invariant bulk CFT's. In this case, $g$ is indeed an isomorphism, or equivalently, spatial fusion anomaly vanishes. 

\medskip
In summary, we have provided an explanation of the formula (\ref{eq:fusing-walls-1}) in terms of RG flow and Principle of Universality. Therefore, the reasoning behind our main result Theorem$^{\mathrm{ph}}$\,\ref{thm:main} is complete. 

%%%%%%%%%%%%%%%%%%%%%%%%%%%%

\section{Conclusions and Outlooks} \label{sec:outlooks}

In this work, for a 2d topological order $(\CC,c)$, we have found a complete and precise mathematical description of all observables on the 1+1D world sheet of a 1d chiral gapless edge. This description also provides a mathematical classification of all chiral gapless edges as summaried in Theorem$^{\mathrm{ph}}$\,\ref{thm:main}. As a consequence, all chiral gapless edges can be obtained from topological Wick rotations.

\medskip
The significance of this work is manifold. As we will show in \cite{kz4}, the mathematical theory of chiral gapless edges immediately implies a theory of non-chiral gapless edges and a generalization of the old boundary-bulk relation \cite[Theorem\ 3.3.7.]{kz1} to include gapless edges. Besides these immediate generalizations, there are more surprising and exciting implications of this work. 
\bnu

\item It provides a new and systematic way of describing and classifying all gapless edges of 2d symmetry protected/enriched topological orders. 

\item It provides a new and systematic way of studying all 1+1D phase transitions among gapped or gapless 1d phases. 

\enu

More importantly, this work also suggests that we can study higher dimensional gapless boundaries similarly because many of the physical arguments used in this works, such as dimensional reductions, topological Wick rotations and Principle of Universality at RG fixed points, should work automatically in any dimensions. 
\bnu

\item As we mentioned in Section\,\ref{sec:gapped-to-gapless}, the main results of \cite{kong-wen-zheng-2} says that a topological order in any dimension is determined by one of its boundaries by taking center regardless if this boundary is gapped or gapless (see \cite[Remark\, 5.7]{kong-wen-zheng-2}). As a consequence, if an $n+$1d topological order has a precise mathematical description based on higher categories (see \cite{kong-wen, kong-wen-zheng-1}), so does its $n$d gapless boundary. This prediction is already highly non-trivial and rather surprising. In particular, this work can be viewed as a consequence of this prediction. 

\item This work also suggests that we should be able to construct gapless boundaries of higher dimensional topological orders via the topological Wick rotations together with additional information of local quantum symmetries such as the chiral symmetry in this work. Note that the topological Wick rotation depicted in (\ref{pic:wick-2}) automatically makes sense in all dimensions. This provides a new way to construct and describe potentially anomalous gapless phases in higher dimensions by certain enriched higher categories. Although the topological Wick rotation alone can not determine local quantum symmetries, it does provide a severe constraint on the possible local quantum symmetries. Moreover, many results can be derived without knowing local quantum symmetries. If this surprising speculation is indeed correct, it will be a very exciting progress for a systematic study of gapless phases in higher dimensions. Moreover, this work also suggests that a phase transition between two $n+$1D gapless phases can be defined by a gap-closing $n+$2D topological phase transitions via topological Wick rotations as illustrated in Figure\,\ref{fig:phase-transition}, which automatically makes sense in any dimensions. 
\enu

We summarize above observations by a correspondence between gapped and gapless phases. It will be called Gapped-Gapless Correspondence, and will serve as a guiding principle for our future studies. 
\begin{quote}
\noindent {\bf Gapped-Gapless Correspondence}: 
All $n$d gapless boundaries (or potentially anomalous $n$d gapless phases), including higher codimensional domain walls, of an $n+$1d topological order can be obtained from topological Wick rotations together with additional information of local quantum symmetries. Moreover, a phase transition between two gapless boundaries (without altering the bulk topological order) can be defined by a gap-closing topological phase transition via a topological Wick rotation (see Figure\,\ref{fig:phase-transition}). 
\end{quote}
The typical structures of local quantum symmetries for QFT's in higher dimension will be briefly discussed in \cite[Remark\ 3.1]{kz4}.

\appendix

\section{Appendix}

\subsection{Enriched monoidal categories}

We first recall the notion of a category enriched in a monoidal category \cite{Ke}. Let $\CA$ be a monoidal category with tensor unit $\one_\CA$ and tensor product $\otimes:\CA\times\CA\to\CA$.

\begin{defn} \label{defn:en-cat}
A {\em category $\CCs$ enriched in $\CA$}, or an {\it $\CA$-enriched category}, consists of a set of objects $Ob(\CC^\sharp)$, an object $\hom_{\CC^\sharp}(x,y)$ in $\CA$ for every pair $x,y\in\CC^\sharp$, a morphism $\id_x:\one_\CA\to\hom_{\CC^\sharp}(x,x)$ for every $x\in\CC^\sharp$, and a morphism $\circ:\hom_{\CC^\sharp}(y,z)\otimes\hom_{\CC^\sharp}(x,y)\to\hom_{\CC^\sharp}(x,z)$ for $x,y,z\in\CC^\sharp$, such that the following diagrams commute for $x,y,z,w\in\CC^\sharp$:
\be \label{diag:right-unit}
\xymatrix@!C=15ex{
  & \hom_{\CC^\sharp}(x,y)\otimes\hom_{\CC^\sharp}(x,x) \ar[rd]^\circ \\
  \hom_{\CC^\sharp}(x,y) \ar[rr]^1 \ar[ru]^{1\otimes\id_x} && \hom_{\CC^\sharp}(x,y), \\
}
\ee
\be \label{diag:left-unit}
\xymatrix@!C=15ex{
  & \hom_{\CC^\sharp}(y,y)\otimes\hom_{\CC^\sharp}(x,y) \ar[rd]^\circ \\
  \hom_{\CC^\sharp}(x,y) \ar[rr]^1 \ar[ru]^{\id_y\otimes 1} && \hom_{\CC^\sharp}(x,y), \\
}
\ee
\be \label{ax:asso-circ}
\xymatrix{
  \hom_{\CC^\sharp}(z,w)\otimes\hom_{\CC^\sharp}(y,z)\otimes\hom_{\CC^\sharp}(x,y) \ar[r]^-{1\otimes\circ} \ar[d]_{\circ\otimes1} & \hom_{\CC^\sharp}(z,w)\otimes\hom_{\CC^\sharp}(x,z) \ar[d]^\circ \\
  \hom_{\CC^\sharp}(y,w)\otimes\hom_{\CC^\sharp}(x,y) \ar[r]^-\circ & \hom_{\CC^\sharp}(x,w). \\
}
\ee
\end{defn}

\begin{rem}
We distinguish the notation $\id_x$ and $1_x$, where the former one is the identity morphism in an enriched category, and $1_x$ is the identity morphism in an ordinary category. 
\end{rem}

Note that an ordinary category is a category enriched in the category of sets. We will call an element $f\in \hom_\CA(\one, \hom_{\CC^\sharp}(x,y))$ a morphism from $x$ to $y$, denoted by $f: x\to y$. A morphism $f: x\to y$ is called an isomorphism (or invertible) if there is a morphism $g: y\to x$ such that $g\circ f=\id_x$ and $f\circ g=\id_y$. 

\medskip
Now we define the {\it underlying category} of $\CCs$, denoted by $\CC$.  The category $\CC$ consists of the same objects as those in $\CC^\sharp$ and 
\begin{itemize}
\item $\hom_\CC(x,y):=\hom_\CA(\one, \hom_\CCs(x,y))$ for $x,y\in Ob(\CCs)$;
\item the identity morphism $1_x \in \hom_\CC(x,x)$ is just $\id_x: \one \to \hom_\CCs(x,x)$; 
\item the composition map $\hom_\CC(y,z) \times \hom_\CC(x,y) \xrightarrow{\circ} \hom_\CC(x,z)$ is defined by the following composed map 
\begin{align}
\hom_\CA(\one, \hom_\CCs(y,z)) \times \hom_\CA(\one, \hom_\CCs(x,y)) &\xrightarrow{\otimes} \hom_\CA(\one, \hom_\CCs(y,z) \otimes \hom_\CCs(x,y)) \nn 
&\to \hom_\CA(\one, \hom_\CCs(x,y)). \nonumber
\end{align}
\end{itemize}

\begin{expl} ({\bf Canonical Construction}) \label{exam:SC}
Let $\CM$ be a left $\CA$-module category which has internal homs in $\CA$, i.e. the functor $-\odot x:\CA\to\CM$ admits a right adjoint $[x,-]:\CM\to\CA$ for every $x\in\CM$. Then $\CM$ can be promoted to an $\CA$-enriched category $\CM^\sharp$ which has the same objects as $\CM$ and $\hom_{\CM^\sharp}(x,y)=[x,y]$. The composition $\circ$ is given by the canonical morphism $[y,z]\otimes[x,y]\to[x,z]$ and $\id_x$ is given by the canonical morphism $\one\to[x,x]$. We also denote this enriched category by ${}^\CA\CM$. Note that its underlying category is just $\CM$ because $\hom_\CM(x,y)\simeq \hom_\CA(\one, [x,y])$. 
\end{expl}

\begin{expl}
If $\CA$ is rigid, then it can be promoted to an $\CA$-enriched category $\CA^\sharp$. In this case, $\hom_{\CA^\sharp}(x,y) = y\otimes x^\ast$ where $x^\ast$ is the left dual of $x$, $\hom_{\CA^\sharp}(\one,\one) = \one$, $\circ:(z\otimes y^\ast)\otimes(y\otimes x^\ast)\to z\otimes x^\ast$ is induced by the counit map $v_y:y^\ast\otimes y\to\one$, and $\id_x$ is given by the unit map $u_x: \one\to x\otimes x^\ast$.
\end{expl}

\begin{defn}
An {\em enriched functor} $F:\CC^\sharp\to\CD^\sharp$ between $\CA$-enriched categories consists of a map $F:Ob(\CC^\sharp)\to Ob(\CD^\sharp)$ and a morphism $F:\hom_{\CC^\sharp}(x,y)\to\hom_{\CD^\sharp}(F(x),F(y))$ for every pair $x,y\in\CC^\sharp$ such that the following diagrams commute for $x,y,z\in\CC^\sharp$:
$$
\xymatrix{
  & \one \ar[ld]_{\id_x} \ar[rd]^{\id_{F(x)}} \\
  \hom_{\CC^\sharp}(x,x) \ar[rr]^-F && \hom_{\CD^\sharp}(F(x),F(x)), \\
}
$$
\be \label{diag:fun-composition}
\xymatrix{
  \hom_{\CC^\sharp}(y,z)\otimes\hom_{\CC^\sharp}(x,y) \ar[r]^-\circ \ar[d]_{F\otimes F} & \hom_{\CC^\sharp}(x,z) \ar[d]^F \\
  \hom_{\CD^\sharp}(F(y),F(z))\otimes\hom_{\CD^\sharp}(F(x),F(y)) \ar[r]^-\circ & \hom_{\CD^\sharp}(F(x),F(z)). \\
}
\ee
\end{defn}
It is clear that the composition of two enriched functor is again an enriched functor. The enriched functor $F: \CCs \to \CDs$ naturally induces an ordinary functor $F: \CC \to \CD$. 

\begin{expl}
Let ${}^\CA\CM$ and ${}^\CA\CN$ be two $\CA$-enriched categories obtained from the canonical construction in Example\,\ref{exam:SC}. Then an $\CA$-module functor $F: \CM \to \CN$ naturally defines an enriched functor $F: \CMs \to \CNs$. 
\end{expl}

\begin{defn}
An {\em enriched natural transformation} $\xi:F\to G$ between two enriched functors $F,G:\CC^\sharp\to\CD^\sharp$ consists of a morphism $\xi_x: F(x)\to G(x)$ for every $x\in \CC$ such that the following diagram commutes for $x,y\in\CC^\sharp$:
\be \label{eq:en-natural-transformation}
\xymatrix@!C=30ex{
  \hom_{\CC^\sharp}(x,y) \ar[r]^-{G} \ar[d]_{F} & \hom_{\CD^\sharp}(G(x),G(y)) \ar[d]^{-\circ\xi_x} \\
  \hom_{\CD^\sharp}(F(x),F(y)) \ar[r]^-{\xi_y\circ-} & \hom_{\CD^\sharp}(F(x),G(y)). \\
}
\ee
An enriched natural transformation $\xi$ is called an enriched natural isomorphism if each $\xi_x$ is an isomorphism. 
\end{defn}
%Note that the composition of two enriched natural transformations $\xi: F\to G$ and $\eta: G \to H$ is defined by $(\eta\circ \xi)_x=\eta_x \circ \xi_x: \one \to \hom_{\CD^\sharp}(F(x), H(x))$. 

\medskip
Now we assume $\CA$ is a braided monoidal category equipped with braiding $c_{x,y}:x\otimes y\to y\otimes x$ for $x,y\in \CA$.
Let $\CC^\sharp,\CD^\sharp$ be $\CA$-enriched categories. The {\em Cartesian product} $\CC^\sharp\times\CD^\sharp$  is an $\CA$-enriched category defined as follows:
\begin{itemize}
  \item $Ob(\CC^\sharp\times\CD^\sharp)=Ob(\CC^\sharp)\times Ob(\CD^\sharp)$;
  \item $\hom_{\CC^\sharp\times\CD^\sharp}((x,y),(x',y')) = \hom_{\CC^\sharp}(x,x')\otimes\hom_{\CD^\sharp}(y,y')$;
  \item the composition
$$
\circ: \hom_{\CC^\sharp\times\CD^\sharp}((x',y'),(x'',y'')) \otimes \hom_{\CC^\sharp\times\CD^\sharp} 
((x,y),(x',y')) \to \hom_{\CC^\sharp\times\CD^\sharp}((x,y),(x'',y''))
$$
is given by 
\begin{align}
&\hom_{\CC^\sharp}(x',x'')\otimes\hom_{\CD^\sharp}(y',y'')\otimes\hom_{\CC^\sharp}(x,x')\otimes\hom_{\CD^\sharp}(y,y') \nn
&\hspace{1cm} \xrightarrow{1\otimes c^{-1}\otimes1}  \hom_{\CC^\sharp}(x',x'')\otimes\hom_{\CC^\sharp}(x,x')\otimes\hom_{\CD^\sharp}(y',y'')\otimes\hom_{\CD^\sharp}(y,y') \\
&\hspace{1cm} \xrightarrow{~~\circ\otimes\circ~~}  \hom_{\CC^\sharp}(x,x'')\otimes\hom_{\CD^\sharp}(y,y''). \nonumber
\end{align} 
\end{itemize}

\begin{defn} \label{def:emc}
An {\em $\CA$-enriched monoidal category} consists of a category $\CC^\sharp$ enriched in $\CA$, an object $\one_{\CC^\sharp}$, an enriched functor $\otimes:\CC^\sharp\times\CC^\sharp\to \CC^\sharp$, and enriched natural isomorphisms 
$$
\lambda:\one_{\CC^\sharp}\otimes- \to \id_{\CC^\sharp}, \quad \rho: -\otimes\one_{\CC^\sharp} \to \id_{\CC^\sharp}, \quad
\alpha: \otimes\circ(\otimes\times\id_{\CC^\sharp}) \to \otimes\circ(\id_{\CC^\sharp}\times\otimes)
$$ 
such that the underlying category $\CC$, together with $\otimes$, $\lambda, \rho, \alpha$, defines a monoidal category. 
\end{defn}

\begin{rem}
An enriched monoidal category is {\em strict} if $\lambda,\rho,\alpha$ are the identity natural transformations \cite[Definition 2.1]{MP}. In the strict case, an $\CA$-enriched monoidal category defined here is equivalent to ``an $\overline{\CA}$-enriched monoidal category'' defined in \cite{MP}. 
\end{rem}

\end{document}